\documentclass[aps,prb,superscriptaddress,amsmath,amsfonts,showpacs]{revtex4}

\usepackage{graphicx}
\usepackage{amssymb,amsmath}

\begin{document}



\title{Non-Abelian Topological Orders and Majorana Fermions in
Spin-Singlet Superconductors} 


\author{Masatoshi Sato}
\affiliation{The Institute for Solid State Physics, The University of
Tokyo, Kashiwanoha 5-1-5, Kashiwa-shi, Chiba 277-8581, Japan}
\author{Yoshiro Takahashi}
\affiliation{Department of Physics, Kyoto University, Kyoto 606-8502, Japan}
\author{Satoshi Fujimoto}
\affiliation{Department of Physics, Kyoto University, Kyoto 606-8502, Japan}



\date{\today}

\begin{abstract}
The non-Abelian topological order for superconductors is characterized
 by the existence of zero-energy Majorana fermions
in edges of systems and in a vortex of a macroscopic condensate, which obey
the non-Abelian statistics.
This paper is devoted to an extensive study on
the non-Abelian topological phase of 
spin-singlet superconductors with the Rashba spin-orbit interaction
proposed in our previous letter [M. Sato, Y. Takahashi, and S. Fujimoto, Phys. Rev. Lett. 103, 020401 (2009)].
We mainly consider the $s$-wave pairing state and the $d+id$ pairing state.
In the case of $d+id$-wave pairing, Majorana fermions appear in almost all parameter 
regions of the mixed state under
an applied magnetic field, provided that the Fermi level crosses $k$-points in the vicinity of 
the $\Gamma$ point or the M point in the Brillouin zone,
while in the case of $s$-wave pairing, a strong magnetic field, the Zeeman energy of which is larger than
the superconducting gap is required to realize the topological phase.
We clarify that Majorana fermions in Rashba spin-singlet superconductors
are much more stable than those realized in spin-triplet $p+ip$
superconductors in certain parameter regions.
We also investigate the topological number which ensures the topological stability of the phase in detail.
Furthermore, as a byproduct, we found that topological order is also
realized in conventional spin (or charge) density wave states with 
the Rashba spin-orbit interaction, for which massless Dirac fermions appear in the edge of the systems
and charge fractionalization occurs.
\end{abstract}



\maketitle

\section{Introduction}

Topological states of condensed matter systems are characterized by a bulk topological number such as the Chern number (or the TKNN number) which represents a topologically non-trivial structure of the many-body Hilbert space.
In such phases, topologically protected surface states and fractionalized quasiparticles, e.g.
anyons, appear.\cite{wen,wen2,moore,nayak,read,ivanov,stone1,stern2,stone2,SRFL08,Volovik01}
In particular, when topological order is realized in a certain class of superconductors, 
this topological phase supports the existence of
chiral Majorana edge modes and a Majorana fermion in a vortex 
core.\cite{moore,nayak,read,ivanov,stone1,stern2,stone2}
Vortices with Majorana fermion modes are neither fermions nor bosons, but
non-Abelian anyons, obeying the non-Abelian statistics for which
the exchange operations of particles are not commutative.\cite{moore,nayak,read,ivanov,stone1,stern2}
Because of this remarkable feature, 
a vortex with a Majorana fermion may be utilized as a decoherence-free qubit,
and plays an important role for the realization of fault-tolerant topological quantum computation.\cite{freedman,kitaev2,das}
The state with non-Abelian anyons, which is called the non-Abelian topological phase, has been discussed 
to be realized
in the fractional quantum Hall effect state with $\nu=5/2$ and
$12/5$.\cite{moore,nayak,read,rezayi}
It has been also known that spin-triplet superconductors such as  chiral $p+ip$
superconductors,\cite{ivanov,stone1,kopnin,tewari2,ZTLDS,machida,cooper} and noncentrosymmetric
(NCS) $p$-wave superconductors with broken time-reversal symmetry,\cite{SF08} possess a zero energy Majorana mode, and realizes a non-Abelian
topological phase.
In general, fully-gapped spin-triplet superconductors support non-Abelian
anyons if the number of the connected Fermi surfaces are odd, in the case without time-reversal symmetry.\cite{Sato10}
For the spin-singlet $s$-wave superconducting state, it was pointed out by Fu and Kane that non-Abelian anyons 
are realized in the proximity with a topological insulator.\cite{FK08}
Also, 
the non-Abelian anyons in the $s$-wave pairing state was discussed before in the context of Axion strings in cosmological systems.\cite{Sato03,com1}

Recently, the present authors proposed 
another scenario of a non-Abelian topological phase in NCS $s$-wave
superfluids or superconductors. 
We pointed out that in the presence of the Rashba spin-orbit (SO)
interaction, $s$-wave superconducting states show a
transition to the non-Abelian topological phase with non-zero Chern
number, under an applied strong Zeeman magnetic field.\cite{STF09}. 
Also, independently, it was proposed by Sau {\it et al.} that such systems can be realized in heterostructure semiconductor devices.\cite{SAU}  The idea was subsequently generalized by Alicea.\cite{ALICEA}

In this paper, we explore extensively properties of the non-Abelian topological phase realized in NCS
spin-singlet superconductors with the Rashba SO
interaction, which was considered in our previous letter.\cite{STF09}
There are two main purposes.
The first one is to present the detail analysis of chiral Majorana edge states 
and a Majorana fermion mode in a vortex core
in the case of the NCS $s$-wave superconductor,
and the calculation of the topological number, which are omitted in ref.\cite{STF09}.
The topological order for time-reversal symmetry broken systems in two
dimensions is characterized by the first Chern number.
We present a formulation for the calculation of the Chern number with the use of
a winding number, which makes the estimation of the topological number easier.
Using this formulation, we explore the non-Abelian topological order realized
in NCS spin-singlet superconductors.
 We, furthermore, analyze the vortex core state by solving the
 Bogoliubov-de-Gennes (BdG) equation. We obtain the zero energy Majorana fermion solution when
the Zeeman energy due to an applied magnetic field is larger than 
the superconducting gap $\Delta$. 
We also discuss that, in some parameter regions, the Majorana fermion in NCS spin-singlet superconductors
is remarkably stable compared to that in chiral $p+ip$ superconductors.
This feature is crucially important for the application to the topological quantum computation.
The superior stability of Majorana mode in a vortex core of NCS spin-singlet superconductors
stems from the fact that when the Zeeman energy $\mu_{\rm B}H_z$ satisfies the condition $\Delta>\mu_{\rm B}H_z-\Delta>0$, the Majorana fermion is mainly formed by the superposition of 
quasiparticles in the vicinity of the $\Gamma$ point (or the M point) with the Fermi momentum $k_F\sim 0$
(or $(\pi,\pi)$), in the long-distance asymptotic regime sufficiently far from
the center of the vortex core.
Because of this property,
the excitation energy $E_0$ in a vortex core which separates
the zero energy Majorana mode and the first excited state is much larger
than a typical energy scale of the Andreev bound state of vortex cores $\sim \Delta^2/E_F$.
Furthermore, the vanishing Fermi momentum for a Majorana fermion implies that
decoherence due to quantum oscillations of quasiparticle energy with a period $\sim 1/k_F$ raised by inter-vortex tunneling,\cite{cheng} which may be an obstruction for
the implementation of the topological quantum computation, is substantially suppressed.

The second purpose of this paper is to extend the scenario of the non-Abelian topological order 
for the case of $s$-wave pairing state to other spin-singlet pairing states. In particular, we consider the cases of $d+id$-wave pairing,
for which there is a full-gap in the energy spectrum, ensuring the nonzero 
 Chern number.
It is demonstrated that in the $d+id$-wave pairing state with the Rashba SO interaction,
when the Fermi level crosses $k$-points in the vicinity of the $\Gamma$ point or the M point in the Brillouin zone,
the non-Abelian topological order, which supports the existence of chiral Majorana edge states
and a Majorana fermion mode in a vortex core,  appears under an applied magnetic field.
In contrast to the case of $s$-wave pairing considered in ref.\cite{STF09,SAU,ALICEA},
for which Majorana fermions appear only when there is
Zeeman splitting larger than the superconducting gap, a small magnetic field 
larger than the lower critical field suffices
for the realization of the non-Abelian topological order in the $d+id$-wave pairing state.
Thus, it may be easier to realize Majorana fermions in
the NCS $d+id$-wave superconductor than in the NCS $s$-wave superconductor.

Furthermore, we consider another direction of the extension of the scenario for the non-Abelian topological order. Our results for Rashba $s$-wave superconductors imply that the topological order is also
realizable in the conventional spin density wave (SDW) state or the charge density wave (CDW) state
with the Rashba SO interaction.
We demonstrate that in these density wave states, the Abelian topological order appear under applied magnetic fields, leading to the existence of gapless edge states described by the Dirac fermion,
which is analogous to the surface states of the topological insulator.
We also discuss the scenario of charge fractionalization in the topological density wave states.

The organization of this paper is as follows. From Sec.\ref{sec:model} to Sec.\ref{sec:gapclosing}, we introduce the model for superconductors with the Rashba SO interaction
in two dimensions, upon which our analysis is focused, 
and, as a first step of our analysis, 
classify the parameter regions of the model, in which different topological phases may be realized.
In Sec.\ref{sec:duality}, we explain the duality relation which holds for our model Hamiltonian. 
This duality relation
was utilized for the analysis of topological properties in ref.\cite{STF09}.
In the most part of this paper, we do not use the duality relation, but instead, confirm the argument based on it developed in ref.\cite{STF09}
by adopting a more direct approach to this issue.
In Sec.\ref{sec:topologicalnumber},
we analyze and discuss the topological number characterizing the non-Abelian topological phases realized in NCS spin-singlet superconductors. 
In particular, we prove the relation between the Chern number and the winding number, which is
useful for the investigation of the topological order.
Using this relation, we elucidate the general condition for the realization of the non-Abelian topological order.
On the basis of the analysis of the topological number, we obtain the phase diagram of the topological order
for spin-singlet NCS superconductors.
We also give some physical arguments on the origin of the topological order in spin-singlet NCS superconductors.
In Sec.\ref{sec:edge}, the numerical results for chiral Majorana edge modes are presented
for the cases of $s$-wave pairing and $d+id$-wave pairing. 
In Sec.\ref{sec:vortex}, we consider an approximated but analytical solution of 
the BdG equation for Majorana zero energy modes in vortex cores. 
We, also, discuss the superior stability of the Majorana mode in
NCS spin-singlet superconductors compared to that in chiral $p+ip$ superconductors.
In Sec.\ref{sec:densitywave}, topological density wave states in which
gapless Dirac fermions on the edge of systems appear and charge fractionalization occurs
are considered.
In Sec.\ref{sec:discussion}, we give a summary of our results, and also discuss
possible realization of the NCS spin-singlet superconductors with the non-Abelian topological order
in real systems.

Some technical details are presented in Appendices.
In Appendices \ref{appendix:a} and \ref{appendix:b}, we derive useful formulas for
the Chern numbers and the winding numbers, which were used for the discussion
on the topological number in Sec.\ref{sec:topologicalnumber}. 
Supplementary discussions related to the topological argument given 
in Sec.\ref{sec:topologicalnumber} are presented in
Appendix \ref{sec:chirality0}.
The details of the derivation of the BdG equation for a singlet vortex are given in
Appendix \ref{appendix:d}.
In Appendix \ref{appendix:c}, we discuss another mechanism of
non-Abelian anyons in spin-singlet superconductors, which was first
discussed in ref.\cite{Sato03}, where non-Abelian
anyons are realized in time-reversal invariant $s$-wave
superconducting state without a Zeeman magnetic field. This discussion is relevant to
the non-Abelian topological order realized in an interface between an $s$-wave superconductor
and a time-reversal invariant topological insulator proposed by Fu and Kane.\cite{FK08}

\section{Model}
\label{sec:model}

In this paper, we consider spin-singlet superconductors with the Rashba
SO interaction\cite{rash} in two dimensions. 
For concreteness, we define our model in the square lattice, while the
following argument does not rely on the particular choice of the crystal
structure. It is also noted that our analysis and results are also generalized straightforwardly to
other type of anti-symmetric SO interactions raised by
the lack of inversion center of systems.
The Hamiltonian is given by
\begin{eqnarray}
{\mathcal H}&=&\sum_{{\bm k},\sigma}\varepsilon({\bm k})
 c_{{\bm k}\sigma}^{\dagger}c_{{\bm k}\sigma}
-\mu_{\rm B}H_z \sum_{{\bm k},\sigma,\sigma'}(\sigma_z)_{\sigma\sigma'}
 c_{{\bm k}\sigma}^{\dagger}c_{{\bm k}\sigma'}
+\alpha\sum_{{\bm k},\sigma,\sigma'}{\bm {\mathcal L}}_0({\bm k})
\cdot{\bm \sigma}_{\sigma\sigma'}c_{{\bm k}\sigma}^{\dagger}c_{{\bm k}\sigma'} 
\nonumber\\
&+&\frac{1}{2}\sum_{{\bm k},\sigma,\sigma'}
\Delta_{\sigma\sigma'}({\bm k})c_{{\bm k}\sigma}^{\dagger}
c_{-{\bm k}\sigma'}^{\dagger}
+\frac{1}{2}\sum_{{\bm k},\sigma,\sigma'}
\Delta_{\sigma'\sigma}^*({\bm k})c_{-{\bm k}\sigma}c_{{\bm k}\sigma'}, 
\label{eq:Hamiltonian}
\end{eqnarray}
where $c_{{\bm k}\sigma}^{\dagger}$ ($c_{{\bm k}\sigma}$) is a creation (an
annihilation) operator for an electron with momentum ${\bm
k}=(k_x,k_y)$, spin $\sigma$.
The energy band dispersion is $\varepsilon({\bm k})=-2t(\cos k_x+\cos
k_y)-\mu$ with the hopping parameter $t$ and the chemical potential
$\mu$, and the Rashba SO coupling is $\alpha{\bm {\mathcal L}}_0({\bm
k})=\alpha(\sin k_y, -\sin k_x)$ $(\alpha>0)$.
We also introduce the Zeeman coupling $-\mu_{\rm B}H_z\sum_{\bm k,\sigma\sigma'}
(\sigma_z)_{\sigma\sigma'}c_{{\bm k}\sigma}^{\dagger} c_{{\bm k}\sigma'}$ 
in the Hamiltonian.

For the spin-singlet superconductors, 
the gap function $\Delta_{\sigma\sigma'}({\bm k})$ is written as 
\begin{eqnarray}
\Delta_{\sigma\sigma'}({\bm k})=i\Delta({\bm k})(\sigma_y)_{\sigma\sigma'} 
\end{eqnarray}
with the $y$-component of the Pauli matrices $\sigma_i$ $(i=x,y,z)$.
In the following, we assume two different full-gapped spin-singlet
superconductors: 
the first one is the $s$-wave pairing, $\Delta({\bm k})=\Delta_s$, and
the other is
the $d+id$-wave pairing. For the $d+id$-pairing, we consider two possible
realization on the lattice, $\Delta({\bm k})=\Delta^{(1)}_d(\cos k_y-\cos k_x)
+i\Delta_d^{(2)}\sin k_x\sin k_y$, or $\Delta({\bm k})=\Delta^{(1)}_d(\sin^2
k_x-\sin^2 k_y) +i\Delta_d^{(2)}\sin k_x\sin k_y$.  
The amplitudes $\Delta_s$ and $\Delta_d^{(i)}$ $(i=1,2)$ are chosen as
real and positive.
The second type of the $d+id$-wave pairing includes higher harmonic
contributions, which may arise depending on detailed structures of
electronic bands and the pairing interactions. 
We use these two types of the $d+id$-wave gap to clarify that, although both of them
support the non-Abelian topological order, the precise condition for
the non-Abelian phase slightly depends on the detail of the gap structure.

For a noncentrosymmetric superconductor, the parity mixing of
the gap function generally occurs.\cite{Edelstein,Gorkov,Frigeri,fuji4,fuji3}
Therefore, in addition to the spin-singlet component of the gap function,
the spin-triplet one is induced generally. 
However, if the spin-singlet amplitude dominates the
gap function, the topological nature is not affected by the
spin-triplet one.
We neglect the spin-triplet component 
in the following.

In the following, we mainly consider the case that SO interaction is much larger than the Zeeman energy;
i.e. $\alpha |\bm{\mathcal{L}}_0(\bm{k})| \gg \mu_{\rm B}H_z$, which is an important condition for the stability
of the superconducting state against the Pauli depairing effect due to the magnetic fields.

\section{gap closing condition}
\label{sec:gapclosing}

In general, continuous topological phase transitions between topologically distinct phases
occur only when the energy gap of the bulk spectrum closes.
Thus, to identify parameter regions for which different topological phases are realized, we first examine the bulk spectrum of the system. To obtain the bulk
spectrum, we rewrite the Hamiltonian as
\begin{eqnarray}
{\mathcal H}=\frac{1}{2}\sum_{{\bm k},\sigma,\sigma'}
\left(
\begin{array}{cc}
c^{\dagger}_{{\bm k}\sigma}, &c_{-{\bm k}\sigma} \\
\end{array}
\right) 
{\cal H}({\bm k})
\left(
\begin{array}{c}
c_{{\bm k}\sigma'} \\
c^{\dagger}_{-{\bm k}\sigma'}
\end{array}
\right),
\end{eqnarray}
where the BdG Hamiltonian ${\mathcal H}({\bm k})$ is
given by
\begin{eqnarray}
{\mathcal H}({\bm k})=\left(
\begin{array}{cc}
\varepsilon({\bm k})-\mu_{\rm B}H_z\sigma_z+\alpha{\bm {\mathcal L}}_0({\bm
 k})\cdot{\bm \sigma} &  i\Delta({\bm k})\sigma_y\\
-i\Delta({\bm k})^*\sigma_y & -\varepsilon({\bm k})+\mu_{\rm
 B}H_z\sigma_z+\alpha{\bm {\mathcal L}}_0({\bm k})\cdot{\bm \sigma}^*
\end{array}
\right). 
\label{eq:BdG}
\end{eqnarray}
Diagonalizing the BdG Hamiltonian, we find 
\begin{eqnarray}
E({\bm k})=\sqrt{\varepsilon({\bm k})^2+\alpha^2{\bm {\mathcal L}}_0({\bm
 k})^2+\mu_{\rm B}^2H_z^2+|\Delta({\bm k})|^2
\pm2\sqrt{\varepsilon({\bm k})^2\alpha^2{\bm {\mathcal L}}_0({\bm k})^2
+(\varepsilon({\bm k})^2+|\Delta({\bm k})|^2)\mu_{\rm B}H_z^2}}. 
\end{eqnarray}

In our model, the gap of the system closes only when the following
condition is satisfied, 
\begin{eqnarray}
\varepsilon({\bm k})^2+\alpha^2{\bm {\mathcal L}}_0({\bm k})^2
+\mu_{\rm B}^2H_z^2+|\Delta({\bm k})|^2=
2\sqrt{\varepsilon({\bm k})^2\alpha^2{\bm {\mathcal L}}_0({\bm
k})^2+(\varepsilon({\bm k})^2+|\Delta({\bm k})|^2)\mu_{\rm B}^2H_z^2}.
\end{eqnarray}
From a straightforward calculation\cite{Sato06}, it is found that this
condition is equivalent to 
\begin{eqnarray}
\varepsilon({\bm k})^2+|\Delta({\bm k})|^2
=\mu_{\rm  B}^2H_z^2+\alpha^2{\bm {\mathcal L}}_0({\bm k})^2,
\quad
|\Delta({\bm k})|\alpha{\bm {\mathcal L}}_0({\bm k})=0. 
\label{eq:gapclosing}
\end{eqnarray}
We examine the gap closing condition using (\ref{eq:gapclosing})
for the $s$-wave pairing state and the $d+id$-wave pairing state in the following.

\subsection{$s$-wave Rashba superconductor}
\label{sec:gapclosingd+id1}

For the $s$-wave pairing, the second equation in (\ref{eq:gapclosing})
is met only when ${\bm {\mathcal L}}_0({\bm k})=0$.
Therefore, the gap closes at ${\bm k}=(0,0), (0,\pi), (\pi, 0), (\pi,\pi)$. 
Substituting those values into the first equation in
(\ref{eq:gapclosing}), we have three different gap closing conditions.
( In the square lattice, the condition at ${\bm k}=(\pi,0)$ and that
at ${\bm k}=(0,\pi)$ are the same, so we have only three conditions.)
\begin{eqnarray}
(4t+\mu)^2+\Delta_s^2=(\mu_{\rm B}H_z)^2,
\quad
\mu^2+\Delta_s^2=(\mu_{\rm B}H_z)^2,
\quad 
(4t-\mu)^2+\Delta_s^2=(\mu_{\rm B}H_z)^2.
\label{eq:sgapclosing}
\end{eqnarray}
When one of these equations (\ref{eq:sgapclosing}) is satisfied, the energy gap closes.
From these conditions, we find that there are at least 7 regions of parameter space, which
may be topologically distinct, as shown in Fig \ref{fig:phase}.
We will explore the topological numbers associated with these different regions 
and classify the topological phases of our system in Sec.\ref{sec:topologicalnumber}.

\subsection{$d+id$-wave Rashba superconductor}

\subsubsection{Case of $\Delta({\bm
   k})=\Delta_d^{(1)}(\cos k_y-\cos k_x)+i\Delta_d^{(2)}\sin k_x \sin k_y$}

For this $d+id$-wave gap,
the second equation in
(\ref{eq:gapclosing}) is met 
either when ${\bm {\mathcal L}}_0({\bm k})=0$ or 
when $\Delta({\bm k})=0$.
This condition is satisfied
at ${\bm k}=(0,0), (\pi,0), (0,\pi), (\pi,\pi)$.
Substituting these ${\bm k}$s' in the first equation in
(\ref{eq:gapclosing}), we have 
\begin{eqnarray}
(4t+\mu)^2=(\mu_{\rm B}H_z)^2, 
\quad 
\mu^2+4(\Delta_d^{(1)})^2=(\mu_{\rm B}H_z)^2,
\quad
(4t-\mu)^2=(\mu_{\rm B}H_z)^2.
\label{eq:dgapclosing2}
\end{eqnarray}

The conditions obtained here are very similar to those for
$s$-wave pairing. 
However, there is an important difference between them.
The first and the last equations of the gap closing condition (\ref{eq:dgapclosing2}) 
do not depend on the amplitude of the pairing gap. 
Because of this feature, even for a relatively weak Zeeman field 
where the orbital depairing effect is negligible, the gap can close, and the topological phase transition occurs, provided that the chemical potential $\mu$ is properly tuned as $\mu\sim \pm 4t$.
As will be seen later, this point is crucially important for
the feasibility of the realization of the non-Abelian topological order
in the $d+id$-wave pairing case, compared to the $s$--wave pairing state.

\subsubsection{Case of $\Delta({\bm
   k})=\Delta_d^{(1)}(\sin^2k_x-\sin^2k_y)+i\Delta_d^{(2)}\sin k_x \sin k_y$}

As in the previous case,
the second equation in (\ref{eq:gapclosing})
is met either when ${\bm {\mathcal L}}_0({\bm k})=0$ or 
when $\Delta({\bm k})=0$.
Accidently, both of them are satisfied at the same momenta ${\bm
k}=(0,0), (0,\pi), (\pi, 0), (\pi,\pi)$. 
Substituting those values into the first equation in
(\ref{eq:gapclosing}), we have
\begin{eqnarray}
(4t+\mu)^2=(\mu_{\rm B}H_z)^2,
\quad
\mu^2=(\mu_{\rm B}H_z)^2,
\quad 
(4t-\mu)^2=(\mu_{\rm B}H_z)^2.
\label{eq:dgapclosing}
\end{eqnarray}

In this case, all of
the gap closing conditions (\ref{eq:dgapclosing}) do not depend
on the pairing gap, and thus, the topological phase transition can occur
even for a weak magnetic field for $\mu\sim \pm 4t, 0$.
The difference between the second equation of (\ref{eq:dgapclosing2})
and that of (\ref{eq:dgapclosing}) yields a slight difference of the parameter regions
where a topological order occurs.

\section{duality relation in BdG Hamiltonian}
\label{sec:duality}

As discussed in ref.\cite{STF09}, an underlying mechanism of the realization of
the non-Abelian topological order in the Rashba $s$-wave superconductor is
understood in terms of the duality relation satisfied by the model (\ref{eq:Hamiltonian});
i.e. the BdG Hamiltonian ${\cal
H}({\bm k})$ is unitary equivalent to the following dual Hamiltonian
${\cal H}_{\rm D}({\bm k})$
\begin{eqnarray}
{\mathcal H}_{\rm D}({\bm k})=D{\mathcal H}({\bm k})D^{\dagger} 
=\left(
\begin{array}{cc}
{\rm Re}\Delta({\bm k})-\mu_{\rm B}H_z\sigma_z & -i[\varepsilon({\bm
 k})-i{\rm Im}\Delta({\bm k})]\sigma_y-i\alpha{\bm {\mathcal
 L}}_0({\bm k})\cdot{\bm \sigma}\sigma_y \\
i[\varepsilon({\bm k})+i{\rm Im}\Delta({\bm k})]\sigma_y+i\alpha{\bm
 {\mathcal L}}_0({\bm k})\sigma_y\cdot{\bm \sigma}
&-{\rm Re}\Delta({\bm k})+\mu_{\rm B}H_z\sigma_z
\end{array}
\right),
\label{eq:duality}
\end{eqnarray}
where $D$ is the constant unitary matrix given by
\begin{eqnarray}
D=\frac{1}{\sqrt{2}}
\left(
\begin{array}{cc}
1 & i\sigma_y\\
i\sigma_y & 1
\end{array}
\right). 
\label{eq:D}
\end{eqnarray}
As is shown in Sec.\ref{appendix:a}, the dual transformation accomplished by
the constant unitary matrix does not change the first Chern number of the
system. 
Therefore, the original Hamiltonian has the same topological properties
as the dual one.

It should be remarked here that the Rashba spin orbit interaction 
$\alpha{\bm {\cal L}}_0({\bm k})\cdot {\bm \sigma}$ in the original BdG
Hamiltonian ${\cal H}({\bm k})$ induces ``the $p$-wave gap function''
$-\alpha{\bm {\cal L}}_0({\bm k})\cdot{\bm \sigma}\sigma_y$ in the dual
BdG Hamiltonian. 
However, this does not necessary means that the topological properties
of our system is the same as those of a $p$-wave superconductor, since
${\cal H}_{\rm D}({\bm k})$ has a non standard kinetic term given by ${\rm
Re}\Delta_{s}({\bm k})$. 
Nevertheless, we will show in the following sections that the topological order
similar to a chiral $p+ip$-wave superconductor  emerges under a large
Zeeman field. 
Furthermore, the topological order in our system is much more robust than
that of a chiral $p+ip$-wave superconductor.

In the most part of this paper,  we do not use the dual Hamiltonian $\mathcal{H}_{\rm D}({\bm k})$, but,
instead, analyze the original Hamiltonian $\mathcal{H}({\bm k})$ directly.
Our analysis using $\mathcal{H}({\bm k})$ in this paper confirm the correctness of the argument based on the dual Hamiltonian developed in ref.\cite{STF09}.

\section{topological numbers}
\label{sec:topologicalnumber}

As a conventional long-range order such as magnetic order is characterized by the existence of a nonzero local order parameter, a topological phase is also specified by a characteristic quantity similar to an order parameter; 
this is a topological number.
In this section, we evaluate the topological number characterizing
the non-Abelian topological order realized in Rashba spin-singlet superconductors. 

For two-dimensional time-reversal symmetry broken (TRB) superconductors, 
on which our discussion is focused,
an important topological number is
the TKNN number (equivalent to the first Chern number) $I_{\rm TKNN}$, which is defined as follows. 
Let us consider the BdG equation
\begin{eqnarray}
{\cal H}({\bm k})|\phi_n({\bm k})\rangle=E_n({\bm k})|\phi_n({\bm k})\rangle. 
\end{eqnarray}
By using the normalized occupied states, ``the gauge field''
$A_i^{(-)}({\bm k})$ is defined as
\begin{eqnarray}
A_i^{(-)}({\bm k})=i\sum_{E_n<0}\langle 
\phi_n({\bm k})|\partial_{k_i}\phi_n({\bm k})\rangle. 
\end{eqnarray}
Then the TKNN number is given by
\begin{eqnarray}
I_{\rm TKNN}=\frac{1}{2\pi}\int_{T^2} dk_x dk_y {\cal F}^{(-)}({\bm k}), 
\label{eq:TKNN0}
\end{eqnarray} 
where $T^2$ is the first Brillouin zone in the momentum space, ${\cal
F}^{(-)}({\bm k})$ is the ``field strength of the gauge field''
$A_i^{(-)}({\bm k})$, that is ${\cal F}^{(-)}({\bm
k})=\epsilon^{ij}\partial_{k_i}A_j({\bm k})$.

The nonzero TKNN number implies the existence of topological order in the system under consideration.
In general, the nonzero TKNN number allows both the Abelian topological order, for which there are no non-Abelian anyons, and the non-Abelian topological order, which is characterized by the non-Abelian statistics.
For the Rashba superconductor, the Hamiltonian of which (\ref{eq:BdG}) is a $4\times 4$ matrix,
we can calculate $I_{\rm TKNN}$ directly from the above equations.
However, here, we exploit a different method for the evaluation of the topological number, which is practically easier to be carried out.
Furthermore, this method is quite
useful for the elucidation of the realization of the non-Abelian topological order.
A key idea of our method is to utilize another topological number
specific to the Rashba superconductors,
which is called the winding number.\cite{SF08} 

The winding number is introduced as follows.\cite{SF08}
Let us consider the particle-hole symmetry of the BdG
Hamiltonian,
\begin{eqnarray}
\Gamma {\cal H}({\bm k})\Gamma^{\dagger}=-{\cal H}^{*}(-{\bm k}), 
\end{eqnarray}
where $\Gamma$ is given by
\begin{eqnarray}
\Gamma=\left(
\begin{array}{cc}
0 & {\bm 1}_{2\times 2}\\
{\bm 1}_2\times 2 & 0
\end{array}
\right). 
\end{eqnarray} 
For $k_y=0$ or $\pi$, it is found that the BdG Hamiltonian (\ref{eq:BdG}) of
our model
satisfies ${\cal H}^{*}(-{\bm k})={\cal H}({\bm k})$. Thus the
particle-hole symmetry yields that 
\begin{eqnarray}
[\Gamma, {\cal H}({\bm k})]_+=0 
\end{eqnarray}
for $k_y=0,\,\pi$.
From this relation, it is found that if we take the basis where $\Gamma$
has the diagonal form as
\begin{eqnarray}
\Gamma=\left(
\begin{array}{cc}
{\bm 1}_{2\times 2} & 0\\
0 & -{\bm 1}_{2\times 2}
\end{array}
\right), 
\end{eqnarray}
then ${\cal H}({\bm k})$ at $k_y=0,\,\pi$ becomes off-diagonal,
\begin{eqnarray}
{\cal H}({\bm k})=
\left(
\begin{array}{cc}
0 & q({\bm k})\\
q^{\dagger}({\bm k}) & 0
\end{array}
\right). 
\end{eqnarray}
By using $q({\bm k})$ in the above, 
the topological number $I(k_y)$ is defined as
\begin{eqnarray}
I(k_y)=\frac{1}{4\pi i}\int_{-\pi}^{\pi} dk_x {\rm tr}
\left[q^{-1}({\bm k})\partial_{k_x}q({\bm k})
-q^{\dagger -1}({\bm k})\partial_{k_x}q^{\dagger}({\bm k})\right].  
\label{eq:defwinding}
\end{eqnarray}
We call $I(k_y)$ as the winding number in the following.

\begin{figure}[h]
\begin{center}
\includegraphics[width=7cm]{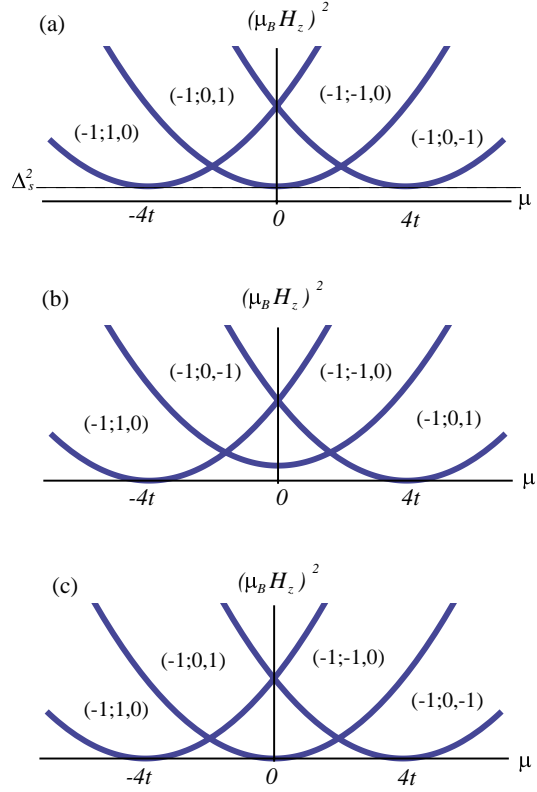}
\caption{
The diagrams of the non-Abelian topological phase for spin-singlet NCS
 superconductors. 
(a) $s$-wave case. (b) $d+id$-wave case with $\Delta({\bm
k})=\Delta_d^{(1)}(\cos k_y-\cos k_x)+i\Delta_d^{(2)}\sin k_x \sin k_y$.
(c) $d+id$-wave case with $\Delta({\bm
k})=\Delta_d^{(1)}(\sin^2 k_x-\sin^2 k_y)+i\Delta_d^{(2)}\sin k_x \sin k_y$. 
The topological numbers $((-1)^{\nu_{\rm Ch}}; I(0),I(\pi))$ are given
 only in the phases supporting a non-Abelian topological order.
In each case, there are four different non-Abelian topological phases.
In the $s$-wave case, the non-Abelian topological phase is realized only
when the Zeeman magnetic field satisfies $(\mu_{\rm B}H_z)^2>(\Delta_s)^2$, but
in the  $d+id$-wave cases, the non-Abelian topological phase can be realized
even for a small but nonzero $H_z$. 
}
\label{fig:phase}
\end{center}
\end{figure}

As is shown in Appendix \ref{appendix:b3}, the winding number and the TKNN number
satisfies 
\begin{eqnarray}
(-1)^{I_{\rm TKNN}}=(-1)^{I(0)-I(\pi)}. 
\label{eq:TKNNwinding}
\end{eqnarray}
Therefore, from the winding number, we can determine $(-1)^{I_{\rm
TKNN}}$. 
The index $(-1)^{I_{\rm TKNN}}$ is of particular interest since it give
a hallmark of the non-Abelian topological phase:
when $(-1)^{I_{\rm TKNN}}=-1$, 
there are an odd 
number of
Majorana zero modes in a vortex, which implies the vortex is a non-Abelian
anyon. 
In the following subsections, we calculate the winding number and the TKNN number
for the cases of the $s$-wave pairing state and the $d+id$ pairing state.
The obtained phase diagrams for the non-Abelian topological phases are
summarized in Fig.\ref{fig:phase}. 

\subsection{$s$-wave pairing}
\label{sec:topnum-swave}

We consider an $s$-wave NCS superconductor with the gap function 
\begin{eqnarray}
\Delta({\bm k})=\Delta_s. 
\end{eqnarray}
For the $s$-wave NCS superconductor, $q({\bm k})$ is given by
\begin{eqnarray}
q({\bm k})=-\left[\varepsilon({\bm k})-\mu_{\rm B}H_z\sigma_z-\alpha
	     \sin k_x\sigma_y\right]+i\Delta_s\sigma_y,
\end{eqnarray}
thus its determinant is 
\begin{eqnarray}
{\rm det}q({\bm k})=
\varepsilon({\bm k})^2-(\mu_{\rm B}H_z)^2-\alpha^2\sin^2 k_x 
+\Delta_s^2
-2i\alpha\Delta_s\sin k_x. 
\end{eqnarray}
Denoting the real (imaginary) part of ${\rm det}q({\bm k})$ as $m_1({\bm
k})$ ($m_2({\bm k})$), we have  
\begin{eqnarray}
m_1({\bm k})=\varepsilon({\bm k})^2-(\mu_{\rm B}H_z)^2-\alpha^2\sin^2 k_x 
+\Delta_s^2, 
\quad
m_2({\bm k})=-2\alpha\Delta_s\sin k_x.  
\end{eqnarray}
From the formula (\ref{eq:winding1}) in Appendix \ref{appendix:b2}, the
winding number is evaluated as
\begin{eqnarray}
I(k_y)=\frac{1}{2}\left[
-{\rm sgn}\left[\varepsilon(0,k_y)^2-(\mu_{\rm B}H_z)^2+\Delta_s^2\right]
+{\rm sgn}\left[\varepsilon(\pi,k_y)^2-(\mu_{\rm B}H_z)^2+\Delta_s^2\right]
\right]. 
\end{eqnarray}
In Table \ref{table:swave}, we summarize the winding number $I(k_y)$ calculated 
from this equation.
We also list $(-1)^{I_{\rm TKNN}}$ obtained from the formula
(\ref{eq:TKNNwinding}).
Note that the topological numbers can change only when one of the gap
closing conditions (\ref{eq:sgapclosing}) is met.
From Table \ref{table:swave}, we obtain the diagram of the non-Abelian
topological phase for $s$-wave
pairing shown in Fig.\ref{fig:phase} (a).

To get a better understanding of the origin of the non-Abelian topological order,
we, here, present some physical discussions about the phase diagram shown in Fig.\ref{fig:phase} (a).
We consider two different but complementary arguments.
The first one is based on mapping from the $s$-wave pairing state to an effective spinless $p$-wave pairing state in the chirality basis. 
The second one is the argument based on the duality relation introduced in Sec.\ref{sec:duality}.
The former is applicable to the case of $\mu_{\rm B}H_z\gg \Delta_s$,
while the latter is 
particularly useful in the vicinity of the topological phase transition point 
$\mu_{\rm B}H_z\sim \Delta_s$.
In this sense, these two arguments are complementary.

We, first, present the first argument which utilizes the mapping onto
an effective $p$-wave pairing state in the chirality basis representation, and
is applicable to the case of $\mu_{\rm B}H_z\gg \Delta_s$.
In the chirality basis, the SO
coupling term and the Zeeman coupling one in the Hamiltonian are diagonalized,
leading to the two SO split bands.
As is shown in Appendix \ref{sec:chirality0}, 
in the parameter region where the
non-Abelian topological phase is realized, we can map our model (\ref{eq:BdG}) into a spinless chiral $p$-wave superconductor in the chirality basis,
provided that $\mu_{\rm B}H_z\gg \Delta_s$. 
For $\mu<0$, the low-energy effective Hamiltonian in this case is, 
\begin{eqnarray}
\tilde{\cal H}_-({\bm k})=
\left(
\begin{array}{cc}
\varepsilon({\bm k})-\Delta\varepsilon({\bm k}) & 
(\alpha {\cal L}_{0 x}({\bm k})-i\alpha {\cal L}_{0 y}({\bm
 k}))(\Delta({\bm k})/\Delta\varepsilon({\bm k}))
\\
(\alpha {\cal L}_{0 x}({\bm k})+i\alpha {\cal L}_{0 y}({\bm
 k}))(\Delta^*({\bm k})/\Delta\varepsilon({\bm k}))
&-\varepsilon({\bm k})+\Delta\varepsilon({\bm k})
\end{array}
\right),
\label{eq:BdGchirality}
\end{eqnarray}
and for $\mu>0$, it is
\begin{eqnarray}
\tilde{\cal H}_+({\bm k})=
\left(
\begin{array}{cc}
\varepsilon({\bm k})+\Delta\varepsilon({\bm k}) & 
(\alpha {\cal L}_{0 x}({\bm k})+i\alpha {\cal L}_{0 y}({\bm
 k}))(\Delta({\bm k})/\Delta\varepsilon({\bm k}))
\\
(\alpha {\cal L}_{0 x}({\bm k})-i\alpha {\cal L}_{0 y}({\bm
 k}))(\Delta^*({\bm k})/\Delta\varepsilon({\bm k}))
&-\varepsilon({\bm k})-\Delta\varepsilon({\bm k})
\end{array}
\right),
\label{eq:BdGchirality2}
\end{eqnarray}
with 
$
\Delta\varepsilon({\bm k})=\sqrt{(\alpha{\cal L}_0({\bm k}))^2+(\mu_{\rm
 B}H_z)^2}. 
$
For the $s$-wave pairing state in the original Hamiltonian, the gap function in
(\ref{eq:BdGchirality}) or (\ref{eq:BdGchirality2}) is given by
\begin{eqnarray}
(\alpha {\cal L}_{0 x}({\bm k})\mp i\alpha {\cal L}_{0 y}({\bm
 k}))(\Delta({\bm k})/\Delta\varepsilon({\bm k}))
\sim i\alpha (\sin k_x \mp i\sin k_y)(\Delta_s/\mu_{\rm B}H_z),
\end{eqnarray}
thus, for both $\mu$'s, the chiral $p+ip$ wave superconductors are
realized in the chirality basis.
By using the effective Hamiltonian,
the phase diagram of our system can be understood more intuitively.
The above effective Hamiltonian (\ref{eq:BdGchirality}) or
(\ref{eq:BdGchirality2}) is obtained by using the fact that,
when $\mu_{\rm B}H_z\gg \Delta_s$, and $\mu$ is in the region of the
phase diagram where the non-Abelian topological order is realized,
only one of the two Fermi surfaces survives and the other
is pushed away by the Zeeman magnetic field.
As a result, the spinless chiral $p+ip$-wave
superconductor is realized effectively.
The non-Abelian topological phase obtained here is
effectively the same as that in the spinless chiral $p+ip$-wave
superconductor. 

We, now, present the second argument on the origin of the topological order which is based on the duality relation.\footnote{The argument here was first outlined in the preprint version
of ref.\cite{STF09}. See ref.\cite{STF09v1}.} 
To grasp the physics shown in the phase diagram, 
let us see what happens at the transition between the
trivial phase ({\it i.e.} the phase with $\mu_{\rm B}H_z=0$) and the
non-Abelian topological one ($(-1)^{I_{\rm TKNN}}=-1$).
From Fig.\ref{fig:phase} (a),  it is found that if we increase the
Zeeman magnetic field, such a phase transition occurs at $\mu=\pm 4t$
when $\mu_{\rm B}H_z=\Delta_s$.   
For simplicity, we consider the case with $\mu=-4t$, where the Fermi
surface is close to the $\Gamma$ point. 
A similar analysis is possible for the transition at $\mu=4t$. 
In the former case,
the gap of the system
closes at ${\bm k}=(0,0)$.  
%

To examine the topological phase transition, it is convenient to use the
dual Hamiltonian instead of the original one.
Using the duality transformation (\ref{eq:duality}), 
we recast the original BdG Hamiltonian into  its unitary equivalent  dual
Hamiltonian ${\cal H}_{\rm D}({\bm k})$. 
Then, we find that for $\mu\sim -4t$, 
the dual Hamiltonian ${\cal H}_{\rm D}({\bm k})$ around ${\bm k}=(0,0)$
is decomposed into the following two 2$\times$2 matrices, 
\begin{eqnarray}
{\cal H}^{D}_{\uparrow\uparrow}({\bm k})=
\left(
\begin{array}{cc}
\Delta_s-\mu_{\rm B}H_z &\alpha (k_y+ik_x) \\
\alpha (k_y-ik_x) & -\Delta_s+\mu_{\rm B}H_z
\end{array}
\right), 
\label{eq:dualRGHamiltonian}\\
{\cal H}^{D}_{\downarrow\downarrow}({\bm k})=
\left(
\begin{array}{cc}
\Delta_s+\mu_{\rm B}H_z &\alpha(-k_y+ik_x) \\
\alpha (-k_y-ik_x) & -\Delta_s-\mu_{\rm B}H_z
\end{array}
\right), 
\label{eq:dualRGHamiltonian2}
\end{eqnarray}
where $\uparrow$ and $\downarrow$ denote the spin in the basis of the
dual Hamiltonian.

We notice here that these Hamiltonians have a close similarity to the
Hamiltonian of the spinless chiral $p+ip$ superconductor discussed in
\cite{read}.
The spinless chiral $p+ip$-wave superconductor shows a phase
transition between the non-Abelian topological phase (or weak-pairing
phase in the terminology used in \cite{read}) and the topologically
trivial phase (strong-pairing phase), and the transition is described by  
the following low energy effective Hamiltonian
\begin{eqnarray}
{\cal H}_{p+ip}({\bm k})= 
\left(
\begin{array}{cc}
\mu_p & \Delta_p(k_x+ik_y)\\
\Delta_p^{*}(k_x-ik_y) &-\mu_p
\end{array}
\right),
\label{eq:RGHamiltonian}
\end{eqnarray}
where $\mu_{p}$ and $\Delta_p$ are the chemical potential and the paring
amplitude for the $p+ip$-wave superconductor, respectively.
For $\mu_p>0$, the state is topologically trivial, and for $\mu_p<0$,
the state supports non-Abelian topological order. The phase transition
occurs at $\mu_p=0$.
By identifying $\Delta_s-\mu_{\rm B}H_z$ and $\alpha$ in
(\ref{eq:dualRGHamiltonian}) with $\mu_p$ and
$\Delta_p$ in (\ref{eq:RGHamiltonian}), respectively, the similarity
between ${\cal H}_{\uparrow\uparrow}({\bm k})$ and ${\cal H}_{p+ip}({\bm
k})$ is evident.
This similarity  immediately implies that the phase transition
at $\mu_{\rm B}H_z=\Delta_s$ is
also accompanied with the emergence of the non-Abelian
topological order.
(We also have a similar phase transition at $\mu_{\rm B}H_z=-\Delta_s$ by
decreasing the Zeeman magnetic field. From the similarity between ${\cal
H}_{\downarrow\downarrow}^{D}({\bm k})$ and ${\cal H}_{p+ip}({\bm k})$,
this transition is also found to be accompanied with the emergence of the
non-Abelian topological order.)

\begin{figure}[h]
\begin{center}
\includegraphics[width=5cm]{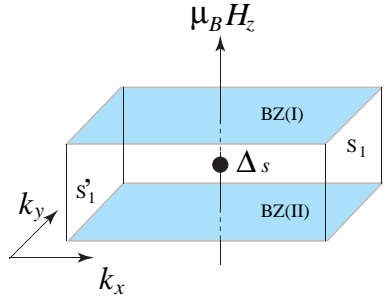}
\caption{Topological phase transition at $\mu_{\rm B}H_z=\Delta_s$.
 BZ(I) and BZ(II) indicate the
 Brillouin zones after and the before the transition. The gap closes at
 $(k_x, k_y)=(0,0)$ when $\mu_{\rm B}H_z=\Delta_s$. ${\rm S}_i$ and
 ${\rm S}_i'$ $(i=1,2)$ are the side faces of the rectangle in which the
 top and bottom faces are BZ(I) and BZ(II). For simplicity, only
 ${\rm S}_1$ and ${\rm S}_1'$ are explicity indicated.} 
\label{fig:monopole}
\end{center}
\end{figure}

Indeed, using the dual Hamiltonian (\ref{eq:dualRGHamiltonian}), we can
see directly that the TKNN number change by $\Delta I_{\rm TKNN}=-1$ at
the transition.
For this purpose, we slightly generalize the gauge field $A_i^{(-)}({\bm
k})$ as follows,
\begin{eqnarray}
A_{\mu}^{(-)}({\bm k})=
\left\{
\begin{array}{ll}
i\sum_{E_n<0}\langle \phi_n({\bm k})|\partial_{k_\mu}\phi_n({\bm k})\rangle 
& \mbox{for $\mu=x,y$},\\
i\sum_{E_n<0}\langle \phi_n({\bm k})|\partial_{\mu_{\rm B}H_z}\phi_n({\bm
 k})\rangle  
& \mbox{for $\mu=z$},
\end{array}
\right.
\end{eqnarray}
where the $z$-component is introduced additionally.
Then, consider the rectangle
illustrated in Fig.\ref{fig:monopole}. 
Here the top face BZ(I) and the bottom one BZ(II) denote the first
Brillouin zones of the system after ($\mu_{\rm B}H_z>\Delta_s$) and
before ($\mu_{\rm B}H_z<\Delta_s$) the transition, respectively.  
Because of the periodicity of the Block wave function $\phi_n({\bm k})$ in
the momentum space, ``the magnetic field'' ${\cal F}^{(-)}({\bm k})$ on
the side faces $S_i$ $(i=1,2)$ are identical to that on the opposite
ones $S_i'$.
Therefore, 
the change of the TKNN number
\begin{eqnarray}
\Delta I_{\rm TKNN}=\frac{1}{2\pi}\int_{\rm BZ(I)}dk_xdk_y
{\cal F}^{(-)}({\bm k})
-\frac{1}{2\pi}\int_{\rm BZ(II)}dk_xdk_y
{\cal F}^{(-)}({\bm k}).
\end{eqnarray}
is rewritten as the total ``magnetic field'' penetrating the surface of the
rectangle $\partial V$,
\begin{eqnarray}
\Delta I_{\rm TKNN}&=&\frac{1}{2\pi}\int_{\rm BZ(I)}dk_xdk_y
{\cal F}^{(-)}({\bm k})
-\frac{1}{2\pi}\int_{\rm BZ(II)}dk_xdk_y
{\cal F}^{(-)}({\bm k})
\nonumber\\
&+&\frac{1}{2\pi}\int_{\rm S_1}dk_y dk_z
{\cal F}^{(-)}({\bm k})
-\frac{1}{2\pi}\int_{\rm S'_1}dk_y dk_z
{\cal F}^{(-)}({\bm k})
\nonumber\\
&+&\frac{1}{2\pi}\int_{\rm S_2}dk_z dk_x
{\cal F}^{(-)}({\bm k})
-\frac{1}{2\pi}\int_{\rm S'_2}dk_z dk_x
{\cal F}^{(-)}({\bm k})
\nonumber\\
&=&\frac{1}{2\pi}\int_{\partial V}dS
{\cal F}^{(-)}({\bm k}),
\label{eq:deltaTKNN}
\end{eqnarray}
with $k_z=\mu_{\rm B}H_z-\Delta_s$.
Therefore, $\Delta I_{\rm TKNN}$ is nonzero only if the ``magnetic
monopole'' of $A_{\mu}^{(-)}({\bm k})$ exists inside the rectangle:
If there is no such a source of the magnetic field, $\Delta I_{\rm
TKNN}$ should be zero from the Gauss-Bonnet theorem.
On the other hand, if the magnetic monopole exists, we have a net magnetic
flux penetrating the surface of the rectangle.

Actually, 
we have a magnetic monopole located at
$(k_x,k_y,k_z(=\mu_{\rm B}H_z-\Delta_s))=(0,0,0)$, where the gap
of the system closes. 
The magnetic charge can be read from the dual Hamiltonian
${\cal H}^{\rm D}_{\uparrow\uparrow}({\bm k})$.
By rewriting ${\cal H}^{\rm D}_{\uparrow\uparrow}({\bm k})$ as ${\cal
H}^{\rm D}_{\uparrow\uparrow}({\bm k})={\bm R}({\bm k})\cdot{\bm \sigma}$  
with $(R_1({\bm
k}),R_2({\bm k}), R_3({\bm k}))=(\alpha k_y, -\alpha k_x, -k_z)$, the
monopole charge is given by,  
\begin{eqnarray}
{\cal Q}&=&\frac{1}{8\pi}\int_{S^2} dS \epsilon_{\mu\nu}
\hat{{\bm R}}\cdot(\partial_{k_{\mu}}\hat{\bm R}
\times \partial_{k_{\nu}}\hat{\bm R})
\label{eq:monopole}
\end{eqnarray}
where $\hat{\bm R}({\bm k})={\bm R}({\bm k})/|{\bm R}({\bm k})|$ and
$S^2$ a small sphere  surrounding the gap-closing point. 
Noting that the right hand side of (\ref{eq:monopole}) counts the number
of times the unit vector $\hat{\bm R}$ wraps around the origin, we
obtain ${\cal Q}=-1$.
Therefore, from (\ref{eq:deltaTKNN}),  we immediately find that $\Delta
I_{\rm TKNN}=-1$.
Before the transition, the system is topologically equivalent to the
ordinary $s$-wave superconductor without the Zeeman magnetic field, thus $I_{\rm
TKNN}=0$. Therefore, we have $I_{\rm TKNN}=-1$ after the
transition.\footnote{Note that the definition of the Chern number in refs. \cite{STF09,
STF09v1} is different from the TKNN number in this paper by a factor $-1$.
}
Again, this result indicates that the system after the transition
belongs to the same topological class as the spinless chiral $p+ip$-wave
superconductor with $I_{\rm TKNN}=-1$. 

As mentioned before, the above two arguments are applicable, respectively, to the different parameter regions,
and thus, they are complementary.
It is noted that these arguments are also straightforwardly applied to the case of the $d+id$ pairing state
discussed in the next subsections.

As seen from Table \ref{table:swave}, the non-Abelian topological order (i.e. $(-1)^{I_{\rm TKNN}}=-1$)
appears
only in the case that the Zeeman energy $\mu_{\rm B}H_z$ is larger than
the superconducting gap $\Delta_s$.
As is well-known, the Rashba superconductors are stable against
the Pauli depairing effect due to applied magnetic fields even for
$\mu_{\rm B}H_z>\Delta_s$
 when the magnetic field is applied
perpendicular to the $xy$-plane,\cite{Frigeri,fuji3}
as long as the Rashba SO interaction is sufficiently strong.
However, there is also the orbital depairing effect due to applied magnetic field.
An important question is how the superconductivity survives the orbital depairing effect
for such a strong magnetic field $\mu_{\rm B}H_z>\Delta_s$.
One possible scenario is to realize this system in the proximity between
a superconductor and a semiconductor as proposed in refs.\cite{SAU,ALICEA}
Another possibility is to realize it in strongly correlated electron systems for which 
the orbital depairing field is large.
Also, one more promising scheme is to utilize ultracold fermionic atom as proposed in \cite{STF09}.
This issue will be discussed in more detail in Sec.\ref{sec:discussion}.

\begin{table}
\begin{center} 
\begin{tabular}[t]{|c|c|c|c|}
\hline
\hline
\multicolumn{4}{c}{a) $\mu \le -2t$} \\ 
 \hline
$(\mu_{\rm B}H_z)^2$ &$(-1)^{I_{\rm TKNN}}$& $I(0)$ & $I(\pi)$\\ 
\hline 
 $0<(\mu_{\rm B}H_z)^2<(4t+\mu)^2+\Delta_s^2$ &  1 &0 &0 \\ 
$(4t+\mu)^2+\Delta_s^2<(\mu_{\rm B}H_z)^2< \mu^2+\Delta_s^2$
& -1 & 1 &0 \\
 $\mu^2+\Delta_s^2<(\mu_{\rm B}H_z)^2<(4t-\mu)^2+\Delta_s^2$
& -1 & 0 &1 \\
 $(4t-\mu)^2+\Delta_s^2 <(\mu_{\rm B}H_z)^2$ & 1& 0 & 0 \\
\hline
\multicolumn{4}{c}{}\\
\multicolumn{4}{c}{b) $-2t<\mu\le 0$} \\ 
 \hline
$(\mu_{\rm B}H_z)^2$ &$(-1)^{I_{\rm TKNN}}$& $I(0)$ & $I(\pi)$\\ 
\hline 
$0<(\mu_{\rm B}H_z)^2<\mu^2+\Delta_s^2$ &  1 &0 &0 \\ 
$\mu^2+\Delta_s^2<(\mu_{\rm B}H_z)^2<(4t+\mu)^2+\Delta_s^2$
& 1 & -1 &1 \\
$(4t+\mu)^2+\Delta_s^2<
(\mu_{\rm B}H_z)^2<(4t-\mu)^2+\Delta_s^2$
& -1 & 0 &1 \\
$(4t-\mu)^2+\Delta_s^2<(\mu_{\rm B}H_z)^2$ & 1& 0 & 0 \\
\hline
\multicolumn{4}{c}{}\\
\multicolumn{4}{c}{c) $0<\mu \le2t$} \\ 
 \hline
$(\mu_{\rm B}H_z)^2$ &$(-1)^{I_{\rm TKNN}}$& $I(0)$ & $I(\pi)$\\ 
\hline 
$0<(\mu_{\rm B}H_z)^2<\mu^2+\Delta_s^2$ &  1 &0 &0 \\ 
$\mu^2+\Delta_s^2<(\mu_{\rm B}H_z)^2<(4t-\mu)^2+\Delta_s^2$
& 1 & -1 &1 \\
$(4t-\mu)^2+\Delta_s^2<
(\mu_{\rm B}H_z)^2<(4t+\mu)^2+\Delta_s^2$
& -1 & -1 &0 \\
$(4t+\mu)^2+\Delta_s^2<(\mu_{\rm B}H_z)^2$ & 1& 0 & 0 \\
\hline
\multicolumn{4}{c}{}\\
\multicolumn{4}{c}{d) $2t < \mu$} \\ 
 \hline
$(\mu_{\rm B}H_z)^2$ &$(-1)^{I_{\rm TKNN}}$& $I(0)$ & $I(\pi)$\\ 
\hline 
$0<(\mu_{\rm B}H_z)^2<(4t-\mu)^2+\Delta_s^2$ &  1 &0 &0 \\ 
$(4t-\mu)^2+\Delta_s^2<(\mu_{\rm B}H_z)^2<\mu^2+\Delta_s^2$
& -1 & 0 &-1 \\
$\mu^2+\Delta_s^2<
(\mu_{\rm B}H_z)^2<(4t+\mu)^2+\Delta_s^2$
& -1 & -1 &0 \\
$(4t+\mu)^2+\Delta_s^2<(\mu_{\rm B}H_z)^2$ & 1& 0 & 0 \\
\hline
\multicolumn{4}{c}{}\\
\hline
\hline
 \end{tabular} 
\end{center}
\caption{The TKNN integer $I_{\rm TKNN}$ and the winding number $I(k_y)$
 for 2D $s$-wave superconductors with the Rashba coupling.
 $(-1)^{I_{\rm TKNN}}=-1$ corresponds to the non-Abelian topological phase.}
\label{table:swave}
\end{table}

\subsection{$d+id$ wave pairing}
\subsubsection{Case of $\Delta({\bm
   k})=\Delta_d^{(1)}(\cos k_y-\cos k_x)+i\Delta_d^{(2)}\sin k_x
  \sin k_y$
}

For the case of the $d+id$-wave superconductor with the gap function
\begin{eqnarray}
\Delta({\bm
   k})=\Delta_d^{(1)}(\cos k_y-\cos k_x)+i\Delta_d^{(2)}\sin k_x
  \sin k_y,
\end{eqnarray}  
$q({\bm k})$ and its determinant are given by
\begin{eqnarray}
q({\bm k})=-[\varepsilon({\bm k})-\mu_{\rm B}H_z\sigma_z-\alpha \sin k_x
 \sigma_y]+i\Delta_d^{(1)}(\cos k_y-\cos k_x)\sigma_y,
\end{eqnarray}
\begin{eqnarray}
{\rm det}q({\bm k})=\varepsilon({\bm k})^2-(\mu_{\rm
 B}H_z)^2-\alpha^2\sin^2 k_x+(\Delta_d^{(1)}(\cos k_y-\cos
 k_x))^2-2i\alpha\sin k_x\Delta_d^{(1)}(\cos k_y-\cos k_x). 
\end{eqnarray}
Therefore, the real and imaginary parts of the determinants are
\begin{eqnarray}
m_1({\bm k})=\varepsilon({\bm k})^2-(\mu_{\rm B}H_z)^2-\alpha^2\sin^2
 k_x+(\Delta_d^{(1)}(\cos k_y-\cos k_x))^2,
\quad
m_2({\bm k})=-2i\alpha \sin k_x \Delta_d^{(1)}(\cos k_y-\cos k_x). 
\end{eqnarray}
To apply the formula (\ref{eq:winding1}), we slightly change $m_2({\bm
k})$ as $-2i\alpha\sin k_x\Delta_d^{(1)}(\cos k_y-\cos k_x)+\delta$
$(\delta \ll 1)$. 
After using the formula (\ref{eq:winding1}), we put $\delta=0$ again. 
Since $I(k_y)$ is a topological number, this procedure does not change
the value of $I(k_y)$.
In this manner, we estimate the winding number $I(k_y)$ as 
\begin{eqnarray}
I(0)=\frac{1}{2}\left[-{\rm sgn}[\varepsilon(0,0)^2-(\mu_{\rm B}H_z)^2]
+{\rm sgn}[\varepsilon(\pi,0)^2-(\mu_{\rm B}H_z)^2+4(\Delta_d^{(1)})^2]
\right], 
\end{eqnarray}
and
\begin{eqnarray}
I(\pi)=\frac{1}{2}\left[-{\rm sgn}[\varepsilon(\pi,\pi)^2-(\mu_{\rm B}H_z)^2]
+{\rm sgn}[\varepsilon(0,\pi)^2-(\mu_{\rm B}H_z)^2+4(\Delta_d^{(1)})^2]
\right].
\end{eqnarray}
We summarize the winding number $I(k_y)$ and
$(-1)^{I_{\rm TKNN}}$ 
in Table \ref{table:d+idwave2}.
The non-Abelian phases are realized in the parameter regions that $(-1)^{I_{\rm TKKN}}=-1$.
From Table II, we obtain the phase diagram shown in Fig.\ref{fig:phase} (b).

In a manner similar to the NCS $s$-wave superconductor, if $\mu_{\rm
B}H_z\gg \Delta({\bm k})$, the obtained phase diagram can be understood
by the effective Hamiltonian (\ref{eq:BdGchirality}) for $\mu<0$ or
(\ref{eq:BdGchirality2}) for $\mu>0$ obtained in the chirality
basis. For simplicity, suppose that
$\Delta_d^{(1)}=\Delta_d^{(2)}\equiv\Delta_d$. The gap function in
(\ref{eq:BdGchirality}) or (\ref{eq:BdGchirality2}) yields
\begin{eqnarray}
(\alpha {\cal L}_{0 x}({\bm k})\pm i\alpha {\cal L}_{0 y}({\bm
 k}))(\Delta({\bm k})/\Delta\varepsilon({\bm k}))
\sim \mp i\alpha (\sin k_x\pm i\sin k_y)(\cos k_y-\cos k_x+i\sin k_x\sin
k_y)(\Delta_d/\mu_{\rm B}H_z),
\label{eq:d+id-eff}
\end{eqnarray}
thus the chiral $f+if$-wave superconductor or the chiral $p+ip$-wave
superconductor is realized both in the effective Hamiltonians $\tilde{\cal
H}_{\mp}({\bm k})$. 
(Whether the $p$-wave state or the $f$-wave state realizes depends on the relative chirality between the $d+id$ order parameter and the $p$-wave factor in (\ref{eq:d+id-eff})
which stems from the SO interaction.)
Therefore, the non-Abelian topological phases obtained here are
effectively the same as that in either the spinless chiral $f+if$-wave
superconductor or the spinless chiral 
$p+ip$-wave superconductor.


One remarkable point observed from Table \ref{table:d+idwave2} (or
Fig.\ref{fig:phase}(b)) is that,
in contrast to the $s$-wave pairing case, for the $d+id$-wave pairing state,
the non-Abelian topological order appears even for small but nonzero magnetic fields,
provided that 
$\mu\sim \pm 4t$.
Thus, in this case, we do not need to worry about the orbital depairing effect due to applied magnetic fields. 
This point makes it easier to realize the Majorana fermion state in the
$d+id$ Rashba superconductor than in the $s$-wave pairing state from the
perspective of the stability against applied magnetic fields, though,
unfortunately, the experimental realization of $d+id$ superconductors has
not yet been established to this date.


Since the $d+id$ pairing state breaks time-reversal symmetry, the TKNN number
is nonzero even for zero magnetic fields.
In the absence of the Zeeman field, we can evaluate the TKNN number
$I_{\rm TKNN}$ directly. 
In this case, we smoothly eliminate the Rashba SO interaction
by setting $\alpha\rightarrow 0$ without gap closing.
This means that the TKNN number of the $d+id$-wave NCS superconductor is the same as that
of the $d+id$ superconductor without the Rashba coupling.

\begin{table}
\begin{center} 
\begin{tabular}[t]{|c|c|c|c|}
\hline
\hline
\multicolumn{4}{c}{a) $\mu \le -2t+(\Delta_{d}^{(1)})^2/2t$} \\ 
 \hline
$(\mu_{\rm B}H_z)^2$ &$(-1)^{I_{\rm TKNN}}$& $I(0)$ & $I(\pi)$\\ 
\hline 
 $0<(\mu_{\rm B}H_z)^2<(4t+\mu)^2$ &  1 &0 &0 \\ 
$(4t+\mu)^2<(\mu_{\rm B}H_z)^2< \mu^2+4(\Delta_d^{(1)})^2$
& -1 & 1 &0 \\
 $\mu^2+4(\Delta_d^{(1)})^2<(\mu_{\rm B}H_z)^2<(4t-\mu)^2$
& -1 & 0 &-1 \\
 $(4t-\mu)^2<(\mu_{\rm B}H_z)^2$ & 1& 0 & 0 \\
\hline
\multicolumn{4}{c}{}\\
\multicolumn{4}{c}{b) $-2t+(\Delta_d^{(1)})^2/2t<\mu\le 0$} \\ 
 \hline
$(\mu_{\rm B}H_z)^2$ &$(-1)^{I_{\rm TKNN}}$& $I(0)$ & $I(\pi)$\\ 
\hline 
$0<(\mu_{\rm B}H_z)^2<\mu^2+4(\Delta_d^{(1)})^2$ &  1 &0 &0 \\ 
$\mu^2+4(\Delta_d^{(1)})^2<(\mu_{\rm B}H_z)^2<(4t+\mu)^2$
& 1 & -1 &-1 \\
$(4t+\mu)^2<
(\mu_{\rm B}H_z)^2<(4t-\mu)^2$
& -1 & 0 &-1 \\
$(4t-\mu)^2<(\mu_{\rm B}H_z)^2$ & 1& 0 & 0 \\
\hline
\multicolumn{4}{c}{}\\
\multicolumn{4}{c}{c) $0<\mu \le2t-(\Delta_d^{(1)})^2/2t$} \\ 
 \hline
$(\mu_{\rm B}H_z)^2$ &$(-1)^{I_{\rm TKNN}}$& $I(0)$ & $I(\pi)$\\ 
\hline 
$0<(\mu_{\rm B}H_z)^2<\mu^2+4(\Delta_d^{(1)})^2$ &  1 &0 &0 \\ 
$\mu^2+4(\Delta_d^{(1)})^2<(\mu_{\rm B}H_z)^2<(4t-\mu)^2$
& 1 & -1 &-1 \\
$(4t-\mu)^2<
(\mu_{\rm B}H_z)^2<(4t+\mu)^2$
& -1 & -1 &0 \\
$(4t+\mu)^2<(\mu_{\rm B}H_z)^2$ & 1& 0 & 0 \\
\hline
\multicolumn{4}{c}{}\\
\multicolumn{4}{c}{d) $2t-(\Delta_d^{(1)})^2/2t < \mu$} \\ 
 \hline
$(\mu_{\rm B}H_z)^2$ &$(-1)^{I_{\rm TKNN}}$& $I(0)$ & $I(\pi)$\\ 
\hline 
$0<(\mu_{\rm B}H_z)^2<(4t-\mu)^2$ &  1 &0 &0 \\ 
$(4t-\mu)^2<(\mu_{\rm B}H_z)^2<\mu^2+4(\Delta_d^{(1)})^2$
& -1 & 0 &1 \\
$\mu^2+4(\Delta_d^{(1)})^2<
(\mu_{\rm B}H_z)^2<(4t+\mu)^2$
& -1 & -1 &0 \\
$(4t+\mu)^2<(\mu_{\rm B}H_z)^2$ & 1& 0 & 0 \\
\hline
\multicolumn{4}{c}{}\\
\hline
\hline
 \end{tabular} 
\end{center}
\caption{The TKNN integer $I_{\rm TKNN}$ and the winding number $I(k_y)$
 for the 2D $d+id$-wave NCS superconductor with $\Delta({\bm
 k})=\Delta_d^{(1)}(\cos k_y-\cos k_x)+i\Delta_d^{(2)}\sin k_x\sin k_y$.
 $(-1)^{I_{\rm TKNN}}=-1$ corresponds to the non-Abelian topological phase.}
\label{table:d+idwave2}
\end{table}

\subsubsection{Case of $\Delta({\bm
   k})=\Delta_d^{(1)}(\sin^2 k_x-\sin^2 k_y)+i\Delta_d^{(2)}\sin k_x
  \sin k_y$
}

For the $d+id$-wave NCS superconductor with the gap function
\begin{eqnarray}
\Delta({\bm k})=\Delta_d^{(1)}(\sin^2 k_x-\sin^2
 k_y)+i\Delta_d^{(2)}\sin k_x \sin k_y, 
\end{eqnarray}
 $q({\bm k})$ is given by
\begin{eqnarray}
q({\bm k})=-\left[\varepsilon({\bm k})-\mu_{\rm B}H_z\sigma_z-\alpha
	     \sin k_x\sigma_y\right]+i\Delta_d^{(1)}\sin^2k_x\sigma_y. 
\end{eqnarray}
Thus its determinant becomes 
\begin{eqnarray}
{\rm det}q({\bm k})=
\varepsilon({\bm k})^2-(\mu_{\rm B}H_z)^2-\alpha^2\sin^2 k_x 
+(\Delta_d^{(2)})^2\sin^4 k_x
-2i\alpha\Delta_d^{(1)}\sin^3 k_x, 
\end{eqnarray}
and the real and imaginary parts of the determinant are  
\begin{eqnarray}
m_1({\bm k})=\varepsilon({\bm k})^2-(\mu_{\rm B}H_z)^2-\alpha^2\sin^2 k_x 
+(\Delta_d^{(1)})^2\sin^4 k_x, 
\quad
m_2({\bm k})=-2\alpha\Delta_d^{(1)}\sin^3 k_x.  
\end{eqnarray}
In a manner similar to the previous $d$-wave case,
we regulate $m_2({\bm
k})$ as $m_2({\bm k})\rightarrow -2\alpha\Delta_d^{(1)}\sin^3
k_x+\delta^3$ ($\delta\ll 1$). 
Then, we obtain
\begin{eqnarray}
I(k_y)=\frac{1}{2}\left[
-{\rm sgn}\left[\varepsilon(0,k_y)^2-(\mu_{\rm B}H_z)^2\right]
+{\rm sgn}\left[\varepsilon(\pi,k_y)^2-(\mu_{\rm B}H_z)^2\right]
\right]. 
\end{eqnarray}
We summarize the winding number $I(k_y)$ and
$(-1)^{I_{\rm TKNN}}$ 
in Table \ref{table:d+idwave}. 
The results are also summarized in the phase diagram Fig.\ref{fig:phase} (c).

When $\mu_{\rm
B}H_z\gg \Delta({\bm k})$, the obtained non-Abelian topological phase
can be understood by using the effective Hamiltonian (\ref{eq:BdGchirality})
or (\ref{eq:BdGchirality2}) in a manner similar to the previous cases.
In the parameter region where the
non-Abelian topological phase is realized,  only one of the
Fermi surfaces survives and the other
is pushed away by the Zeeman magnetic field, 
then, it is found that the non-Abelian topological phases obtained here are
effectively the same as that in either the spinless chiral $f+if$-wave
superconductor or the spinless chiral $p+ip$-wave superconductor. 

As in the previous subsection, 
the non-Abelian topological order is realized even for small but nonzero magnetic fields,
provided that $\mu\sim \pm 4t$.
This is a generic feature of the $d+id$-wave pairing state.
However, in contrast to the case of $\Delta({\bm
   k})=\Delta_d^{(1)}(\cos k_y-\cos k_x)+i\Delta_d^{(2)}\sin k_x\sin k_y$,
the parameters which distinguish different phases do not depend on
the superconducting gap function.

 \begin{table}
\begin{center} 
\begin{tabular}[t]{|c|c|c|c|}
\hline
\hline
\multicolumn{4}{c}{a) $\mu \le -2t$} \\ 
 \hline
$(\mu_{\rm B}H_z)^2$ &$(-1)^{I_{\rm TKNN}}$& $I(0)$ & $I(\pi)$\\ 
\hline 
 $0<(\mu_{\rm B}H_z)^2<(4t+\mu)^2$ &  1 &0 &0 \\ 
$(4t+\mu)^2<(\mu_{\rm B}H_z)^2< \mu^2$
& -1 & 1 &0 \\
 $\mu^2<(\mu_{\rm B}H_z)^2<(4t-\mu)^2$
& -1 & 0 &1 \\
 $(4t-\mu)^2<(\mu_{\rm B}H_z)^2$ & 1& 0 & 0 \\
\hline
\multicolumn{4}{c}{}\\
\multicolumn{4}{c}{b) $-2t<\mu\le 0$} \\ 
 \hline
$(\mu_{\rm B}H_z)^2$ &$(-1)^{I_{\rm TKNN}}$& $I(0)$ & $I(\pi)$\\ 
\hline 
$0<(\mu_{\rm B}H_z)^2<\mu^2$ &  1 &0 &0 \\ 
$\mu^2<(\mu_{\rm B}H_z)^2<(4t+\mu)^2$
& 1 & -1 &1 \\
$(4t+\mu)^2<
(\mu_{\rm B}H_z)^2<(4t-\mu)^2$
& -1 & 0 &1 \\
$(4t-\mu)^2<(\mu_{\rm B}H_z)^2$ & 1& 0 & 0 \\
\hline
\multicolumn{4}{c}{}\\
\multicolumn{4}{c}{c) $0<\mu \le2t$} \\ 
 \hline
$(\mu_{\rm B}H_z)^2$ &$(-1)^{I_{\rm TKNN}}$& $I(0)$ & $I(\pi)$\\ 
\hline 
$0<(\mu_{\rm B}H_z)^2<\mu^2$ &  1 &0 &0 \\ 
$\mu^2<(\mu_{\rm B}H_z)^2<(4t-\mu)^2$
& 1 & -1 &1 \\
$(4t-\mu)^2<
(\mu_{\rm B}H_z)^2<(4t+\mu)^2$
& -1 & -1 &0 \\
$(4t+\mu)^2<(\mu_{\rm B}H_z)^2$ & 1& 0 & 0 \\
\hline
\multicolumn{4}{c}{}\\
\multicolumn{4}{c}{d) $2t < \mu$} \\ 
 \hline
$(\mu_{\rm B}H_z)^2$ &$(-1)^{I_{\rm TKNN}}$& $I(0)$ & $I(\pi)$\\ 
\hline 
$0<(\mu_{\rm B}H_z)^2<(4t-\mu)^2$ &  1 &0 &0 \\ 
$(4t-\mu)^2<(\mu_{\rm B}H_z)^2<\mu^2$
& -1 & 0 &-1 \\
$\mu^2<
(\mu_{\rm B}H_z)^2<(4t+\mu)^2$
& -1 & -1 &0 \\
$(4t+\mu)^2<(\mu_{\rm B}H_z)^2$ & 1& 0 & 0 \\
\hline
\multicolumn{4}{c}{}\\
\hline
\hline
 \end{tabular} 
\end{center}
\caption{The TKNN integer $I_{\rm TKNN}$ and the winding number $I(k_y)$
 for the 2D $d+id$-wave NCS superconductor with $\Delta({\bm
 k})=\Delta_d^{(1)}(\sin^2k_x-\sin^2k_y)+i\Delta_d^{(2)}\sin k_x\sin k_y$.
 $(-1)^{I_{\rm TKNN}}=-1$ corresponds to the non-Abelian topological phase.}
\label{table:d+idwave}
\end{table}

\section{Majorana chiral edge state}
\label{sec:edge}

In this section, we investigate edge states for the 2D spin-singlet NCS superconductors
numerically. From the bulk-edge correspondence, a non-trivial
bulk topological number implies the existence of gapless edge states.
In the case of TRB superconductors, the gapless edge states is a chiral Majorana fermion mode.
We confirm this in the following.  

\subsection{$s$-wave}

To study edge states, we consider the lattice version of Hamiltonian
(\ref{eq:Hamiltonian}).
For an $s$-wave NCS superconductor, the lattice Hamiltonian is given by
\begin{eqnarray}
&&{\cal H}={\cal H}_{\rm kin}+{\cal H}_{\rm SO}+{\cal H}_{\rm s},
\\
&&{\cal H}_{\rm kin}=-t\sum_{\langle {\bm i},{\bm j}\rangle,\sigma}
c_{{\bm i}\sigma}^{\dagger}c_{{\bm j}\sigma}-\mu\sum_{{\bm i},\sigma}
c^{\dagger}_{{\bm i}\sigma}c_{{\bm i}\sigma}
-\mu_{\rm B}H_z\sum_{{\bm
i},\sigma,\sigma'}(\sigma_z)_{\sigma\sigma'}c_{{\bm
i}\sigma}^{\dagger}c_{{\bm i}\sigma'},
\label{eq:kin}
\\
&&{\cal H}_{\rm SO}=-\lambda\sum_{{\bm i}}
\left[(c_{{\bm i}-\hat{x}\downarrow}^{\dagger}c_{{\bm i}\uparrow}
-c^{\dagger}_{{\bm i}+\hat{x}\downarrow}c_{{\bm i}\uparrow})
+i(c_{{\bm i}-\hat{y}\downarrow}^{\dagger}c_{{\bm i}\uparrow}
-c^{\dagger}_{{\bm i}+\hat{y}\downarrow}c_{{\bm i}\uparrow})
+{\rm H.c.}
\right],
\label{eq:SOint}
\\
&&{\cal H}_{\rm s}=\Delta_s
(c_{{\bm i}\uparrow}^{\dagger}c_{{\bm i}\downarrow}^{\dagger}+{\rm H.c.}),
\end{eqnarray}
where ${\bm i}=(i_x,i_y)$ denotes a site on the square lattice,
$c_{{\bm i}\sigma}^{\dagger}$ ($c_{{\bm i}\sigma}$) the creation
(annihilation) operator of an electron with spin $\sigma$ at site
${\bm i}$, and $\lambda=\alpha/2$. 
The sum $\sum_{\langle {\bm i},{\bm j}\rangle}$ is taken between the
nearest-neighbor sites.
Suppose that the system has two open boundary edges at $i_x=0$ and $i_x=N_x$, and impose
the periodic boundary condition in the $y$-direction.
By solving numerically the energy spectrum as a function of the
momentum $k_y$ in the $y$-direction, we study edge states.

\begin{figure}[h]
\begin{center}
\includegraphics[width=7cm]{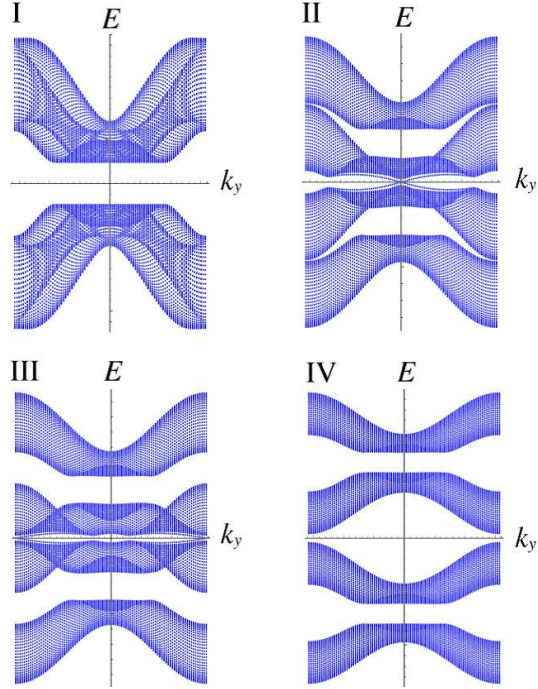}
\caption{The energy spectra of the 2D $s$-wave NCS superconductor with open edges at $i_x=0$
 and $i_x=30$ for $\mu<-2t$. Here $k_y$ denotes the momentum in the
 $y$-direction, and $k_y\in [-\pi,\pi]$. We
 take $t=1$, $\mu=-2.5$, $\lambda=0.5$, and $\Delta_s=1$.
 The Zeeman magnetic field $H_z$ is (I) $\mu_{\rm B}H_z=0$, (II) $\mu_{\rm
 B}H_z=2$, (III) $\mu_{\rm B}H_z=3$, and (IV) $\mu_{\rm B}H_z=7$.
 The cases of (I), (II), (III), and (IV) correspond to, respectively,
 the four regions in Table \ref{table:swave}.a).
 The non-Abelian phases are (II) and (III).
 The two gapless modes found in (II) correspond to edge modes for two open edges, respectively.
 Thus, they are chiral. The same is also true for the gapless modes in (III).}
\label{fig:edgestate_s_a}
\end{center}
\end{figure}

\begin{figure}[h]
\begin{center}
\includegraphics[width=7cm]{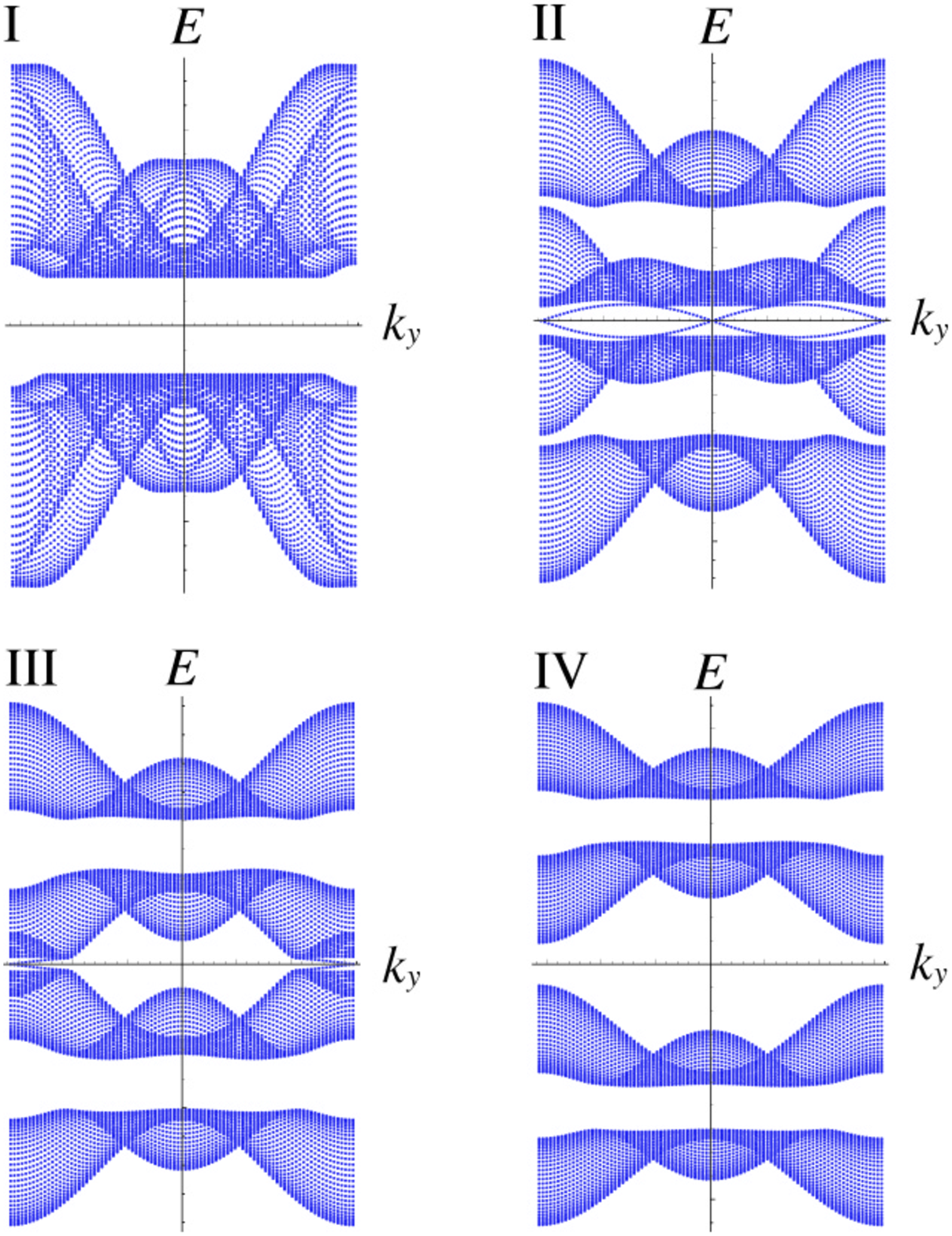}
\caption{The energy spectra of the 2D $s$-wave NCS superconductor with open edges at $i_x=0$
 and $i_x=30$ for $-2t<\mu< 0$. Here $k_y$ denotes the momentum in the
 $y$-direction, and $k_y\in [-\pi,\pi]$. We
 take $t=1$, $\mu=-1$, $\lambda=0.5$, and $\Delta_s=1$.
 The Zeeman magnetic field $H_z$ is (I) $\mu_{\rm B}H_z=0$, (II) $\mu_{\rm
 B}H_z=2$, (III) $\mu_{\rm B}H_z=4$, and (IV) $\mu_{\rm B}H_z=6$.
 The cases of (I), (II), (III), and (IV) correspond to, respectively,
 the four regions in Table \ref{table:swave}.b).
 The non-Abelian phase is (III).}
\label{fig:edgestate_s_b}
\end{center}
\end{figure}

We illustrate the energy spectra for the 2D $s$-wave
NCS superconductor with edges at $i_x=0$ and $i_x=30$ in Figs \ref{fig:edgestate_s_a}
and \ref{fig:edgestate_s_b}.
In Fig. \ref{fig:edgestate_s_a} (Fig. \ref{fig:edgestate_s_b}), $\mu$
satisfies $\mu<-2t$ ($-2t<\mu<0$), and the
corresponding bulk topological numbers are given in Table
\ref{table:swave} (a) (Table \ref{table:swave} (b)).  
We find that if the TKNN number is odd, odd numbers of gapless edge
modes appear.
In this case, the non-Abelian topological order appears, and the edge zero mode
is a chiral Majorana fermion.
It is also found that when the winding number $I(k_y)$ 
is non-zero for $k_y=0$, or $\pi$, 
the energy of the gapless edge mode becomes zero at this value of $k_y$.
Therefore, all the results are consistent with 
the existence of the correspondence between the bulk topological numbers and 
the edge spectra.


\subsection{$d+id$ wave}

\subsubsection{Case of $\Delta({\bm k})=\Delta_d^{(1)}(\cos k_y-\cos
   k_x)+i\Delta_d^{(2)}\sin k_x \sin k_y$}

The lattice Hamiltonian for the 2D $d+id$-wave NCS superconductor with $\Delta({\bm
k})=\Delta_d^{(1)}(\cos k_y-\cos
   k_x)+i\Delta_d^{(2)}\sin k_x \sin k_y$
is given by ${\cal H}={\cal H}_{\rm kin}+{\cal H}_{\rm SO}+{\cal H}_{\rm
s}$ with
\begin{eqnarray}
{\cal H}_{\rm s}&=&-\frac{\Delta_d^{(1)}}{4}
\left[
c_{{\bm i}+\hat{x}\uparrow}^{\dagger}c_{{\bm
 i}\downarrow}^{\dagger}
+c_{{\bm i}-\hat{x}\uparrow}^{\dagger}c_{{\bm
 i}\downarrow}^{\dagger}
-c_{{\bm i}+\hat{y}\uparrow}^{\dagger}c_{{\bm
 i}\downarrow}^{\dagger}
-c_{{\bm i}-\hat{y}\uparrow}^{\dagger}c_{{\bm
 i}\downarrow}^{\dagger}
\right]
\nonumber\\
&&-i\frac{\Delta_d^{(2)}}{4}
\left[
c_{{\bm i}+\hat{x}+\hat{y}\uparrow}^{\dagger}c_{{\bm
 i}\downarrow}^{\dagger}
+c_{{\bm i}-\hat{x}-\hat{y}\uparrow}^{\dagger}c_{{\bm
 i}\downarrow}^{\dagger}
-c_{{\bm i}+\hat{x}-\hat{y}\uparrow}^{\dagger}c_{{\bm
 i}\downarrow}^{\dagger}
-c_{{\bm i}-\hat{x}+\hat{y}\uparrow}^{\dagger}c_{{\bm
 i}\downarrow}^{\dagger}
\right]
+{\rm H.c.}.
\end{eqnarray}
The kinetic term ${\cal H}_{\rm kin}$ and the Rashba SO
interaction ${\cal H}_{\rm SO}$ are the same as (\ref{eq:kin}) and (\ref{eq:SOint}),
respectively. 

\begin{figure}[h]
\begin{center}
\includegraphics[width=7cm]{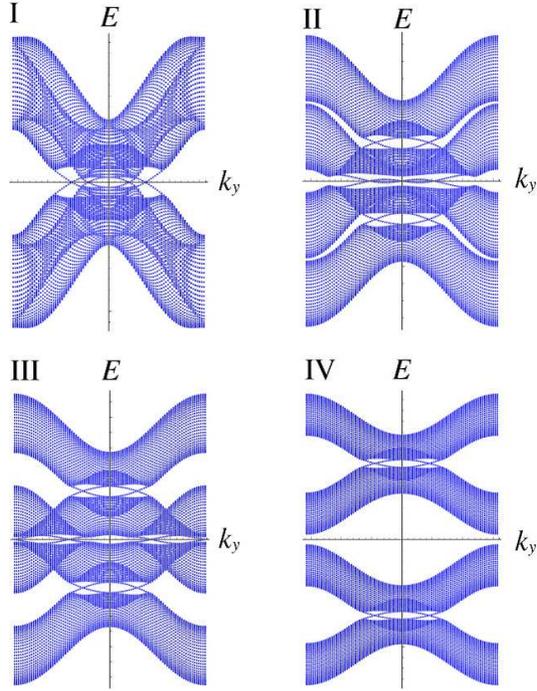}
\caption{The energy spectra of the 2D $d+id$-wave NCS superconductor 
($\Delta({\bm k})=\Delta_d^{(1)}(\cos k_y-\cos
   k_x)+i\Delta_d^{(2)}\sin k_x \sin k_y$)
with open edges at $i_x=0$
 and $i_x=30$ for $\mu<-2t+\Delta_d^{(1)2}/2t$.
Here $k_y$ denotes the momentum in the $y$-direction, and $k_y\in[-\pi,\pi]$. We
 take $t=1$, $\mu=-2.5$, $\lambda=0.5$, $\Delta_d^{(1)}=0.5$, and
 $\Delta_d^{(2)}=0.8$.
 The Zeeman magnetic field $H_z$ is (I) $\mu_{\rm B}H_z=0$, (II) $\mu_{\rm
 B}H_z=2$, (III) $\mu_{\rm B}H_z=3$, and (IV) $\mu_{\rm B}H_z=7$.
  The cases of (I), (II), (III), and (IV) correspond to, respectively,
 the four regions in Table \ref{table:d+idwave2}.a).
 The non-Abelian phases are (II) and (III).}
\label{fig:edgestate_d2_a}
\end{center}
\end{figure}

\begin{figure}[h]
\begin{center}
\includegraphics[width=7cm]{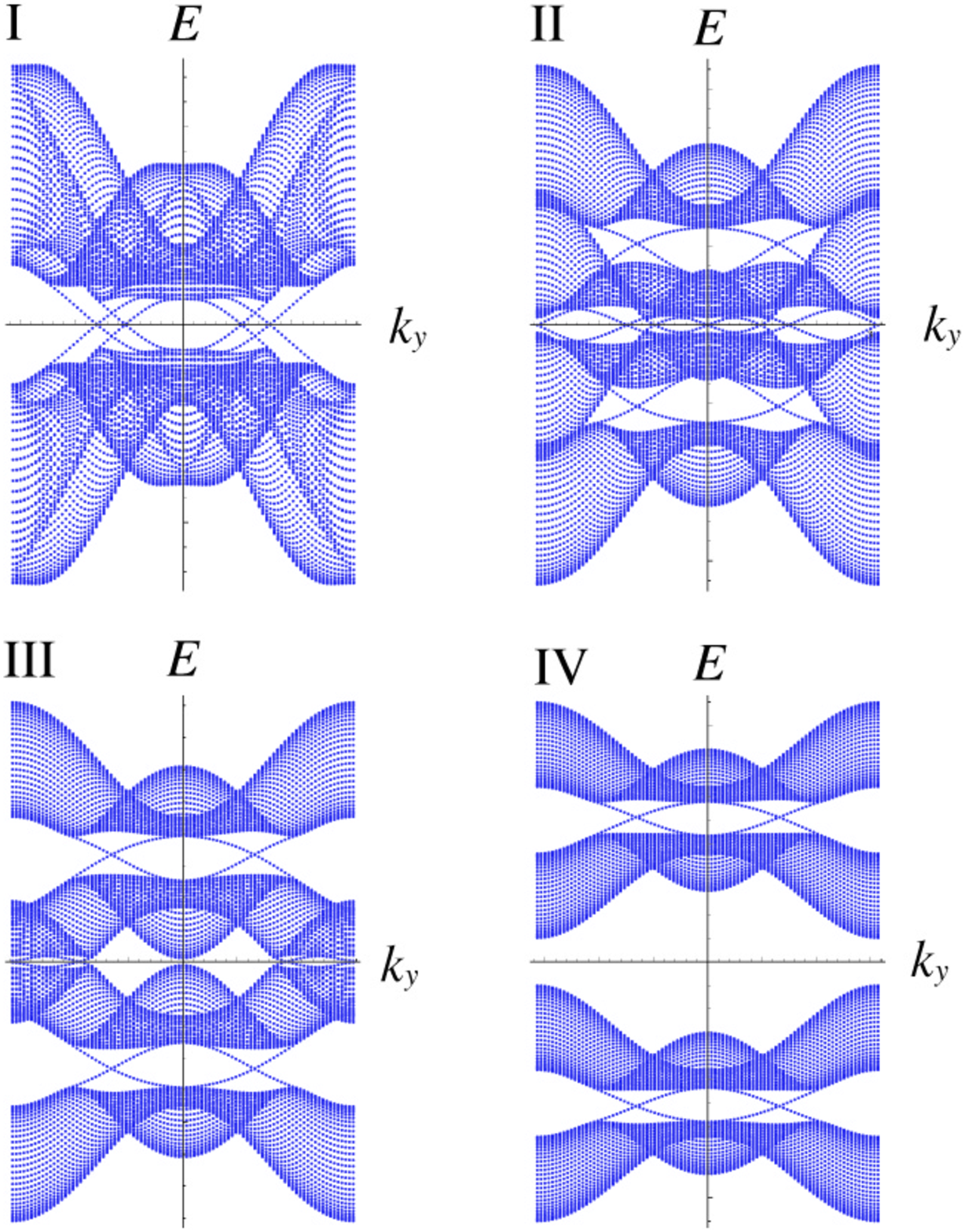}
\caption{The energy spectra of the 2D $d+id$-wave NCS superconductor 
($\Delta({\bm k})=\Delta_d^{(1)}(\cos k_y-\cos
   k_x)+i\Delta_d^{(2)}\sin k_x \sin k_y$)
with open edges at $i_x=0$
 and $i_x=30$ for $-2t+\Delta_d^{(1)2}/2t<\mu<0$. Here $k_y$ denotes the
 momentum in the $y$-direction, and $k_y\in[-\pi,\pi]$. We
 take $t=1$, $\mu=-1$, $\lambda=0.5$, $\Delta_d^{(1)}=0.5$, and
 $\Delta_d^{(2)}=0.8$.
 The Zeeman magnetic field $H_z$ is (I) $\mu_{\rm B}H_z=0$, (II) $\mu_{\rm
 B}H_z=1.6$, (III) $\mu_{\rm B}H_z=3.1$, and (IV) $\mu_{\rm B}H_z=6$.
  The cases of (I), (II), (III), and (IV) correspond to, respectively,
 the four regions in Table \ref{table:d+idwave}.b).
 The non-Abelian phase is (III).}
\label{fig:edgestate_d2_b}
\end{center}
\end{figure}

In a manner similar to the $s$-wave NCS superconductor, we obtain the energy spectra for the system
with edges at $i_x=0$ and $i_x=30$ numerically. 
We illustrate the energy spectra for the 2D $d$-wave
NCS superconductor with the gap function $\Delta({\bm
k})=\Delta_d^{(1)}(\cos k_y-\cos k_x)+i\Delta_d^{(2)}\sin k_x \sin k_y$
in Figs \ref{fig:edgestate_d2_a}
and \ref{fig:edgestate_d2_b}.
In Fig. \ref{fig:edgestate_d2_a} (Fig. \ref{fig:edgestate_d2_b}), $\mu$
satisfies $\mu<-2t+(\Delta_d^{(1)})^2/2t$ ($-2t+(\Delta_d^{(1)})^2/2t<\mu<0$), 
and the corresponding bulk topological numbers are given in Table
\ref{table:d+idwave2} (a) (Table \ref{table:d+idwave2} (b)).  
We find that if the TKNN number is odd, odd numbers of gapless edge
modes appear.
The non-Abelian topological order (and hence the chiral Majorana fermion mode) is realized in this case.
In addition, it is found that if the winding number $I(k_y)$ $(k_y=0,\pi)$ is
non-zero for some $k_y$, 
the energy of the gapless edge state becomes zero at this value of $k_y$.
These results are consistent with the bulk-edge correspondence.

\begin{figure}[h]
\begin{center}
\includegraphics[width=7cm]{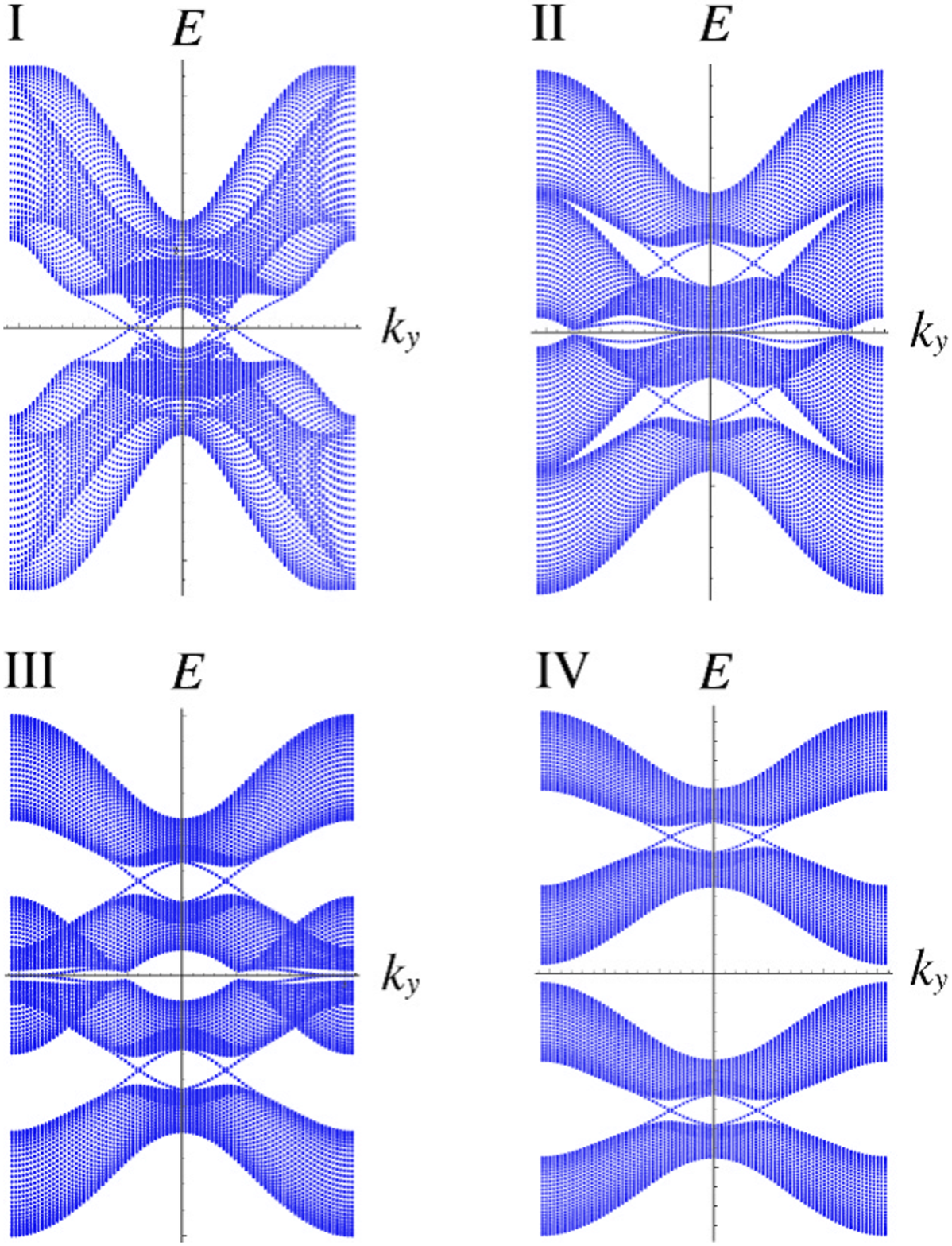}
\caption{The energy spectra of the 2D $d+id$-wave NCS superconductor 
($\Delta({\bm k})=\Delta_d^{(1)}(\sin^2 k_x-\sin^2
   k_y)+i\Delta_d^{(2)}\sin k_x \sin k_y$)
with open edges at $i_x=0$
 and $i_x=30$ for $\mu<-2t$. Here $k_y$ denotes the momentum in the
 $y$-direction, and $k_y\in [-\pi,\pi]$. We
 take $t=1$, $\mu=-2.5$, $\lambda=0.5$, $\Delta_d^{(1)}=1$, and
 $\Delta_d^{(2)}=1$.
 The Zeeman magnetic field $H_z$ is (I) $\mu_{\rm B}H_z=0$, (II) $\mu_{\rm
 B}H_z=2$, (III) $\mu_{\rm B}H_z=3.5$, and (IV) $\mu_{\rm B}H_z=7$.
 The cases of (I), (II), (III), and (IV) correspond to, respectively,
 the four regions in Table \ref{table:d+idwave}.a).
 The non-Abelian phases are (II) and (III).}
\label{fig:edgestate_d1_a}
\end{center}
\end{figure}

\begin{figure}[h]
\begin{center}
\includegraphics[width=7cm]{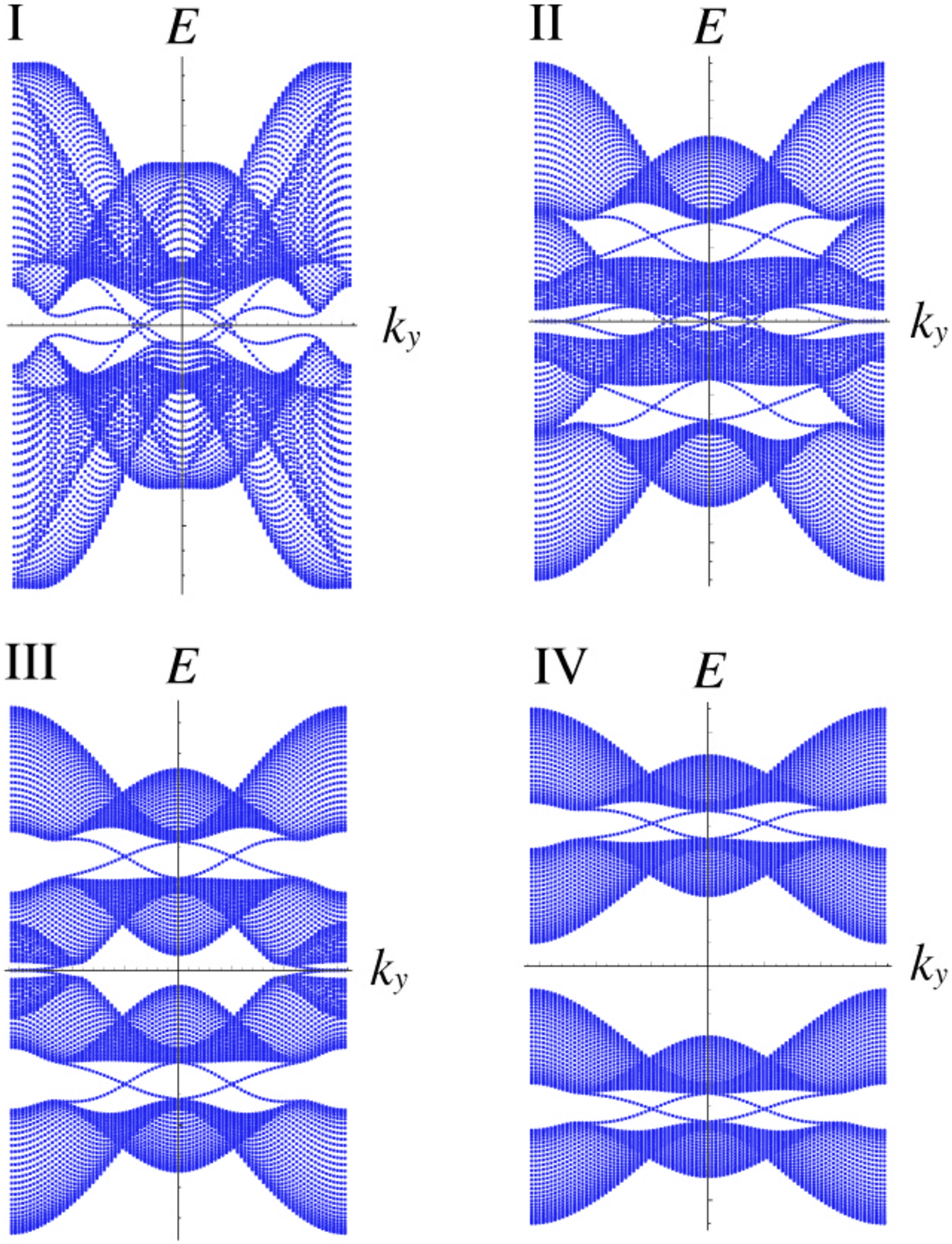}
\caption{The energy spectra of the 2D $d+id$-wave NCS superconductor 
($\Delta({\bm k})=\Delta_d^{(1)}(\sin^2 k_x-\sin^2
   k_y)+i\Delta_d^{(2)}\sin k_x \sin k_y$)
with open edges at $i_x=0$
 and $i_x=30$ for $-2t<\mu<0$. Here $k_y$ denotes the momentum in the
 $y$-direction, and $k_y\in [-\pi,\pi]$. We
 take $t=1$, $\mu=-1$, $\lambda=0.5$, $\Delta_d^{(1)}=1$, and
 $\Delta_d^{(2)}=1$.
 The Zeeman magnetic field $H_z$ is (I) $\mu_{\rm B}H_z=0$, (II) $\mu_{\rm
 B}H_z=2$, (III) $\mu_{\rm B}H_z=3.5$, and (IV) $\mu_{\rm B}H_z=6$.
 The cases of (I), (II), (III), and (IV) correspond to, respectively,
 the four regions in Table \ref{table:d+idwave}.b).
 The non-Abelian phase is (III).}
\label{fig:edgestate_d1_b}
\end{center}
\end{figure}

Since the $d+id$-wave superconductor breaks time-reversal symmetry even in the absence of
a magnetic field, the TKNN number is nonzero for $H_z=0$, and there are four chiral edge modes,
which are seen in Figs.\ref{fig:edgestate_d2_a} (I), \ref{fig:edgestate_d2_b} (I),
\ref{fig:edgestate_d1_a} (I), and \ref{fig:edgestate_d1_b} (I).
Since there are even numbers of gapless edge modes, they do not behave as non-Abelian anyons.
These edge modes crosses the zero energy at some $k_y\neq 0, \pm \pi$.
In contrast, 
the Majorana fermion mode associated with the non-Abelian topological order 
crosses the zero energy at $k_y=0$ or $\pi$ corresponding to
the nonzero values of the winding number $I(k_y)$ at these points.

\subsubsection{Case of $\Delta({\bm k})=\Delta_d^{(1)}(\sin^2 k_x-\sin^2
   k_y)+i\Delta_d^{(2)}\sin k_x \sin k_y$}

In this case, we use the lattice Hamiltonian ${\cal H}={\cal H}_{\rm
kin}+{\cal H}_{\rm SO}+{\cal H}_{\rm s}$ with
\begin{eqnarray}
{\cal H}_{\rm s}&=&-\frac{\Delta_d^{(1)}}{4}
\left[
c_{{\bm i}+2\hat{x}\uparrow}^{\dagger}c_{{\bm
 i}\downarrow}^{\dagger}
+c_{{\bm i}-2\hat{x}\uparrow}^{\dagger}c_{{\bm
 i}\downarrow}^{\dagger}
-c_{{\bm i}+2\hat{y}\uparrow}^{\dagger}c_{{\bm
 i}\downarrow}^{\dagger}
-c_{{\bm i}-2\hat{y}\uparrow}^{\dagger}c_{{\bm
 i}\downarrow}^{\dagger}
\right]
\nonumber\\
&&-i\frac{\Delta_d^{(2)}}{4}
\left[
c_{{\bm i}+\hat{x}+\hat{y}\uparrow}^{\dagger}c_{{\bm
 i}\downarrow}^{\dagger}
+c_{{\bm i}-\hat{x}-\hat{y}\uparrow}^{\dagger}c_{{\bm
 i}\downarrow}^{\dagger}
-c_{{\bm i}+\hat{x}-\hat{y}\uparrow}^{\dagger}c_{{\bm
 i}\downarrow}^{\dagger}
-c_{{\bm i}-\hat{x}+\hat{y}\uparrow}^{\dagger}c_{{\bm
 i}\downarrow}^{\dagger}
\right]
+{\rm H.c.}.
\end{eqnarray}
Here, the kinetic term ${\cal H}_{\rm kin}$ and the Rashba SO
interaction ${\cal H}_{\rm SO}$ are the same as (\ref{eq:kin}) and (\ref{eq:SOint}),
respectively. 

We calculate the energy spectra for the system
with edges at $i_x=0$ and $i_x=30$ numerically. 
The energy spectra for the 2D $d$-wave
NCS superconductor with the gap function $\Delta({\bm
k})=\Delta_d^{(1)}(\sin^2 k_x-\sin^2 k_y)+i\Delta_d^{(2)}\sin k_x \sin k_y$ are shown 
in Figs \ref{fig:edgestate_d1_a}
and \ref{fig:edgestate_d1_b}.
In Fig. \ref{fig:edgestate_d1_a} (Fig. \ref{fig:edgestate_d1_b}), $\mu$
satisfies $\mu<-2t$ ($-2t<\mu<0$), and the
corresponding bulk topological numbers are given in Table
\ref{table:d+idwave} (a) (Table \ref{table:d+idwave} (b)).  
When the TKNN number is odd, odd numbers of gapless edge
states appear, signifying the non-Abelian topological order.
It is also found that if the winding number $I(k_y)$ $(k_y=0,\pi)$ is
non-zero for some $k_y$, 
the energy of the gapless edge state becomes zero at this value of $k_y$.
These results are consistent with the bulk-edge correspondence again.

\section{Majorana zero mode in a vortex of an $s$-wave Rashba superconductor}
\label{sec:vortex}

In this section, we discuss a Majorana fermion mode in a vortex core of
an $s$-wave Rashba superconductor, which is one of the important features of the non-Abelian topological order, and relevant to the application to the topological quantum computation.
We also discuss that, in certain parameter region, the Majorana mode in the $s$-wave Rashba superconductors is strongly stable against thermal noise and
inter-vortex tunneling, which are serious obstructions to the implementation of the topological quantum computation in the case of $p+ip$ superconductors.

\subsection{Majorana solution of the Bogoliubov-de Gennes equation}
\label{sec:vortexsol}

The non-Abelian topological order is characterized by the existence of the Majorana zero energy mode
in a vortex core, which leads to the realization of the non-Abelian statistics.
In this section, we demonstrate that the Majorana fermion mode in a vortex core exists
 for an $s$-wave Rashba superconductor when the Zeeman energy  $\mu_{\rm B}H_z$ is larger than the superconducting gap $\Delta$, 
 by solving the Bogoliubov-de
Gennes (BdG) equation for a
single vortex of the superconducting order parameter; $\Delta(\bm{r})=\Delta \exp(in\theta)$
with $n$ vorticity.
In ref.\cite{STF09}, when $n$ is odd,
the Majorana bound state for a vortex of the SO interaction is obtained by the analysis of the BdG equation for
the dual Hamiltonian (\ref{eq:duality}). 
The existence of the Majorana bound state in a vortex of the SO interaction strongly implies
that there exists the Majorana bound state also in a vortex of the superconducting condensate, because
these two vortex states are related by a singular gauge transformation.
In the following, we will find the zero energy Majorana solution for a vortex core of
the superconducting condensate explicitly in the case that the vorticity $n$ is odd, and
the condition $\mu_{\rm B}H_z>\Delta>0$ is satisfied.
This result is in accordance with the finding of Sau et al.\cite{SAU}
In the following analysis, 
instead of using the dual Hamiltonian (\ref{eq:duality}),
we deal with the BdG equation for the original Hamiltonian (\ref{eq:BdG}) directly.
Since we can not obtain the exact solution of the BdG equation analytically, 
we adopt the following approximation scheme.
As clarified in the previous sections, the non-Abelian topological order appears when
the Fermi level crosses $\bm{k}$-points in the vicinity of the $\Gamma$ point or the M point
in the Brillouin zone.
In this situation, there are two Fermi surfaces split by the Rashba SO interaction.
One is located in the vicinity of the $\Gamma$ (or M) point with 
the  Fermi momentum $\bm{k}_F\sim 0$ (or $(\pi,\pi)$),
and the other has the large Fermi momentum $\bm{k}_F\neq 0, (\pi,\pi)$.
Since the momentum is not a good quantum number in the presence of a vortex core,
a bound state in the vortex core generally consists of a superposition of
quasiparticles from both of these two Fermi surfaces.
However, according to the discussion based on the duality Hamiltonian
given in Sec.\ref{sec:topnum-swave},
it is strongly suggested that
in the vicinity of the topological phase transition point $\mu_{\rm B}H_z\sim \Delta$,
quasiparticles with $\bm{k}\sim 0$ (or $(\pi,\pi)$) play a very important role
in the realization of the non-Abelian topological order.
This implies that when the Zeeman energy is close to the superconducting gap, 
the Majorana fermion mode is mainly formed by
quasiparticles with $\bm{k}_F\sim 0$ or $(\pi,\pi)$ rather than those with $\bm{k}_F\neq 0, (\pi,\pi)$,
in the long-distance asymptotic regime away from the center of the vortex core.
Thus, in the following, 
we try to construct an approximate solution for the zero energy mode in a vortex
from quasiparticles with $k_F\sim 0$ or $(\pi,\pi)$.
As will be shown below, the vortex core for this approximated solution has a characteristic length
$\sim v_F/(\mu_{\rm B}H_z-\Delta)$, while quasiparticles with $\bm{k}_F\neq 0, (\pi,\pi)$ give
contributions to the vortex core bound state with a characteristic length
$\sim v_F/\Delta$.
Thus, the approximated solution presented in the following is valid when $0<\mu_{\rm B}H_z-\Delta <\Delta$
in the long-distance asymptotic regime.

To solve the BdG equation, we choose the gauge for which the gap function is real by
applying the gauge transformation $e\bm{A}\rightarrow e\bm{A}-n\nabla\theta/2$, $\Delta \exp(in\theta)\rightarrow \Delta$.
Then, the BdG equation is
\begin{eqnarray}
\mathcal{H}\tilde{\Psi}=E\tilde{\Psi}
\label{bdg1}
\end{eqnarray}
\begin{eqnarray}
\mathcal{H}=
\left(
\begin{array}{cc}
\varepsilon(\hat{k}-e\bm{A}+\frac{n}{2}\nabla\theta)+\bm{g}(\hat{k}-e\bm{A}+\frac{n}{2}\nabla\theta)\cdot
\bm{\sigma}-h\sigma_z & \Delta i\sigma_y \\
-\Delta i\sigma_y & 
-\varepsilon(\hat{k}+e\bm{A}-\frac{n}{2}\nabla\theta)+\bm{g}(\hat{k}+e\bm{A}-\frac{n}{2}\nabla\theta)\cdot
\bm{\sigma}^{*}+h\sigma_z 
\end{array}
\right).
\label{eq:BDGham}
\end{eqnarray}
Here $\tilde{\Psi}^T=(\tilde{u}_{\uparrow},\tilde{u}_{\downarrow},\tilde{v}_{\uparrow},\tilde{v}_{\downarrow})$,
$\varepsilon(k)=\frac{k^2}{2m}-\mu$, $\bm{g}(k)=2\lambda (k_y,-k_x,0)$, $\hat{k}=-i\nabla$,
and $h=\mu_{\rm B}H_z$.
Assuming $H_z\ll H_{c2}$, we neglect $e\bm{A}$ compared to $n\nabla\theta/2$ in 
$\varepsilon(\hat{k}-e\bm{A}+\frac{n}{2}\nabla\theta)$ and 
$\bm{g}(\hat{k}-e\bm{A}+\frac{n}{2}\nabla\theta)$.\cite{CGM}
We also assume that $\Delta=0$ for $r<r_c$, and $\Delta\neq 0$ for $r>r_c$, where
$r_c \ll \xi$.
For simplicity, we consider the case of $\mu=0$, for which one of the two SO split bands crosses the $\Gamma$
point $k=0$.
This is a typical situation which realizes 
the non-Abelian topological order (and hence
the Majorana zero mode) as discussed in the previous sections.
In the following, we restrict our analysis to the zero energy state with $E=0$.
Furthermore, we impose the condition that the magnitude of the SO interaction is much larger than 
the Zeeman energy scale; i.e. $h \ll m\lambda^2$.
Under this condition, we can find the following approximated solution for the zero energy mode in a vortex core
with the odd vorticity $n$.
For $r<r_c$,
\begin{eqnarray}
\tilde{u}_{\uparrow}(r,\theta)=A_{\uparrow}e^{-i\frac{\theta}{2}}H^{(1)}_{\frac{n-1}{2}}(i\frac{hr}{2\lambda}),
\label{sol1-main1}
\end{eqnarray}
\begin{eqnarray}
\tilde{u}_{\downarrow}(r,\theta)=-i A_{\uparrow}e^{i\frac{\theta}{2}}H^{(1)}_{\frac{n+1}{2}}(i\frac{hr}{2\lambda}),
\label{sol1-main2}
\end{eqnarray}
and for $r>r_c$,
\begin{eqnarray}
\tilde{u}_{\uparrow}(r,\theta)=\sqrt{2}A_{\uparrow}e^{-i\frac{\theta}{2}}e^{\int^r dr'\frac{h-\Delta}{2\lambda}}H^{(1)}_{\frac{n+3}{2}}(i\frac{h-\Delta}{\lambda}r),
\label{sol1-main3}
\end{eqnarray}
\begin{eqnarray}
\tilde{u}_{\downarrow}(r,\theta)=-i \sqrt{2}A_{\uparrow}e^{i\frac{\theta}{2}}e^{\int^r dr'\frac{h-\Delta}{2\lambda}}H^{(1)}_{\frac{n+1}{2}}(i\frac{h-\Delta}{\lambda}r),
\label{sol1-main4}
\end{eqnarray}
where 
$H^{(1)}_{\nu}(z)$ is the first Hankel function, and the constant $A_{\uparrow}$ is determined by the normalization condition. Also, $\tilde{v}_{\sigma}(r,\theta)=\tilde{u}_{\sigma}(r,\theta)$.
The solution for $r>r_c$ and those for $r<r_c$ can be matched at $r=r_c$ by using the asymptotic form of
the Hankel function, as explained in the Appendix \ref{appendix:d}.
Since there is only one zero energy mode, the above solution indicates that there is a Majorana zero energy mode
in a vortex core with odd vorticity.
Note that this approximated solution is constructed from quasiparticles in the vicinity of the $\Gamma$ point
$\bm{k} \sim 0$.
As seen from Eqs.(\ref{sol1-main3}) and (\ref{sol1-main4}), the above solution for the vortex core bound state
decays as $\sim \exp(-\frac{h-\Delta}{2\lambda}r)$ in the long distance regime.
On the other hand, the contribution from quasiparticles with the large Fermi momentum to the vortex core state
decays like $\sim \exp(-\frac{\Delta}{2\lambda}r)$ (note that when $\mu=0$, the Fermi velocity $v_F\sim 2\lambda$). Thus, for $0<h-\Delta<\Delta$, the long distance behavior of
the vortex core state is dominated by quasiparticles with $\bm{k}\sim 0$,
and hence the above approximated solution mainly constructed from quasiparticles with $\bm{k}\sim 0$
is valid under this condition.

In the above derivation of the zero energy Majorana bound state, we have used several approximations.
In particular, at the stage of matching the solutions for $r>r_c$ and $r<r_c$, we have neglected corrections of order
$O(h/(m\lambda^2) )$.
Actually, such approximations are not essential for the realization of the Majorana zero energy mode,
but, rather, required by our approximation method of the construction of the zero mode.
In fact, the exact solution for the zero energy mode in a vortex core should be a superposition of
quasiparticles with $\bm{k}\sim 0$ and those with $k_F\neq 0$, because the momentum is not a good quantum number in the presence of a vortex core.
Since it is quite difficult to obtain the exact solution of the zero energy mode
constructed from both quasiparticles with $\bm{k}\sim 0$ and those with $k_F\neq 0$, we approximate
it by the bound state mainly formed by quasiparticles with $\bm{k}\sim 0$, neglecting
contributions of quasiparticles from the large Fermi surface.
Because of this approximation, we need the additional approximations mentioned above,
when we match the solutions for $r>r_c$ and $r<r_c$.
We expect that for the exact solution of the zero energy Majorana mode,
the weight of quasiparticles from the large Fermi surface may become substantially large 
in the short-distance region in the vicinity of the center of the vortex core,
compared to the contributions from quasiparticles with $\bm{k}\sim 0$.
If we properly include the mixing with quasiparticles from the large Fermi surface,
we may be able to match the solutions without such additional approximations. 
Also, our zero energy solution is not regular at $r=0$, though it is
still normalizable, and physically allowed.
We expect that this singular behavior for $r\sim 0$ is also raised by our approximation neglecting the mixing with
quasiparticles from the large Fermi surface, which should be important for small $r$.
If one takes into account contributions of quasiparticles from the large Fermi surface, it may be possible to cure
this singular behavior of our solution at $r=0$.
Nevertheless, when the conditions $h \ll m\lambda^2$ and $0<h-\Delta <\Delta$ is satisfied,
the zero energy solution may be dominated by quasiparticles with $\bm{k}\sim 0$
in the long-distance asymptotic regime sufficiently 
far away from the center of the vortex core,
and our analytical solution obtained above may become a good approximation.

We would like to note that although the zero energy solution given by (\ref{sol1-main3}) and (\ref{sol1-main4})
with the asymptotic forms of (\ref{fasy1}) and (\ref{fasy2}) in the Appendix D looks like localized
even in the limit of $\Delta\rightarrow 0$,
this never means that there is a zero energy Majorana bound state in the normal state.
In fact, the above zero energy solution is not applicable to the normal state without a vortex,
because of the following reason.
The zero energy solutions in the normal state are indeed given by (\ref{sol1-main1}) and (\ref{sol1-main2}).
However, the Hankel function is not regular at $r=0$.
This singularity is unphysical because
in the absence of a vortex in the normal state, translation invariance is recovered,
and thus, there should not be a special point in the coordinate space at which the wave function is singular. 
This implies that the zero energy solution given by (\ref{sol1-main1}) and (\ref{sol1-main2})
is not allowed in the normal state without a vortex.
Thus, there is no zero energy bound state in the limit of $\Delta\rightarrow 0$.
The Majorana bound state in a vortex core exists only for
$\mu_{\rm B}H_z>\Delta>0$.

The above analysis may be also extended to the case of $d+id$-wave pairing straightforwardly,
since the $d+id$-wave state, which is the superposition of the $d_{x^2-y^2}$-wave state and the
$d_{xy}$-wave state, is an eigen state of the orbital angular momentum operator, as in the case of the $s$-wave pairing, and the treatment
for the gap function given above is applicable.
We obtain the Majorana fermion mode for a nonzero magnetic field $h>0$ in the case of
the $d+id$-wave pairing, which is consistent with the existence of the chiral Majorana edge discussed
in Sec.\ref{sec:edge}. 

We stress again that the Majorana zero-energy mode in a vortex core is formed mainly by
the superposition of an electron and a hole with the vanishing Fermi momentum $k_F\sim 0$
in the long-distance asymptotic regime, provided that the energy scale of the SO interaction is
sufficiently larger than the Zeeman energy and that $0<h-\Delta <\Delta$.
As will be discussed in the following, this property is very important for the
stability of the Majorana fermion against various sources of decoherence which exist in real materials and may 
destroy the Majorana fermion acting as a qubit.

\subsection{Strong stability of the Majorana fermion mode against thermal noise}

For the detection of the non-Abelian anyons and also for
the implementation of the topological quantum computation utilizing them, it is desirable
that the Majorana zero energy state in a vortex core is well separated from
excited states, the interaction with which may cause decoherence.
As was pointed out in ref.\cite{STF09}, in the non-Abelian phase
of $s$-wave Rashba superconductors,
when the energy scale of the SO interaction is much larger than the Zeeman energy
and the condition $0<\mu_{\rm B} H_z-\Delta <\Delta$ is satisfied,
the excitation energy of the vortex core state is of
order $\mu_{\rm B} H_z-\Delta$,
which is much larger than the typical size of the excitation energy in
the vortex core bound state of weak-coupling superconductors, i.e. $\sim \Delta^2/E_F$.\cite{CGM}
Thus, the Majorana fermion mode found here is quite stable against thermal noise
even at moderately low temperatures, to which it is not difficult to access within
standard experimental techniques.
We, here, explain the origin of the strong stability of the Majorana mode in more details.
The excitation energy in the vortex core is due to the kinetic energy of quasiparticles
in the Andreev bound state,
which stems from the derivative term in Eqs.(\ref{bdg3-1}) and (\ref{bdg3-2}) in the Appendix D. 
We restrict the following argument within the case that the magnitude of the SO interaction is much larger than the Zeeman energy, and that the condition $0<\mu_{\rm B} H_z-\Delta <\Delta$ is satisfied.
In this case, the Majorana solution is constructed mainly from quasiparticles with $\bm{k}\sim 0$,
as clarified in the previous section.
Then, the first order derivative terms $2\lambda (\frac{\partial}{\partial r}\pm \frac{i}{r}\frac{\partial}{\partial \theta})$
give leading contributions to the kinetic energy.
On the other hand, from the solution of the BdG equation (\ref{sol1-main3}) and (\ref{sol1-main4}),
we see that the characteristic size of the vortex core is $\xi_{\rm core}\sim 2\lambda/(\mu_{\rm B}H_z-\Delta)$. 
Thus, 
the excitation energy is of the order
$\sim 2\lambda/(2\lambda/(\mu_{\rm B} H_z-\Delta))\sim \mu_{\rm B} H_z-\Delta$.
This large magnitude of the excitation energy
implies that the Majorana zero energy mode in the $s$-wave Rashba superconductor
with $\mu_{\rm B}H_z>\Delta$ is significantly stable against thermal noise, compared to
 chiral $p+ip$ superconductors.
Also, such large excitation energy ensures that 
the experimental detection of the non-Abelian anyons is quite feasible for our system.
The origin of the strong stability of the Majorana fermion mode is deeply related to
the fact that it is mainly constructed from quasiparticles in 
the Dirac cone at the $\Gamma$ point in the Brillouin zone for $0<\mu_{\rm B} H_z-\Delta<\Delta$
 in the long-distance asymptotic regime,
as mentioned above.
Since the Dirac cone has a vanishing Fermi momentum, the kinetic energy is dominated by
the SO interaction 
which is of order $\lambda /\xi_{\rm core}$ rather than 
the standard kinetic energy term of order $\xi_{\rm core}^{-2}/(2m)$.
This feature leads to the strong stability of the Majorana zero energy mode.
It is noted that the robustness of the Majorana fermion mode in the $s$-wave Rashba superconductor
was also pointed out in ref. \cite{SAU2} from a different point of view.

\subsection{Stability against decoherence due to inter-vortex tunneling}

It has been proposed that the Majorana fermion modes in superconductors can be utilized
as decoherence-free qubits, which enable us the construction of the fault-tolerant 
topological quantum computer.\cite{freedman,kitaev2,das,ANB}
Two Majorana fermions, say $\gamma_1$ and $\gamma_2$, constitute one
complex fermion state described by $\psi=\gamma_1+i\gamma_2$, which is occupied or unoccupied.
This doubly-degenerate state stores a qubit non-locally, which is protected against any local perturbations,
as long as the distance between two vortices, each of which contains one Majorana mode is sufficiently large.
One crucial obstruction to this scheme is the decoherence raised by inter-vortex tunneling;
tunneling processes between two Majorana modes in two vortices lift the degeneracy,
leading to decoherence.
In particular, it was pointed out by  Cheng et al.\cite{cheng} that
the energy of the complex fermion $\psi$ exhibits quantum oscillation as a function of the spatial separation $r$ between two vortices as $\sim \cos k_Fr$ with $k_F$ the Fermi momentum, and thus takes both positive and negative values depending on $r$. This rapid change of the sign of the energy seriously flaws 
the initialization and the readout of the qubit.

Here, we discuss the drastic suppression of the quantum oscillation for inter-vortex tunneling in 
the Rashba $s$-wave superconductors in a particular parameter region.
As explained in Sec.\ref{sec:vortexsol}, 
the zero energy Majorana bound state in a vortex core obtained above consists of two contributions; one from
quasiparticles with $\bm{k}\sim 0$ and the other from quasiparticles with
the large $k_F$. 
The former contribution has the characteristic length scale of order $\lambda/(\mu_{\rm B} H_z-\Delta)$,
as clarified by the solution (\ref{sol1-main3}) and (\ref{sol1-main4}).
On the other hand, for the latter contribution, the characteristic length scale is of order
$\lambda/\Delta$, since the Fermi velocity is of order $\lambda$ when $\mu \sim 0$ in Eq.(\ref{eq:BDGham}).
Then, in the case of $0<\mu_{\rm B} H_z-\Delta < \Delta$, 
the Majorana zero mode is mainly formed by quasiparticles with the vanishing
Fermi momentum $k_F\sim 0$ in the long-distance asymptotic regime.
That is, in the tunneling process 
between two vortices separated by the distance $R$,
the overlap of the component with $\bm{k}\sim 0$, which is 
of order $\exp(-R(\mu_{\rm B} H_z-\Delta)/\lambda)$ is much larger in the magnitude than 
the oscillating contribution from the large Fermi surface which is 
of order $ \exp(-R \Delta/\lambda)\cos(k_F R)$ for large $R$.
The inter-vortex tunneling mediated via quasiparticles with $k_F\sim 0$
does not involve the quantum oscillation,
making a sharp contrast to Majorana modes found in $p+ip$ superconductors.
As a result, the energy of a complex fermion made of the two Majorana fermions
is dominated by the non-oscillating part from quasiparticles with $\bm{k}\sim 0$,
and thus, the decoherence due to the quantum oscillation is suppressed.
It is noted that this protection mechanism also works for 
a Majorana fermion mode in the proximity between
a topological insulator and an $s$-wave superconductor,\cite{FK08} as long as the chemical potential is
properly tuned to realize $k_F\sim 0$ for the surface Dirac cone.
Since momentum is not conserved in the vicinity of a vortex,
there may be hybridization between the Majorana state in a vortex core and
quasiparticles with the finite Fermi momentum, which raises inter-vortex tunneling
involving the quantum oscillation.
However, the analysis of the zero energy Majorana mode in a vortex core in
the previous subsections implies that we can construct the zero energy Majorana state
which is mainly formed by quasiparticles with $\bm{k}\sim 0$, provided
that the energy scale of the SO interaction is much larger than the Zeeman energy, and that
$0<\mu_{\rm B} H_z-\Delta<\Delta$.
Thus, it may be possible to realize the Majorana fermion in the vortex core of the Rashba superconductor 
which is stable against decoherence due to the inter-vortex tunneling.

\section{Topological density wave states and charge fractionalization}
\label{sec:densitywave}

The above argument for the non-Abelian topological order in $s$-wave superconductors implies that
topological order is realizable in conventional spin (or charge) density wave states.
The Hamiltonian for the $s$-wave superconductivity on a bipartite lattice is mapped to
the Hamiltonian for the spin density wave (SDW) state or
the charge density wave (CDW) state by changing the basis of fermion fields.
Thus, an $s$-wave superconducting state with the topological order
can be mapped to a density wave state with a certain topological order.
As shown below, the topological order in the density wave state is Abelian, and
quasiparticles in this topological order are not Majorana fermions, but possess U(1) charge.
However, the quasiparticles exhibit charge fractionalization which characterizes the Abelian topological order.

\subsection{Topological spin density wave state}

We, first, consider the SDW state with the order parameter
$\Delta_{\rm S}=\langle c^{\dagger}_{k\uparrow}c_{k+Q\uparrow}\rangle=
-\langle c^{\dagger}_{k\downarrow}c_{k+Q\downarrow}\rangle$, where $\bm{Q}$ is the ordering wave number vector.
We also assume that there are the Rashba-type SO interaction and the Zeeman magnetic field 
$h=\mu_{\rm B}H_z$.
Then, the mean field Hamiltonian is given by
\begin{eqnarray}
\mathcal{H}_{\rm SDW}=\frac{1}{2}\sum_{\bm{k}}\Psi^{\dagger}_{\bm{k}}\mathcal{H}_{\rm SDW}(\bm{k})
\Psi_{\bm{k}},
\end{eqnarray}
with
\begin{eqnarray}
\mathcal{H}_{\rm SDW}(\bm{k})=
\left(
\begin{array}{cc}
\varepsilon({\bm{k}})-h\sigma_z+\alpha\bm{\mathcal{L}}_0(\bm{k})\cdot\bm{\sigma} &
i\Delta_{\rm S}\sigma_y \\
-i\Delta_{\rm S}\sigma_y & 
\varepsilon({\bm{k+Q}})+h\sigma_z+\alpha\bm{\mathcal{L}}_0(\bm{k}+\bm{Q})\cdot\sigma_x\bm{\sigma}\sigma_x 
\end{array}
\right)
\label{hamsdw}
\end{eqnarray}
and $\Psi^T_{\bm{k}}=(c_{\bm{k}\uparrow},c_{\bm{k}\downarrow},c_{\bm{k}+\bm{Q}\downarrow},c_{\bm{k}+\bm{Q}\uparrow})$.
In the following, we assume the perfect nesting condition of the energy band
$\varepsilon({{\bm k}+{\bm Q}})=-\varepsilon({\bm k})$. This situation is realized in the case of the half-filling
electron density for our model on the square lattice with the nearest-neighbor hopping.
The nesting vector is $\bm{Q}=(\pm\pi,\pm\pi)$.
Furthermore, we postulate $\bm{\mathcal{L}}_0(\bm{k}+\bm{Q})=\bm{\mathcal{L}}_0(\bm{k})$.
This condition is satisfied when $\bm{\mathcal{L}}_0(\bm{k})=(\sin 2k_y,-\sin 2k_x)$ in the case of
$\bm{Q}=(\pm\pi,\pm\pi)$.
Then, the Hamiltonian (\ref{hamsdw}) has the same form as that of the 
Rashba $s$-wave superconductor considered in the previous sections, and topological order is realized when
the condition $h>\Delta_{\rm S}$ is fulfilled.

In a similar manner as Sec.\ref{sec:topologicalnumber}, topological numbers of the SDW state
are calculated. 
The energy gap of the system closes one of the following conditions are
satisfied,  
\begin{eqnarray}
\Delta_{\rm S}^2=(\mu_{\rm B}H_z)^2, 
\quad
4t^2+\Delta_{\rm S}^2=(\mu_{\rm B}H_z)^2,
\quad
16t^2+\Delta_{\rm S}^2=(\mu_{\rm B}H_z)^2,
\end{eqnarray}
and the winding number $I(k_y)$ is calculated as
\begin{eqnarray}
I(k_y)&=&\frac{1}{2}\left[{\rm sgn}[\varepsilon(-\pi/2,k_y)^2-(\mu_{\rm
		   B}H_z)^2+\Delta_{\rm S}^2]
-{\rm sgn}[\varepsilon(0,k_y)^2-(\mu_{\rm
		   B}H_z)^2+\Delta_{\rm S}^2]
\right.
\nonumber\\
&&\left.
+{\rm sgn}[\varepsilon(\pi/2,k_y)^2-(\mu_{\rm
		   B}H_z)^2+\Delta_{\rm S}^2]
-{\rm sgn}[\varepsilon(\pi,k_y)^2-(\mu_{\rm
		   B}H_z)^2+\Delta_{\rm S}^2]
\right]. 
\end{eqnarray}
Because ${\cal H}_{\rm
SDW}^{*}(-{\bm k})={\cal H}_{\rm SDW}({\bm k})$ at $k_y=0,\pm\pi/2,\pi$,
the winding number $I(k_y)$ is defined at these values of $k_y$.
We summarize the winding number of the SDW state in Table \ref{table:sdw}.
The TKNN integer in the magnetic Brillouin
zone, which we denote as $I_{\rm mTKNN}$, is also given in Table
\ref{table:sdw}.

\begin{figure}[h]
\begin{center}
\includegraphics[width=7cm]{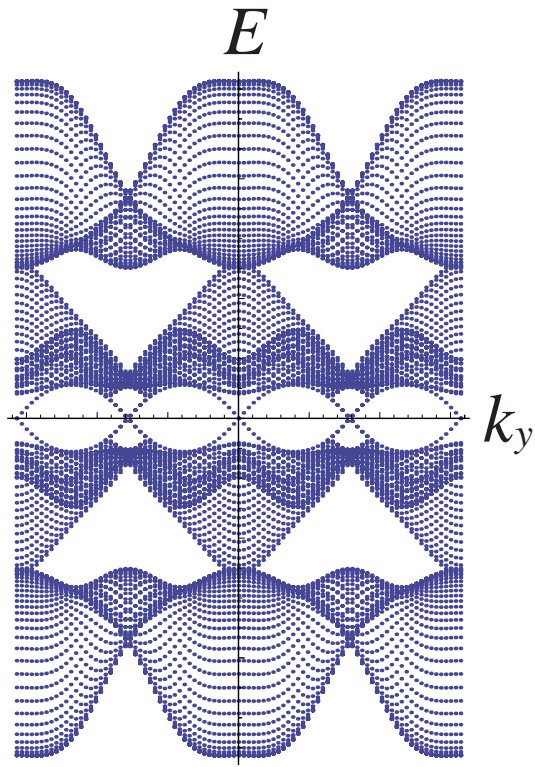}
\caption{The energy spectra of the NCS SDW state with edges at $i_x=0$
 and $i_x=30$. Here $k_y$ denotes the momentum in the $y$-direction, and
 $k_y\in [-\pi,\pi]$. We
 take $t=1=\alpha=1=\Delta_{\rm S}=1$.
 The Zeeman magnetic field $H_z$ is $\mu_{\rm B}H_z=1.5$. 
 At $k_y=0,\pm\pi/2,\pi$, 
 we have 
 gapless states on each edge. These gapless edge modes are chiral.}
\label{fig:edgestate_sdw}
\end{center}
\end{figure}

In a similar manner as $s$-wave NCS
superconductors, gapless edge modes appear in
the NCS SDW state considered above when  $h>\Delta_{\rm S}$.
See Fig.\ref{fig:edgestate_sdw}.
However, in comparison with the case of $s$-wave superconductivity, 
we should taken into account only half of the zero energy modes,  
since 
the Brillouin zone is folded to the magnetic Brillouin zone in the SDW
state, and as a result, the $\bm{k}$-point $\bm{k}=(\pi,0)$ is
equivalent to $\bm{k}=(0,\pi)$.
In other words, 
when $(-1)^{I_{\rm mTKNN}}=-1$, we have odd numbers of zero energy edge
modes in the SDW state. In Fig.\ref{fig:edgestate_sdw}, we see three zero energy modes.
The existence of odd numbers of zero energy edge modes implies that
in the vortex core of the SDW order $\Delta_{\rm S}e^{i\theta}$, there
are odd numbers of zero energy modes, because of the bulk-edge correspondence.
These zero energy modes have U(1) charge.
As in the case of the Su-Schrieffer-Heager model,\cite{SSH} these isolated
zero energy modes with U(1) charge
lead to the charge fractionalization.\cite{jack}
That is, when these three zero modes are occupied by electrons,
the charge carried by the vortex core is $Q=3e/2$.

\begin{table}
\begin{center} 
\begin{tabular}[t]{|c|c|c|c|}
\hline
\hline
\multicolumn{4}{c}{} \\ 
\hline
$(\mu_{\rm B}H_z)^2$ &$(-1)^{I_{\rm mTKNN}}$& $I(0)=I(\pi)$ & $I(\pi/2)=I(-\pi/2)$\\ 
\hline 
 $0<(\mu_{\rm B}H_z)^2<\Delta_{\rm S}^2$ &  1 &0 &0 \\ 
$\Delta_{\rm S}^2<(\mu_{\rm B}H_z)^2< 4t^2+\Delta_{\rm S}^2$
& -1 & 1 &2 \\
 $4t^2+\Delta_{\rm S}^2<(\mu_{\rm B}H_z)^2<16t^2+\Delta_{\rm S}^2$
& -1 & -1 &0 \\
 $16t^2+\Delta_{\rm S}^2 <(\mu_{\rm B}H_z)^2$ & 1& 0 & 0 \\
\hline
\multicolumn{4}{c}{}\\
\hline
\hline
 \end{tabular} 
\end{center}
\caption{The TKNN integer in the magnetic BZ, $I_{\rm mTKNN}$, and the
 winding number $I(k_y)$ for 2D SDW state}
\label{table:sdw}
\end{table}

\subsection{Topological charge density wave state}

\begin{figure}[h]
\begin{center}
\includegraphics[width=5cm]{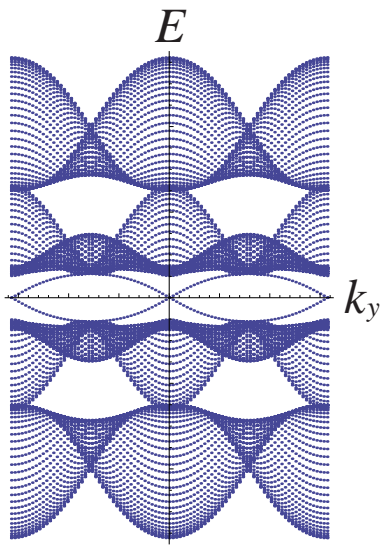}
\caption{The energy spectra of the NCS CDW state with edges at $i_x=0$
 and $i_x=30$. Here $k_y$ denotes the momentum in the $y$-direction, and
 $k_y\in [-\pi,\pi]$. We
 take $t=1$, $\mu=0$, $\lambda=0.5$, and $\Delta_{\rm C}=1$.
 The Zeeman magnetic field $H_z$ is $\mu_{\rm B}H_z=1.5$.}
\label{fig:edgestate_dws}
\end{center}
\end{figure}

The above consideration for the topological SDW state is also applicable to the CDW state.
We consider the CDW state with the order parameter
$\Delta_{\rm C}=\langle c^{\dagger}_{k\uparrow}c_{k+Q\uparrow}\rangle=
\langle c^{\dagger}_{k\downarrow}c_{k+Q\downarrow}\rangle$ in the case with the Rashba SO interaction.
The mean field Hamiltonian for the Rashba CDW state is
\begin{eqnarray}
\mathcal{H}_{\rm CDW}=\frac{1}{2}\sum_{\bm{k}}\Psi^{\dagger}_{\bm{Ck}}\mathcal{H}_{\rm CDW}(\bm{k})
\Psi_{C\bm{k}},
\end{eqnarray}
with
\begin{eqnarray}
\mathcal{H}_{\rm CDW}(\bm{k})=
\left(
\begin{array}{cc}
\varepsilon({\bm{k}})-h\sigma_z+\alpha\bm{\mathcal{L}}_0(\bm{k})\cdot\bm{\sigma} &
i\Delta_{\rm C}\sigma_y \\
-i\Delta_{\rm C}\sigma_y & 
\varepsilon({\bm{k+Q}})+h\sigma_z+\alpha\bm{\mathcal{L}}_0(\bm{k}+\bm{Q})\cdot\bm{\sigma}^*
\end{array}
\right)
\label{hamcdw}
\end{eqnarray}
and $\Psi^T_{C\bm{k}}=(c_{\bm{k}\uparrow},c_{\bm{k}\downarrow},-c_{\bm{k}+\bm{Q}\downarrow},c_{\bm{k}+\bm{Q}\uparrow})$.
We assume the prefect nesting condition $\varepsilon({\bm k}+{\bm
Q})=-\varepsilon({\bm k})$ again, and also assume that
$\bm{\mathcal{L}}_0(\bm{k}+\bm{Q})=-\bm{\mathcal{L}}_0(\bm{k})$.
The condition for $\bm{\mathcal{L}}_0(\bm{k})$ is satisfied when
$\bm{\mathcal{L}}_0(\bm{k})=(\sin k_y,-\sin k_x)$ in the case of
$\bm{Q}=(\pm\pi,\pm\pi)$.
Eq.(\ref{hamcdw}) is formally equivalent to the Hamiltonian of the Rashba $s$-wave superconductor.
In this CDW state, the topological ordered phase with a single gapless
edge mode realizes.
We depict an example of the energy spectra for the system with open boundary edges 
in the case of $h>\Delta_{\rm C}$
in Fig.\ref{fig:edgestate_dws}.
There are two zero energy edge modes at $k_y=0$ and $\pi$.
Since the Brillouin zone is folded by the CDW order with the ordering vector $\bm{Q}=(\pi,\pi)$,
the zero mode at $k_y=\pi$ is equivalent to that at $k_y=0$. Thus there is only one zero mode. 
In this Abelian topological phase, the charge fractionalization can occur, as in the case of the NCS SDW state.
When there is a vortex of the CDW order, the zero mode of which is occupied by one electron,
the vortex carries the fractional charge $e/2$.

\section{Discussion and Conclusion}
\label{sec:discussion}

\subsection{Realization scheme for the non-Abelian topological order in spin-singlet superconductors}

In this section, we discuss possible realization schemes for the non-Abelian topological order
considered in the previous sections. 
Recently, it was discussed by several authors that $s$-wave Rashba superconductors with the condition
$\mu_{\rm B}H_z>\Delta$
are realizable in heterostructure semiconductor devices.\cite{SAU,ALICEA}
We, here, present two more possible schemes; one
based on quasi-two-dimensional bulk superconductors, and the other
utilizing ultracold atoms.

\subsubsection{Possible realization in strongly correlated electron systems}

The most crucial hurdle for the realization of the non-Abelian topological order in the Rashba superconductors
is to stabilize the superconducting state under applied strong magnetic fields $\mu_{\rm B}H_z>\Delta$.
As mentioned before, the Pauli depairing effect due to magnetic fields is absent
when the SO splitting of the Fermi surface is sufficiently larger than the superconducting gap and
the Zeeman energy.\cite{Frigeri,fuji3} 
Such large SO interactions are typically realized in noncentrosymmetric materials.
Then, the issue which should be addressed here is how to suppress
the orbital depairing effect of magnetic fields.
The typical size of the orbital limiting field $H^{\rm orb}_{c2}$  is 
given by
$\mu_{\rm B}H^{\rm orb}_{c2}\sim \Delta^2/(zE_F)$. Here $z$ is the mass normalization factor.
If the mass enhancement factor $1/z$ is sufficiently large, it is possible to attain
$\mu_{\rm B}H^{\rm orb}_{c2}>\Delta$.
For instance, for strongly correlated electron systems such heavy fermion superconductors,
$1/z$ can reach to be $\sim 100$.
Thus, it is not difficult to fulfill the condition $\mu_{\rm B}H^{\rm orb}_{c2}>\Delta$ even when
$\Delta/E_F\sim 0.01$.
The above argument implies that
noncentrosymmetric heavy fermion superconductors may  be good candidates
for the realization of the non-Abelian topological order.
Unfortunately, to this date, there are no noncentrosymmetric heavy fermion superconductors with no gap-nodes. Experimental studies suggest that all heavy fermion superconductors without inversion symmetry
discovered so far such as CePt$_3$Si and CeRh(Iir)Si$_3$ possess nodes of the superconducting gap,\cite{CePtSi1,CePtSi2,CeRhSi,CeIrSi,mukuda,tada} excitations from which may destabilize
the topological order.

\subsubsection{Possible realization in ultracold fermionic atoms}

We now discuss an experimental scheme for the realization of
 the non-Abelian topological order in ultracold fermionic atoms. 
 We can use the Feshbach resonance in the $s$-wave channel for the formation of
the $s$-wave Cooper pairs in this system \cite{sbcs,chin}.
For superfluid states of charge neutral atoms, there is no orbital depairing effect
 due to applied magnetic fields. 
Moreover, it was recently proposed by several authors that effective
SO interactions acting on charge neutral atoms
can be generated by spatially varying laser fields \cite{ost,ruse,zhu,stan}.
When the effective SO interaction is the Rashba type (or more generally, anti-symmetric),
and its magnitude is sufficiently larger than the Zeeman energy due to an applied magnetic field,
the Pauli depairing effect is suppressed, as mentioned before.
Thus, the condition for the non-Abelian topological order
$\mu_{\rm B}H_z>\Delta$ can be easily fulfilled.
As discussed in the previous subsections, 
the non-Abelian anyons are stable for sufficiently low energies
$\ll {\rm min}\{\mu_{\rm B}H_z-\Delta,\Delta\}$.
Since the BCS gap $\Delta$ can be tuned to be large, i.e. $\Delta\sim E_F$, 
by using the $s$-wave Feshbach resonance,
the realization of the non-Abelian anyons in this scheme is quite promising.
In the following, we present two schemes for the realization of the 
topological phase in ultracold fermionic atoms with laser generated SO interaction.

The first scheme utilizes an optical lattice with laser-assisted tunneling of fermionic atoms.
We consider 
fermionic atoms loaded in a two-dimensional (2D) periodic optical lattice.\cite{ost}
The atoms occupy doubly-degenerate Zeeman levels of the hyperfine 
ground state manifolds, which correspond to the spin up and spin down states of electrons. 
We introduce the Zeeman field to lift the degeneracy.
We assume that standard tunneling of atoms between sites due to
kinetic energy
is suppressed by the large depth of the optical lattice potential.
Tunneling of atoms between neighboring sites along the $\nu$-direction
($\nu=x,y$)
which conserves spins 
is caused by laser beams 
via optical Raman transitions.\cite{ost,jak}
In addition, tunneling which accompanies spin flip
is also driven by two Raman lasers.\cite{ost}
The Rabi frequency of the laser 
$\Omega_{\nu 1}$ ($\Omega_{\nu 2}$) 
is resonant for transition from the spin-up state to the spin-down state
for the tunneling between neighboring sites in the forward (backward) 
$\nu$-direction. 
Furthermore,
the confining optical potential is tilted along both the $x$-direction 
and the $y$-direction to assure that the forward and backward
tunneling processes are
respectively induced by the lasers with the different Rabi frequencies
$\Omega_{\nu 1}$ and $\Omega_{\nu 2}$.
The tilting potential for the $x$-direction $\Delta_x$ is different from that for $y$-direction
$\Delta_y$
to ensure that tunneling accompanying spin flip along the $y(x)$-direction is not raised by
the lasers with $\Omega_{x(y)1,2}$.
It is also assumed that the detuning from excited states for optical Raman
transitions is much larger than $\Delta_{x(y)}$, and thus
the spatial variation of the amplitudes of the Rabi frequencies due to
the tilting potential is negligible.
To realize the Rashba SO interaction for the two Zeeman levels, 
we choose the phases of the lasers as follows.
The lasers are propagating along the $z$-direction with an oscillating factor
$e^{ik_zz}$.
The Rabi frequency $\Omega_{x 2}$ is expressed as 
$\Omega_{x 2}=|\Omega_{x 2}|e^{ik_zz}$.
The phase of the laser $\Omega_{x 1}$ is shifted by $\pi$ 
from that of $\Omega_{x 2}$, and
$\Omega_{x 2}=-\Omega_{x 1}$ holds. 
The phase of $\Omega_{y 1}$ ($\Omega_{y 2}$) is shifted by $-\pi/2$ ($\pi/2$), 
and $\Omega_{y 2}=-i\Omega_{x 1}$, $\Omega_{y 2}=-\Omega_{y 1}$.
Then, the laser-induced tunneling term which accompanies spin flip
is expressed by
\begin{eqnarray}
\mathcal{H}_{\rm SO}&=&\sum_{i}[\alpha_x(c^{\dagger}_{{\bm i}-\hat{x}\downarrow}
c_{{\bm i}\uparrow}-
c^{\dagger}_{{\bm i}+\hat{x}\downarrow}c_{{\bm i}\uparrow}) \nonumber \\
&+&i\alpha_y(c^{\dagger}_{{\bm i}-\hat{y}\downarrow}c_{{\bm i}\uparrow}-
c^{\dagger}_{{\bm i}+\hat{y}\downarrow}c_{{\bm i} \uparrow})+{\rm H.c.}],
\nonumber\\
\alpha_{\nu}&=&c_{\nu}\int d\mbox{\boldmath $r$}\psi^{*}_{\downarrow}
(\mbox{\boldmath $r$}
-\mbox{\boldmath $r$}_{i-\hat{\mu}})\Omega_{\nu 2}(\mbox{\boldmath $r$})
\psi_{\uparrow}(\mbox{\boldmath $r$}-\mbox{\boldmath $r$}_i), 
\label{so}
\end{eqnarray}
with $\nu=x$, $y$, and 
$c_x=1$ and $c_y=-i$.
Since we consider the 2D $xy$-plane with $z=0$,
$\alpha_{\nu}$ is real.
For $\alpha_x=\alpha_y$, Eq.(\ref{so}) is the Rashba SO interaction.

In the second scheme, we employ the idea that an effective SO interaction 
is created by utilizing the dark states
generated by spatially-varying laser fields, as proposed in refs. \cite{ruse,stan}
We can introduce a vortex of the superfluid order parameter by
changing the topology of the shape of a trapping potential.
Since the underlying topological effective 
field theory is the SU(2)$_2$ Chern-Simons theory,\cite{moore,nayak,fradkin}
a vortex with a zero energy Majorana mode is equivalent to
a hole pierced in the system provided that 
the total number of the holes in the system are odd.
The trapping potential with holes can be prepared by using
a holographically engineered laser technique.\cite{HG1}
Furthermore, the dynamical motions of holes are possible through
a computer-generated hologram; i.e. the positions of holes can be moved by changing temporally 
the shape of the confining potential which may be
achieved by preprogrammed hologram.
Remarkably, this enables
the braiding of vortices (holes), which amounts to a quantum gate
for fault-tolerant quantum computation based upon the manipulation
of non-Abelian anyons.\cite{kitaev2,freedman}

\subsection{Conclusion}

We have verified that the non-Abelian topological order, which is characterized by
the existence of chiral Majorana edge modes and a Majorana fermion in a vortex core,
can be realized in almost all classes of fully-gapped spin-singlet superconductors with anti-symmetric SO interactions such as the Rashba SO interaction in the case with the Zeeman magnetic field.
Majorana fermions in superconductors have been, recently, attracting much attention in connection
with the topological quantum computation.
The idea of the topological quantum computation has been examined for
the non-Abelian fractional quantum Hall state,\cite{freedman,das} spin-triplet chiral $p$-wave superconductors,\cite{tewari2} and
the proximity between a topological insulator and a conventional superconductor.\cite{FK08,ANB}
Our results in the present paper provide another category of promising systems for the realization of the topological quantum computation. 
We would like to stress that our systems in the normal state are topologically trivial; i.e.
the combination of the conventional superconducting order and the conventional anti-symmetric SO interaction with the Zeeman field gives rise to a highly nontrivial topological phase.
In our scenario, the existence of the Zeeman energy, the magnitude of which
exceeds the size of the superconducting gap opened at time-reversal invariant $\bm{k}$-points 
in the Brillouin zone where
the SO interaction vanishes, is important.
At these $\bm{k}$-points, there are Dirac cones in the absence of magnetic fields.
When the magnetic field is switched on, the effective gap of electrons in the vicinity of the Dirac cone
is given by $\Delta-\mu_{\rm B}H_z$. 
As the magnetic field increases, and $\mu_{\rm B}H_z$ exceeds $\Delta$, 
the sign of the effective gap changes, and
topological phase transition occurs.
As indicated by the analysis of Majorana fermion modes in open boundary edges and in a vortex core 
in Sec.\ref{sec:edge} and Sec.\ref{sec:vortex}, and in the topological argument based on
the winding number in Sec.\ref{sec:topologicalnumber},
quasiparticles in the vicinity of the Dirac cone, i.e. time-reversal invariant $\bm{k}$-points, 
play an important role for the realization of the topological quantum phase transition.
In the case of the $s$-wave pairing state, the large magnetic field satisfying $\mu_{\rm B}H_z>\Delta$ is required to realize the non-Abelian topological order.
For conventional bulk weak-coupling superconductors, it is difficult to fulfill this condition,
though there are some proposals to realize this system in heterostructure devices, for which
$\mu_{\rm B}H_z>\Delta$ is attainable.\cite{SAU,ALICEA} 
We have proposed promising schemes for the realization of the condition $\mu_{\rm B}H_z>\Delta$
in the bulk $s$-wave pairing
state; one based on strongly correlated electron systems, and the other utilizing ultracold fermionic atoms.
In contrast to the $s$-wave pairing state, in the case of the $d+id$ wave pairing state, 
a small magnetic field larger than the lower critical field is sufficient to fulfill the condition $\mu_{\rm B}H_z>\Delta_k$ in the vicinity of
time-reversal invariant $\bm{k}$-points, and thus the realization of the non-Abelian topological phase is easier than the $s$-wave case.
However, unfortunately, up to our knowledge, there is no superconductor for which the $d+id$-wave pairing state is experimentally established, though it was suggested by some authors that this pairing state may be
realized in sodium cobalt oxide superconductors Na$_x$CoO$_2\cdot$yH$_2$O.\cite{cobalt1,cobalt2}
It is an interesting open issue to pursue the experimental detection of Majorana fermions in such
superconductors mentioned above.
We have also clarified that, in certain parameter regions, the Majorana fermion realized in Rashba spin-singlet superconductors
is stable against various sources of decoherence such as thermal fluctuations which exist in real systems generically. Because of this feature, the Rashba spin-singlet superconductor may be a promising candidate for the realization of the topological quantum computation.

As a byproduct of our analysis, 
we have also found that an Abelian topological order, which supports the existence of gapless Dirac fermions 
in edges and the charge fractionalization, is realizable in the conventional SDW or CDW state
with an anti-symmetric SO interaction.

\acknowledgments
The authors thank S. Das Sarma for discussions.
M.S. would like to thank Y. Tanaka for discussions. 
This work is partly supported by the Sumitomo Foundation (M.S.) and the
Grant-in-Aids for Scientific Research
(Grants No.19052003, No.21102510 (S.F.) and No.22540383 (M.S.) ).

\appendix

\section{TKNN number and the dual transformation}
\label{appendix:a}
In this appendix, we show that the dual transformation (\ref{eq:duality})
does not change the TKNN number, and hence
the topological properties of the original Hamiltonian (\ref{eq:BdG}) and the dual
Hamiltonian (\ref{eq:duality}) are the same.
Let us consider the BdG equations for the original Hamiltonian ${\cal
H}({\bm k})$ and that for the dual one ${\cal H}_{\rm D}({\bm k})$,
\begin{eqnarray}
{\cal H}({\bm k})|\phi_n({\bm k})\rangle=E_n({\bm k})|\phi_n({\bm k})\rangle,
\quad
{\cal H}_{\rm D}({\bm k})|\phi^{\rm D}_n({\bm k})\rangle=E_n({\bm
k})|\phi^{\rm D}_n({\bm k})\rangle.  
\end{eqnarray}
Since the dual transformation is a unitary transformation, the energy
spectrum of the dual Hamiltonian is the same as that of the original one. 
Moreover, the eigen state $|\phi_n^{\rm D}({\bm k})\rangle$ for the dual
Hamiltonian is related to the eigen state $|\phi_n({\bm k})\rangle$ for
the original one as
\begin{eqnarray}
|\phi_n^{\rm D}({\bm k})\rangle= D|\phi_n({\bm k})\rangle,
\end{eqnarray}
with $D$ in (\ref{eq:D}).
As $D$ is a constant unitary matrix, the gauge field constructed from
the occupied states for the dual Hamiltonian is the same as that for the
original one,
\begin{eqnarray}
A_i^{(-){\rm D}}({\bm k})=i\sum_{E_n({\bm k})<0}\langle \phi_n^{\rm
 D}({\bm k})|\partial_{k_i}\phi_n^{\rm D}({\bm k})\rangle
= i\sum_{E_n({\bm k})<0}
\langle \phi_n({\bm k})|\partial_{k_i}\phi_n({\bm k})\rangle=A_i^{(-)}({\bm k}).
\end{eqnarray}
This relation immediately means that the TKNN number (\ref{eq:TKNN0}) is
invariant under the dual transformation.

\section{TKNN number and winding number}
\label{appendix:b}

\subsection{TKNN number for TRB topological superconductors}

In this appendix, we summarize useful properties of the TKNN number for TRB
topological superconductors.\cite{Sato10}
Let us first consider the BdG equation,
\begin{eqnarray}
{\cal H}({\bm k})|\phi_n({\bm k})\rangle=E_n({\bm k})|\phi_n({\bm k})\rangle, 
\end{eqnarray}
where the BdG Hamiltonian ${\mathcal H}({\bm k})$ has the particle-hole
symmetry,
\begin{eqnarray}
\Gamma{\cal H}({\bm k})\Gamma^{\dagger}=-{\cal H}(-{\bm k})^*, 
\end{eqnarray}
with the $4\times 4$ matrix $\Gamma$ 
\begin{eqnarray}
\Gamma=\left(
\begin{array}{cc}
0 & {\bm 1}_{2\times 2}\\
{\bm 1}_{2\times 2} & 0
\end{array}
\right). 
\end{eqnarray}
From the particle-hole symmetry, we can say that if $|\phi_n({\bm
k})\rangle$ is a positive energy state with $E_n({\bm k})>0$, then 
$\Gamma|\phi_n^*(-{\bm k})\rangle$ is a negative energy state with
$-E_n(-{\bm k})$. 
Therefore, in the following, we use a positive (negative) $n$ to
represent a positive (negative) energy state, and set 
\begin{eqnarray}
|\phi_{-n}({\bm k})\rangle=\Gamma|\phi_n^{*}(-{\bm k})\rangle. 
\label{eq:phphi}
\end{eqnarray}
For the ground state in a superconductor, the negative energy states
are occupied. 

Now we introduce the following
``gauge fields'' $A_i^{(\pm)}({\bm k})$,
\begin{eqnarray}
A_i^{(\pm)}=i\sum_{n \gtrless 0}\langle \phi_n({\bm k})|
\partial_{k_i}\phi_n({\bm k})\rangle.
\end{eqnarray}
From (\ref{eq:phphi}), the gauge fields $A_i^{(\pm)}({\bm k})$ satisfy
\begin{eqnarray}
A_i^{(+)}({\bm k})=A_i^{(-)}(-{\bm k}). 
\label{eq:phsA}
\end{eqnarray}
It is also found that their sum $A_i({\bm k})\equiv A_i^{(+)}({\bm
k})+A_i^{(-)}({\bm k})$ is given by a total derivative
of a function. To see this, we rewrite $\phi_n({\bm k})$ in the
components, 
\begin{eqnarray}
|\phi_n({\bm k})\rangle
=\left(
\begin{array}{c}
\phi_n^1({\bm k}) \\
\phi_n^2({\bm k}) \\
\phi_n^3({\bm k}) \\
\phi_n^4({\bm k}) 
\end{array}
\right), 
\end{eqnarray}
and introduce the $4\times 4$ unitary matrix $W({\bm k})$ as
\begin{eqnarray}
W_{a,n}({\bm k})=\phi_n^a({\bm k}). 
\end{eqnarray}
Then $A_i({\bm k})$ is rewritten as the total derivative of $i{\rm
ln}[{\rm det}W({\bm k})]$,
\begin{eqnarray}
A_i({\bm k})=i{\rm tr}\left[W^{\dagger}({\bm k})\partial_{k_i}W({\bm k}) \right]
=i\partial_{k_i}{\rm ln}[{\rm det}W({\bm k})].
\end{eqnarray}
This equation implies that the field strength of $A_i({\bm k})$ is
identically zero, 
\begin{eqnarray}
{\cal F}({\bm k})=\epsilon^{ij}\partial_{k_i}A_j({\bm k})=0. 
\end{eqnarray}
Combining this with (\ref{eq:phsA}), we find that the field strength
${\cal F}^{(\pm)}({\bm k})$ of $A_i^{(\pm)}({\bm k})$ satisfies 
\begin{eqnarray}
{\cal F}^{(\pm)}({\bm k})={\cal F}^{(\pm)}(-{\bm k}).
\label{eq:phsf}
\end{eqnarray}

By using the field strength of the occupied state, the TKNN number is define as
\begin{eqnarray}
I_{\rm TKNN}=\frac{1}{2\pi}\int_{T^2}d^2k {\cal F}^{(-)}({\bm k}),
\end{eqnarray}
where $T^2$ is the first Brillouin zone in the momentum space. Using the
relation (\ref{eq:phsf}) and the Stokes' theorem, we obtain 
\begin{eqnarray}
I_{\rm TKNN}&=&\frac{1}{\pi}\int_{T^2_{+}} d^2k {\cal F}^{(-)}({\bm k})
\nonumber\\
&=&\frac{1}{\pi}\left[\int_{-\pi}^{\pi}dk_x A_{x}^{(-)}(k_x,0)
-\int_{-\pi}^{\pi}dk_x A_x^{(-)}(k_x,\pi)\right],
\label{eq:TKNNA}
\end{eqnarray}
where $T_+^2$ is the upper half plane of $T^2$.
As is shown in Appendix \ref{appendix:b3}, this formula enables us to connect
the Chern number to the winding number defined in (\ref{eq:defwinding}).

\subsection{winding number}
\label{appendix:b2}

As was discussed in Sec.\ref{sec:topologicalnumber}, the winding number $I(k_y)$ is defined
by taking the basis where the BdG Hamiltonian has the following
particular form,\cite{SF08}
\begin{eqnarray}
{\cal H}({\bm k})=\left(
\begin{array}{cc}
0 & q({\bm k})\\
q^{\dagger}({\bm k}) & 0
\end{array}
\right). 
\label{eq:qqhamiltonian}
\end{eqnarray}
By using $q({\bm k})$, the winding number is defined as
\begin{eqnarray}
I(k_y)=\frac{1}{4\pi i}\int_{-\pi}^{\pi}dk_x {\rm tr}
\left[q^{-1}({\bm k})\partial_{k_x}q({\bm k})-q^{\dagger -1}({\bm k})
\partial_{k_x}q^{\dagger}({\bm k})\right], 
\end{eqnarray}
which is equivalently rewritten as 
\begin{eqnarray}
I(k_y)
&=&
-\frac{1}{2\pi i}\int_{-\pi}^{\pi}dk_x {\rm
 tr}\left[
q({\bm k})\partial_{k_i}q^{-1}({\bm k})
\right]
\nonumber\\
&=&
\frac{1}{2\pi i}\int_{-\pi}^{\pi}dk_x \partial_{k_i}{\rm ln}[{\rm
det}q({\bm k})].
\end{eqnarray}

Now we derive a useful formula to evaluate the winding number $I(k_y)$.
Denote the real and imaginary parts of ${\rm det}q({\bm k})$ as
$m_1({\bm k})$ and $m_2({\bm k})$, respectively,  
\begin{eqnarray}
{\rm det}q({\bm k})=m_1({\bm k})+im_2({\bm k}). 
\end{eqnarray}
Then, $I({\bm k})$ is rewritten as
\begin{eqnarray}
I({\bm k})=\frac{1}{2\pi }\int_{-\pi}^{\pi}dk_x
 \epsilon^{ij}\hat{m}_i({\bm k})\partial_{k_x}\hat{m}_j({\bm k}),  
\label{eq:windingm}
\end{eqnarray}
where $\hat{m}_i({\bm k})$ is given by
\begin{eqnarray}
\hat{m}_i({\bm k})=\frac{m_i({\bm k})}{\sqrt{m_1({\bm k})^2+m_2({\bm k})^2}}. 
\end{eqnarray}
To evaluate (\ref{eq:windingm}), we use the technique developed in
\cite{Sato09}. 
From the topological nature of $I(k_y)$, we can rescale one of $m_i({\bm
k})$s, say $m_1({\bm k})$, as $m_1({\bm k})\rightarrow a m_1({\bm k})$
$(a\ll 1)$ without changing the value of $I(k_y)$.
Then it is found that only neighborhoods of $k_x^0$ satisfying
$m_2(k_x^0,k_y)=0$ contribute to $I(k_y)$ if $a$ is small enough. 
By expanding $m_i({\bm k})$ as
\begin{eqnarray}
m_1({\bm k})=m_1(k_x^0,k_y)+\cdots,
\quad
m_2({\bm k})=\partial_{k_x}m_2(k_x^0,k_y)(k_x-k_x^0)+\cdots, 
\end{eqnarray}
the contribution from $k_x^0$ is calculated as
\begin{eqnarray}
\frac{1}{2}{\rm sgn}[m_1(k_x^0,k_y)]
\cdot {\rm sgn}[\partial_{k_x} m_2(k_x^0,k_y)].
\end{eqnarray}
Summing up the contribution of all zeros, we obtain
\begin{eqnarray}
I(k_y)=\sum_{\{k_x^0: m_2(k_x^0,k_y)=0\}} \frac{1}{2}{\rm sgn}[m_1(k_x^0,k_y)]
\cdot {\rm sgn}[\partial_{k_x} m_2(k_x^0,k_y)].
\label{eq:winding1}
\end{eqnarray}
Exchanging $m_1({\bm k})$ with $m_2({\bm k})$ in the above argument, we
also have
\begin{eqnarray}
I(k_y)=-\sum_{\{k_x^0: m_1(k_x^0,k_y)=0\}} \frac{1}{2}{\rm sgn}[\partial_{k_x} m_1(k_x^0,k_y)].
\cdot 
{\rm sgn}[m_2(k_x^0,k_y)].
\label{eq:winding2}
\end{eqnarray}

\subsection{The relation between the TKNN number and the winding number}
\label{appendix:b3}

Here we prove the relation (\ref{eq:TKNNwinding}) between the TKNN
number and the winding number.
As was shown in (\ref{eq:TKNNA}), the TKNN number for a TRB
superconductor is evaluated by the line integral
\begin{eqnarray}
\frac{1}{\pi}\int_{-\pi}^{\pi}dk_x A_i^{(-)}(k_x,k_y),
\end{eqnarray} 
with $k_y=0$ or $k_y=\pi$.
From the particle-hole symmetry, this line integral itself is a
${\bm Z}_2$ topological number.\cite{QHZ08, Sato10}
In this section, we relate this line integral to the winding number
defined in the previous subsection.

First, consider the eigen equation
\begin{eqnarray}
q({\bm k})q^{\dagger}({\bm k})|u_{n}({\bm
 k})\rangle=E_n({\bm k})^2|u_{n}({\bm k})\rangle, 
\end{eqnarray}
where $q({\bm k})$ is given by an off-diagonal component of the Hamiltonian
(\ref{eq:qqhamiltonian}). 
Using the components of $|u_n({\bm k})\rangle$,
\begin{eqnarray}
|u_n({\bm k})\rangle=\left(
\begin{array}{c}
u_n^1({\bm k}) \\
u_n^2({\bm k})
\end{array}
\right), 
\end{eqnarray}
we define the $2\times 2$ unitary matrix $U({\bm k})$,
\begin{eqnarray}
U({\bm k})_{a,n}=u_n^a({\bm k}). 
\end{eqnarray}
Then the eigen equation is recast into
\begin{eqnarray}
q({\bm k})q^{\dagger}({\bm k})U({\bm k})=U({\bm k})\Lambda({\bm k}), 
\label{eq:qqdagger}
\end{eqnarray}
where $\Lambda({\bm k})$ is given by
\begin{eqnarray}
\Lambda({\bm k})={\rm diag}(E_{1}({\bm k})^2,E_{2}({\bm k})^2). 
\end{eqnarray}
Equation (\ref{eq:qqdagger}) yields 
\begin{eqnarray}
q^{-1}({\bm k})=q^{\dagger}({\bm k})U({\bm k})
\Lambda^{-1}({\bm k})U^{\dagger}({\bm k}). 
\label{eq:qinverse}
\end{eqnarray}

By using $q({\bm k})$ and $|u_n({\bm k})\rangle$ in the above, 
the occupied state $|\phi_n^{(-)}({\bm k})\rangle$ of the Hamiltonian
(\ref{eq:qqhamiltonian}) is given by
\begin{eqnarray}
|\phi_n^{(-)}({\bm k})\rangle
=\frac{1}{\sqrt{2}}
\left(
\begin{array}{c}
|u_n({\bm k})\rangle \\
q^{\dagger}({\bm k})|u_n({\bm k})\rangle/E_n({\bm k})
\end{array}
\right), 
\end{eqnarray}
with negative $n$. Here $|u_n({\bm k})\rangle$ is normalized as $\langle
u_n({\bm k})|u_n({\bm k})\rangle=1$.
Thus the gauge field $A_i^{(-)}({\bm k})$ is calculated as
\begin{eqnarray}
A_i^{(-)}({\bm k})&=&i\sum_{n<0}\langle \phi_n^{(-)}({\bm
 k})|\partial_{k_i}\phi_n^{(-)}({\bm k})\rangle
\nonumber\\
&=&i\sum_{n<0}\langle u_n({\bm k})|\partial u_n({\bm k})\rangle
+i\sum_{n<0}\frac{1}{2E_n({\bm k})^2}u_n^{a*}({\bm k})q_{ab}({\bm k})
\partial_{k_i}q^{\dagger}_{bc}({\bm k})u_n^{c}({\bm k})
+i\sum_{n<0}\frac{1}{2}E_n({\bm k})\partial_{k_i}
\left(\frac{1}{E_n({\bm k})}\right)
\nonumber\\
&=&i{\rm tr}\left[U^{\dagger}({\bm k})\partial_{k_i}U({\bm k}))\right]
+i\frac{1}{2}{\rm tr}\left[q({\bm k})\partial_{k_i}q^{\dagger}U({\bm k})
\Lambda^{-1}({\bm k})U^{\dagger}({\bm k})\right]
+i\sum_{n<0}\frac{1}{2}E_n({\bm k})\partial_{k_i}
\left(\frac{1}{E_n({\bm k})}\right)
\nonumber\\
&=&
i{\rm tr}\left[U^{\dagger}({\bm k})\partial_{k_i}U({\bm k}))\right]+
i\frac{1}{2}{\rm tr}\left[q({\bm k})
\partial_{k_i}q^{-1}({\bm k})\right]
+i\frac{1}{2}\sum_{n<0}\partial_{k_i}{\rm ln}E_n({\bm k}),
\end{eqnarray}
where we  have used (\ref{eq:qinverse}) in the last line of the above
equation. 
We also notice here that the unitary transformation used to
obtain the Hamiltonian
in the form (\ref{eq:qqhamiltonian}) is accomplished by a constant
unitary matrix, so  it does not change the value of $A_i^{(-)}({\bm k})$. 
Substituting this into (\ref{eq:TKNNA}), we obtain
\begin{eqnarray}
I_{\rm TKNN}=I(0)-I(\pi)
+\frac{i}{\pi}\left.\int_{-\pi}^{\pi}dk_x {\rm tr}
\left[U^{\dagger}({\bm k})\partial_{k_x}U({\bm k})\right]\right|_{k_y=0} 
-\frac{i}{\pi}\left.\int_{-\pi}^{\pi}dk_x {\rm tr}
\left[U^{\dagger}({\bm k})\partial_{k_x}U({\bm k})\right]\right|_{k_y=\pi}. 
\label{eq:TKNNwinding2}
\end{eqnarray}
Since we have
\begin{eqnarray}
\frac{i}{\pi}\int_{-\pi}^{\pi}dk_x{\rm tr}
\left[U^{\dagger}({\bm k})\partial_{k_x}U({\bm k})\right]
=
\frac{i}{\pi}\int_{-\pi}^{\pi}dk_x
\partial_{k_x}{\rm ln}{\rm det}U({\bm k})=2N,
\end{eqnarray}
with an integer $N$, the last two terms in the right-hand side of
(\ref{eq:TKNNwinding2}) are even integers.
Therefore, we obtain
\begin{eqnarray}
(-1)^{I_{\rm TKNN}}=(-1)^{I(0)-I(\pi)}. 
\end{eqnarray}

\section{TKNN number and chirality basis}
\label{sec:chirality0}

In Sec.\ref{sec:topologicalnumber}, it is discussed that
the origin of the non-Abelian topological order in spin-singlet Rashba
superconductors
is understood in terms of mapping to spinless odd-parity superconductors
which is derived from 
the chirality basis representation. 
This mapping was first considered in ref.\cite{fuji1}, though the
parameter region in which this mapping is applicable was not fully
elucidated in ref.\cite{fuji1}.  
Actually, this mapping can be used when the Zeeman energy is much larger
than the superconducting gap. 
In this appendix, we will show that  this mapping does not change the
topological number of the original system. 
This property is not trivial, since the mapping depends on wave numbers
and hence the band structure of electrons in a non-trivial way.

\subsection{chirality basis}
\label{sec:chirality}

For simplicity, we suppose that $\mu_{\rm B}H_z>0$ in the following.
In our Hamiltonian (\ref{eq:BdG}), the normal dispersion of electron is
determined by 
\begin{eqnarray}
{\cal E}({\bm k})=\varepsilon({\bm k})-\mu_{\rm B}H_z\sigma_z+\alpha{\cal
 L}_0({\bm k})\cdot{\bm \sigma}. 
\end{eqnarray}
In the chirality basis, it is diagonalized as
\begin{eqnarray}
{\cal E}({\bm k})=U({\bm k})
\left(
\begin{array}{cc}
\varepsilon_{+}({\bm k})& 0\\
0 & \varepsilon_{-}({\bm k})
\end{array}
\right) 
U^{\dagger}({\bm k}),
\end{eqnarray}
where $\varepsilon_{\pm}(\bm{k})=\varepsilon({\bm k})\pm\Delta\varepsilon({\bm k}) $, and
$\Delta\varepsilon({\bm k})$ is given  by
\begin{eqnarray}
\Delta\varepsilon({\bm k})=\sqrt{
(\alpha{\cal
 L}_0({\bm k}))^2+(\mu_{\rm B}H_z)^2}, 
\end{eqnarray}
and $U({\bm k})$ is
\begin{eqnarray}
U({\bm k})=\frac{1}{\sqrt{2\Delta\varepsilon({\bm
 k})(\Delta\varepsilon({\bm k})+\mu_{\rm B}H_z)} }
\left(
\begin{array}{cc}
\alpha{\cal L}_{0 x}({\bm k})-i\alpha{\cal L}_{0 y}({\bm k}) 
&-\Delta\varepsilon({\bm k})-\mu_{\rm B}H_z \\
\Delta\varepsilon({\bm k})+\mu_{\rm B}H_z 
& \alpha{\cal L}_{0 x}({\bm k})+i\alpha{\cal L}_{0 y}({\bm k}) 
\end{array}
\right). 
\end{eqnarray}

From the following unitary transformation, 
\begin{eqnarray}
{\cal H}({\bm k})=G({\bm k})^{\dagger}\tilde{\cal H}({\bm k})G({\bm k}),  
\label{eq:unitarychirality}
\end{eqnarray}
with 
\begin{eqnarray}
G({\bm k})=\left(
\begin{array}{cc}
U^{\dagger}({\bm k}) & 0 \\
0 & U^{T}(-{\bm k})
\end{array}
\right), 
\end{eqnarray}
it is found that the BdG Hamiltonian $\tilde{\cal H}({\bm k})$ in the
chirality basis is given by
\begin{eqnarray}
\tilde{\cal H}({\bm k})=
\left(
\begin{array}{cc}
\varepsilon({\bm k})+\Delta\varepsilon({\bm k})\sigma_z &
 i\Delta({\bm k})\left[U^{\dagger}({\bm k})\sigma_y U^{*}(-{\bm k})\right] \\
-i\Delta^{*}({\bm k}) \left[U^{T}(-{\bm k})\sigma_y U({\bm k})\right]&
-\varepsilon({\bm k})-\Delta\varepsilon({\bm k})\sigma_z 
\end{array}
\right). 
\end{eqnarray}
Therefore, the gap function $\tilde{\Delta}_{\sigma\sigma'}({\bm k})$ in
the chirality basis becomes
\begin{eqnarray}
&&\tilde{\Delta}_{\sigma\sigma'}({\bm k})
\nonumber\\
&=&i\Delta({\bm k})\left[U^{\dagger}({\bm k})\sigma_y U^{*}(-{\bm
		    k})\right]_{\sigma\sigma'}
\nonumber\\
&=&
\frac{1}{\Delta\varepsilon({\bm k})}
\left(
\begin{array}{cc}
(\alpha {\cal L}_{0 x}({\bm k})+i\alpha {\cal L}_{0 y}({\bm
 k}))\Delta({\bm k}) & 
\mu_{\rm B}H_z \Delta({\bm k})\\
-\mu_{\rm B}H_z \Delta({\bm k})
&(\alpha {\cal L}_{0 x}({\bm k})-i\alpha{\cal L}_{0 y}({\bm
k}))\Delta({\bm k}) 
\end{array}
\right).
\label{eq:chiralgap}
\end{eqnarray}
This equation indicates that when the original gap function
$\Delta(\bm{k})$ is even-parity, the odd-parity gap functions (the
diagonal terms of (\ref{eq:chiralgap}))
are induced in the chirality basis due to the existence of the SO
interaction. 
However, this never means that the topological class is always
the same as that of spin-triplet superconductors, because the
off-diagonal terms of (\ref{eq:chiralgap}) corresponding to the
inter-band pairing may change its topological property. 
As a matter of fact, the non-Abelian topological order appears
when one of the two bands $\varepsilon_{\pm}(\bm{k})$ is gapped by the large Zeeman field
satisfying $\mu_{\rm B}H_z \gg |\Delta(\bm{k})|$, and only one band survives in the low energy region.
More precisely, in the parameter regions of the non-Abelian phase shown in Tables I, II, and III
in Sec.\ref{sec:topologicalnumber}, when $\mu<0$,
the Zeeman term generates a large gap in the band
$\varepsilon_{+}(\bm{k})$, leaving only one band
$\varepsilon_{-}(\bm{k})$ in the low energy region,  
and hence a spinless odd-parity superconductor is realized for the
band $\varepsilon_{-}(\bm{k})$.
The effective Hamiltonian $\tilde{\mathcal{H}}_{-}(\bm{k})$ for the
spinless chiral odd-parity superconductor is obtained by
integrating out fermion fields for the high energy massive band:
\begin{eqnarray}
\tilde{\cal H}_-({\bm k})=
\left(
\begin{array}{cc}
\varepsilon_{-}(\bm{k})
& 
(\alpha {\cal L}_{0 x}({\bm k})-i\alpha {\cal L}_{0 y}({\bm
 k}))(\Delta({\bm k})/\Delta\varepsilon({\bm k}))
\\
(\alpha {\cal L}_{0 x}({\bm k})+i\alpha {\cal L}_{0 y}({\bm
 k}))(\Delta^*({\bm k})/\Delta\varepsilon({\bm k}))
&-\varepsilon_{-}({\bm k})
\end{array}
\right).
\end{eqnarray}
In a similar manner, in the non-Abelian phase with $\mu>0$,
the Zeeman term generates a large gap in the band
$\varepsilon_{-}(\bm{k})$, leaving only one band
$\varepsilon_{+}(\bm{k})$ in the low energy region,  
and thus, the following spinless odd-parity superconductor is realized for the
band $\varepsilon_{+}(\bm{k})$,
\begin{eqnarray}
\tilde{\cal H}_+({\bm k})=
\left(
\begin{array}{cc}
\varepsilon_+({\bm k})
& 
(\alpha {\cal L}_{0 x}({\bm k})+i\alpha {\cal L}_{0 y}({\bm
 k}))(\Delta({\bm k})/\Delta\varepsilon({\bm k}))
\\
(\alpha {\cal L}_{0 x}({\bm k})-i\alpha {\cal L}_{0 y}({\bm
 k}))(\Delta^*({\bm k})/\Delta\varepsilon({\bm k}))
&-\varepsilon_+({\bm k})
\end{array}
\right).
\end{eqnarray}
In both cases, the Rashba $s$-wave ($d+id$-wave)
superconductor in this situation is mapped into spinless chiral $p$-wave
($f$-wave or $p$-wave) superconductor. 

The above consideration based on the chirality basis is also useful for understanding the origin of
the Abelian topological order presented in Sec.\ref{sec:densitywave}.
The Abelian order realizes in the vicinity of the half-filling $\mu\approx 0$.
In this case, there are one particle-like band and one hole-like band. In addition, there are Dirac cones
at $\bm{k}=(\pm \pi, 0)$, $(0, \pm \pi)$.
When $\mu_{\rm B}H_z \gg |\Delta(\bm{k})|$, the Dirac cone bands have a large gap $\sim \mu_{\rm B}H_z$,
and can be integrated out in the low energy region.
Then, there remain one particle-like band and one hole-like band.
Furthermore, when $\alpha |\mathcal{\bm{L}}_0(\bm{k})|\gg \mu_{\rm B}H_z \gg |\Delta(\bm{k})|$,
the inter-band pairs (off-diagonal terms of Eq.(\ref{eq:chiralgap})) are negligibly small 
compared to the intra-band pairs.
Since the chiral gapless edge state associated with $\tilde{H}_{+}(\bm{k})$ which corresponds to the particle-like band and that with $\tilde{H}_{-}(\bm{k})$ corresponding to the hole-like band 
have the same chirality, propagating in the same direction,
the perturbation due to the inter-band pairs does not raise gap in the two edge modes.
This implies that the system is mapped to two decoupled spinless chiral odd-parity superconductors
for $\alpha |\mathcal{\bm{L}}_0(\bm{k})|\gg \mu_{\rm B}H_z \gg |\Delta(\bm{k})|$.

By contrast, in the parameter regions where there is no topological order, there are two particle-like bands ($\mu<0$) 
or two hole-like bands ($\mu>0$).
In such cases, the edge modes associated with $\tilde{H}_{+}(\bm{k})$ and $\tilde{H}_{-}(\bm{k})$
propagate, respectively, in the opposite directions.
Thus, even a small perturbation due to the inter-band pairing terms of (\ref{eq:chiralgap}) gives rise to
a gap  in the edge excitations, and hence there is no topological order.

\subsection{TKNN number in the chirality basis}
\label{sec:TKNNchirality}

Here we will show that the unitary transformation
(\ref{eq:unitarychirality}) for the chirality basis does not change the
TKNN number in the presence of the Zeeman magnetic field, and hence the
topological properties of the original Hamiltonian ${\cal H}({\bm k})$ and the
Hamiltonian $\tilde{\cal H}({\bm k})$ in the chirality basis are the same.

Let us first rewrite the formula (\ref{eq:TKNNA}) in a more convenient form.
Using (\ref{eq:phsA}), the TKNN number is recast into 
\begin{eqnarray}
I_{\rm TKNN}=\frac{1}{\pi}\int_0^{\pi}dk_x 
\left[A_x(k_x,0)-A_x(k_x,\pi)\right].
\label{eq:TKNNAtotal}
\end{eqnarray}
We will use this formula to prove the above statement.

As was shown in Sec.\ref{sec:chirality}, $\tilde{\cal H}({\bm k})$ is
related to the the original Hamiltonian ${\cal H}({\bm k})$ as follows,
\begin{eqnarray}
{\cal H}({\bm k})=G({\bm k})^{\dagger}\tilde{\cal H}({\bm k})G({\bm k}) 
\end{eqnarray}
Therefore,  the eigenstate $|\tilde{\phi}_n({\bm k})\rangle$ for $\tilde{\cal
H}({\bm k})$ is also related to the eigenstate $|\phi_n({\bm k})\rangle$  for ${\cal H}({\bm k})$ as
\begin{eqnarray}
|\tilde{\phi}_n({\bm k})\rangle=G({\bm k})|\phi_n({\bm k})\rangle. 
\end{eqnarray}
When $\mu_{\rm B}H_z\neq 0$, $G({\bm k})$ is non-singular, thus we have
\begin{eqnarray}
A_i({\bm k})&=&i\sum_n\langle \phi_n({\bm k})|\partial_{k_i}\phi_n({\bm
 k})\rangle
\nonumber\\
&=&i \sum_n\langle \tilde{\phi}_n({\bm k})|\partial_{k_i}\tilde{\phi}_n({\bm
 k})\rangle-i{\rm tr}\left[
G^{\dagger}\partial_{k_i}G({\bm k}) 
\right]
\nonumber\\
&=& \tilde{A}_i({\bm k})-i\partial_{k_i}{\rm ln}\left[
{\rm det}G({\bm k})\right]
= \tilde{A}_i({\bm k}),
\end{eqnarray}
where we have used  ${\rm det}G({\bm k})=1$. 
Therefore, from (\ref{eq:TKNNAtotal}), it is found that the TKNN number
remains the same in the chirality basis.

\section{An approximated solution for the Majorana zero energy mode in a vortex core}
\label{appendix:d}

We, here, present a derivation of an approximated solution for the Majorana zero energy mode in a vortex core considered in Sec.\ref{sec:vortex}.
Zero energy solutions of the BdG equation generally satisfy the condition
$\tilde{u}_{\sigma}=\tilde{v}^{*}_{\sigma}$, because of the particle-hole symmetry of the BdG Hamiltonian. 
Thus, the BdG equation (\ref{bdg1}) for $E=0$ 
can be recast into the following two equations for $\tilde{u}_{\uparrow}$ and $\tilde{u}_{\downarrow}$.
\begin{eqnarray}
\left[-\frac{1}{2m}\left(\frac{\partial^2}{\partial r^2}+\frac{1}{r}\frac{\partial}{\partial r}+\frac{1}{r^2}\frac{\partial^2}{\partial \theta^2}+i\frac{n}{r^2}\frac{\partial}{\partial\theta}-\frac{n^2}{4r^2}\right)-\mu-h\right]\tilde{u}_{\uparrow}
+2\lambda e^{-i\theta}\left(\frac{\partial}{\partial r}-\frac{i}{r}\frac{\partial}{\partial \theta}+\frac{n}{2r}\right)\tilde{u}_{\downarrow}+\Delta\tilde{u}_{\downarrow}^{*}=0 
\label{bdg3-1}
\end{eqnarray}
\begin{eqnarray}
\left[-\frac{1}{2m}\left(\frac{\partial^2}{\partial r^2}+\frac{1}{r}\frac{\partial}{\partial r}+\frac{1}{r^2}\frac{\partial^2}{\partial \theta^2}+i\frac{n}{r^2}\frac{\partial}{\partial\theta}-\frac{n^2}{4r^2}\right)-\mu+h\right]\tilde{u}_{\downarrow}
-2\lambda e^{i\theta}\left(\frac{\partial}{\partial r}+\frac{i}{r}\frac{\partial}{\partial \theta}-\frac{n}{2r}\right)\tilde{u}_{\uparrow}-\Delta\tilde{u}_{\uparrow}^{*}=0
\label{bdg3-2}
\end{eqnarray}
Here $\Delta=0$ for $r<r_c$, and $\Delta\neq 0$ for $r>r_c$ with
$r_c \ll \xi$.

We first consider the solution for $r<r_c$.  Then, we put $\Delta=0$ in Eqs.(\ref{bdg3-1}) and (\ref{bdg3-2}).
We examine the following form of the solution,
\begin{eqnarray}
\tilde{u}_{\uparrow}(r,\theta)=e^{-i\frac{n\theta}{2}}e^{i(\beta-1)\theta}f^{<}_{\uparrow}(r),
\qquad
\tilde{u}_{\downarrow}(r,\theta)=e^{-i\frac{n\theta}{2}}e^{i\beta\theta}f^{<}_{\downarrow}(r).
\label{sol1}
\end{eqnarray}
Here $\beta$ is a constant which will be determined later.
Substituting (\ref{sol1}) into (\ref{bdg3-1}) and (\ref{bdg3-2}), we have
the equations for $f_{\uparrow(\downarrow)}^{<}$,
\begin{eqnarray}
\left[\frac{1}{2m}\left(\frac{\partial^2}{\partial r^2}+\frac{1}{r}\frac{\partial}{\partial r}-\frac{(\beta-1)^2}{r^2}\right)+\mu+h\right]f_{\uparrow}^{<}-2\lambda \left(\frac{\partial}{\partial r}+\frac{\beta}{r}\right)f_{\downarrow}^{<}=0,
\label{diff1}
\end{eqnarray}
\begin{eqnarray}
\left[\frac{1}{2m}\left(\frac{\partial^2}{\partial r^2}+\frac{1}{r}\frac{\partial}{\partial r}-\frac{\beta^2}{r^2}\right)+\mu-h\right]f_{\downarrow}^{<}+2\lambda \left(\frac{\partial}{\partial r}-\frac{\beta -1}{r}\right)f_{\uparrow}^{<}=0.
\label{diff2}
\end{eqnarray}
We search for the solutions of (\ref{diff1}) and (\ref{diff2}) in the form
$f_{\uparrow}^{<}(r)=A_{\uparrow}Z_{\beta -1}(\alpha r)$ and 
$f_{\downarrow}^{<}(r)=A_{\downarrow}Z_{\beta}(\alpha r)$, where $Z_{\nu}(\alpha r)$ is the Bessel function
and $\alpha$ is a constant. 
Substituting these expressions  into Eqs.(\ref{diff1}) and (\ref{diff2})
and using the following relations for the Bessel functions $Z_{\nu}(\alpha r)$,
\begin{eqnarray}
\frac{\partial Z_{\nu}(\alpha r)}{\partial r}=\frac{\nu}{r}Z_{\nu}(\alpha r)-\alpha Z_{\nu+1}(\alpha r)
=\alpha Z_{\nu-1}(\alpha r)-\frac{\nu}{r}Z_{\nu}(\alpha r),
\label{bes}
\end{eqnarray}
we obtain,
\begin{eqnarray}
\left[\frac{\partial^2}{\partial r^2}+\frac{1}{r}\frac{\partial}{\partial r}-\frac{(\beta-1)^2}{r^2}+2m(\mu+h)
-4m\lambda\alpha \frac{A_{\downarrow}}{A_{\uparrow}}\right]Z_{\beta -1}(\alpha r)=0,
\label{diff3}
\end{eqnarray}
\begin{eqnarray}
\left[\frac{\partial^2}{\partial r^2}+\frac{1}{r}\frac{\partial}{\partial r}-\frac{\beta^2}{r^2}+2m(\mu-h)
-4m\lambda\alpha \frac{A_{\uparrow}}{A_{\downarrow}}\right]Z_{\beta}(\alpha r)=0.
\label{diff4}
\end{eqnarray}
Eqs.(\ref{diff3}) and (\ref{diff4}) are actually the Bessel differential equations with the solutions
$Z_{\beta -1}(\alpha r)$ and $Z_{\beta}(\alpha r)$, respectively, provided that
the following relation is satisfied.
\begin{eqnarray}
2m(\mu+h)
-4m\lambda\alpha \frac{A_{\downarrow}}{A_{\uparrow}}
=2m(\mu-h)
-4m\lambda\alpha \frac{A_{\uparrow}}{A_{\downarrow}}=\alpha^2.
\label{cond1}
\end{eqnarray}
$\alpha$ and $A_{\uparrow}/A_{\downarrow}$ are determined from Eq.(\ref{cond1}).
For simplicity, we consider the case of $\mu=0$, for which one of the two SO split bands crosses the $\Gamma$
point $k=0$.
This is a typical situation which realizes 
the non-Abelian topological order (and hence
the Majorana zero mode) as discussed in the previous sections.
We obtain two sets of solutions of (\ref{cond1});
\begin{enumerate}
\item[(i)] $\alpha=\gamma_{+}$, and $A_{\downarrow}/A_{\uparrow}=(\gamma_{-}-\gamma_{+})/(4m\lambda)$.
\item[(ii)] $\alpha=i\gamma_{-}$, and $A_{\downarrow}/A_{\uparrow}=-i(\gamma_{-}+\gamma_{+})/(4m\lambda)$.
\end{enumerate}
Here
$\gamma_{\pm}=\sqrt{\sqrt{64m^4\lambda^4+4m^2h^2}\pm8m^2\lambda^2}$.
Thus, we have two solutions of the BdG equation for $r<r_c$;
\begin{eqnarray}
\tilde{u}_{\uparrow}=e^{-i\frac{n\theta}{2}}e^{i(\beta-1)\theta}A_{\uparrow}Z_{\beta -1}(\gamma_{+}r),
\qquad
\tilde{u}_{\downarrow}=e^{-i\frac{n\theta}{2}}e^{i\beta\theta}\frac{\gamma_{-}-\gamma_{+}}{4m\lambda}A_{\uparrow}Z_{\beta}(\gamma_{+}r)
\qquad \mbox{Solution I}
\label{sol01}
\end{eqnarray}
\begin{eqnarray}
\tilde{u}_{\uparrow}=e^{-i\frac{n\theta}{2}}e^{i(\beta-1)\theta}A_{\uparrow}Z_{\beta -1}(i\gamma_{-}r),
\qquad
\tilde{u}_{\downarrow}=e^{-i\frac{n\theta}{2}}e^{i\beta\theta}\frac{-i}{4m\lambda}(\gamma_{+}+\gamma_{-})A_{\uparrow}Z_{\beta}(i\gamma_{-}r).
\qquad \mbox{Solution II}
\label{sol02}
\end{eqnarray}
Note that the solution I corresponds to the contribution from electrons with
the finite Fermi momentum $k_F=\gamma_{+}$, while the solution II is dominated by
electrons in the vicinity of the $\Gamma$ point $\bm{k}\sim 0$.
In the following,
we try to construct a solution for
the Majorana zero energy mode which is mainly formed by quasiparticles
with $\bm{k}\sim 0$ corresponding to the solution II above.

We now proceed to analyze the solution of the BdG equations (\ref{bdg3-1}) and (\ref{bdg3-2}) for $r>r_c$.
The dependence on $\theta$ of the wave function 
can be separated by assuming the following form of the solution.
\begin{eqnarray}
\tilde{u}_{\uparrow}(r,\theta)=e^{-i\frac{\theta}{2}}f_{\uparrow}^{>}(r), \qquad
\tilde{u}_{\downarrow}(r,\theta)=e^{i\frac{\theta}{2}}f_{\downarrow}^{>}(r).
\label{sol2}
\end{eqnarray}
Since $\tilde{u}_{\sigma}(r,\theta)$ is multiplied by $(-1)^n$ 
when $\theta$ is changed from $0$ to $2\pi$
in this gauge,\cite{CGM}
the solution (\ref{sol2}) satisfies the correct boundary condition with respect to $\theta$ only when
the vorticity $n$ is odd.
Thus, we will find the zero energy solution only for odd $n$.

Substituting (\ref{sol2}) into Eqs.(\ref{bdg3-1}) and (\ref{bdg3-2}), we obtain,
\begin{eqnarray}
\left[\frac{1}{2m}
\left(\frac{\partial^2}{\partial r^2}
+\frac{1}{r}\frac{\partial}{\partial r}-\frac{(n-1)^2}{4r^2}\right)
+\mu+h\right]f_{\uparrow}^{>}-2\lambda
\left(\frac{\partial}{\partial
 r}+\frac{n+1}{2r}\right)f_{\downarrow}^{>} 
-\Delta f_{\downarrow}^{>*}=0, 
\label{diff5}
\end{eqnarray}
\begin{eqnarray}
\left[\frac{1}{2m}
\left(\frac{\partial^2}{\partial r^2}
+\frac{1}{r}\frac{\partial}{\partial
r}-\frac{(n+1)^2}{4r^2}\right)+\mu-h
\right]f_{\downarrow}^{>}
+2\lambda \left(\frac{\partial}{\partial r}
-\frac{n-1}{2r}\right)f_{\uparrow}^{>}
+\Delta f_{\uparrow}^{>*}=0.
\label{diff6}
\end{eqnarray}
In the following, we set $\mu=0$, as mentioned before.
We postulate that the solutions of (\ref{diff5}) and (\ref{diff6}) consist of
a slowly-varying function of $r$, $g_{\uparrow(\downarrow)}(r)$ and the Bessel function; i.e. 
\begin{eqnarray}
f_{\uparrow}^{>}(r)=g_{\uparrow}(r)Z_{\frac{n-1}{2}}(\alpha' r), \qquad
f_{\downarrow}^{>}(r)=g_{\downarrow}(r)Z_{\frac{n+1}{2}}(\alpha' r)
\label{fl}
\end{eqnarray}
with $\alpha'$ a constant.
Substituting these expressions into (\ref{diff5}) and (\ref{diff6}), we have,
\begin{eqnarray}
\frac{1}{m}\frac{\partial Z_{\frac{n-1}{2}}}{\partial r}\frac{\partial g_{\uparrow}}{\partial r}
-2\lambda Z_{\frac{n+1}{2}}\frac{\partial g_{\downarrow}}{\partial r}-
2\lambda\alpha'Z_{\frac{n-1}{2}}g_{\downarrow}-\Delta g^{*}_{\downarrow}Z^{*}_{\frac{n+1}{2}}
+hg_{\uparrow} Z_{\frac{n-1}{2}}-\frac{\alpha'^2}{2m}g_{\uparrow}Z_{\frac{n-1}{2}}=0,
\label{diff7}
\end{eqnarray}
\begin{eqnarray}
\frac{1}{m}\frac{\partial Z_{\frac{n+1}{2}}}{\partial r}\frac{\partial g_{\downarrow}}{\partial r}
-2\lambda Z_{\frac{n-1}{2}}\frac{\partial g_{\uparrow}}{\partial r}-
2\lambda\alpha'Z_{\frac{n+1}{2}}g_{\uparrow}+\Delta g^{*}_{\uparrow}Z^{*}_{\frac{n-1}{2}}
-hg_{\downarrow} Z_{\frac{n+1}{2}}-\frac{\alpha'^2}{2m}g_{\downarrow}Z_{\frac{n+1}{2}}=0.
\label{diff8}
\end{eqnarray}
To derive the third terms of the left-hand sides of Eqs.(\ref{diff7}) and (\ref{diff8}), 
we have used the relations (\ref{bes}).
We assume that the Bessel functions $Z_{\nu}(\alpha' r)$ appearing in the solution of $f_{\uparrow,\downarrow}$ are  the first Hankel function $H^{(1)}_{\nu}(\alpha' r)$.
To solve Eqs.(\ref{diff7}) and (\ref{diff8}) for $g_{\uparrow}$ and $g_{\downarrow}$, we use the asymptotic form of
the Hankel function $H^{(1)}_{\nu}(z)\sim\sqrt{\frac{2}{\pi z}}\exp[i(z-\frac{\pi}{4}(2\nu+1))]$, and the asymptotic relations,
\begin{eqnarray}
\frac{1}{Z_{\nu}(\alpha' r)}\frac{\partial Z_{\nu}(\alpha' r)}{\partial r}\rightarrow i\alpha'+\frac{\nu}{r},
\qquad Z_{\nu-1}/Z_{\nu}\rightarrow i.
\label{asymptotic}
\end{eqnarray}
In the following analysis, it will be revealed that the parameter $\alpha'$ is determined 
as a pure imaginary number. (see Eq.(\ref{alcon}).)
Thus, we approximate $Z^{*}_{\frac{n+1}{2}}/Z_{\frac{n-1}{2}}\rightarrow \exp(i\frac{\pi}{2}(n+2))$
in the asymptotic regime.
Then, Eqs.(\ref{diff7}) and (\ref{diff8}) are rewritten into
\begin{eqnarray}
\frac{i\alpha'}{m}\frac{d g_{\uparrow}}{d r}
+2i\lambda \frac{d g_{\downarrow}}{d r}
-2\lambda\alpha' g_{\downarrow}
-\Delta g^{*}_{\downarrow}e^{i\frac{\pi}{2}(n+2)}+hg_{\uparrow}
-\frac{\alpha'^2}{2m}g_{\uparrow}=0,
\label{eqg1}
\end{eqnarray}
\begin{eqnarray}
\frac{i\alpha'}{m}\frac{d g_{\downarrow}}{d r}
+2i\lambda \frac{d g_{\uparrow}}{d r}
-2\lambda\alpha' g_{\uparrow}
+\Delta g^{*}_{\uparrow}e^{i\frac{\pi}{2}(n+2)}-hg_{\downarrow}
-\frac{\alpha'^2}{2m}g_{\downarrow}=0.
\label{eqg2}
\end{eqnarray}
We here introduce new functions $g_{\pm}(r)=g_{\uparrow}\pm i g_{\downarrow}$. From (\ref{eqg1})
and (\ref{eqg2}), we have,
\begin{eqnarray}
\frac{i\alpha'}{m}\frac{d g_{+}}{d r}-2\lambda \frac{d g_{-}}{d r}
-2i\lambda\alpha' g_{-}+i\Delta g^{*}_{-}e^{i\frac{\pi}{2}(n+2)}+hg_{-}
-\frac{\alpha'^2}{2m}g_{+}=0,
\label{eqg3}
\end{eqnarray}
\begin{eqnarray}
\frac{i\alpha'}{m}\frac{d g_{-}}{d r}+2\lambda \frac{d g_{+}}{d r}
+2i\lambda\alpha' g_{+}-i\Delta g^{*}_{+}e^{i\frac{\pi}{2}(n+2)}+hg_{+}
-\frac{\alpha'^2}{2m}g_{-}=0.
\label{eqg4}
\end{eqnarray}
To solve Eqs.(\ref{eqg3}) and (\ref{eqg4}), we examine the following two possible solutions;
\begin{enumerate}
\item[(a)] $g_{-}\equiv 0$,
and $g_{+}$ gives a nontrivial solution.
\item[(b)] $g_{+}\equiv 0$, and $g_{-}$ gives a nontrivial solution.
\end{enumerate}
We, first, consider the solution (a). We postulate that $g_{+}=\tilde{g}_{+}\exp[i\frac{\pi}{4}(n-1)]$
with $\tilde{g}_{+}$ a real function. Then, Eqs.(\ref{eqg3}) and (\ref{eqg4}) are recast into
\begin{eqnarray}
\frac{d \tilde{g}_{+}}{dr}=-i\frac{\alpha'}{2}\tilde{g}_{+},
\label{gpeq1}
\end{eqnarray}
\begin{eqnarray}
\frac{d \tilde{g}_{+}}{d r}=(-i\alpha'-\frac{h-\Delta}{2\lambda})\tilde{g}_{+}.
\label{gpeq2}
\end{eqnarray}
These two equations are equivalent to each other provided that
\begin{eqnarray}
\alpha'=i\frac{h-\Delta}{\lambda}. \label{alcon}
\end{eqnarray}
Thus, we obtain $\alpha'$ as a pure imaginary, as already noticed above (below Eq.(\ref{asymptotic})).
Therefore, the approximation used for the derivation of (\ref{eqg1}) and (\ref{eqg2}) from (\ref{diff7}) and (\ref{diff8})
is justified.
Solving Eq.(\ref{gpeq1}) or (\ref{gpeq2}) with (\ref{alcon}), we obtain the solution for $\tilde{g}_{+}$,
\begin{eqnarray}
\tilde{g}_{+}=C_{+}e^{\int^r dr'\frac{h-\Delta}{2\lambda}}.
\end{eqnarray}
Then, from Eqs.(\ref{fl}), and $g_{\uparrow}=g_{+}/2$, $g_{\downarrow}=-ig_{+}/2$,
we obtain the radial part of the wave functions 
$f_{\uparrow}^{>}$ and $f_{\downarrow}^{>}$ for the solution (a);
\begin{eqnarray}
f_{\uparrow}^{>}(r)=\frac{C_{+}}{2}e^{i\frac{\pi}{4}(n-1)}e^{\int^rdr'\frac{h-\Delta}{2\lambda}}
H^{(1)}_{\frac{n+3}{2}}\left(i\frac{h-\Delta}{\lambda}r\right),
\label{solf1}
\end{eqnarray}
\begin{eqnarray}
f_{\downarrow}^{>}(r)=-i\frac{C_{+}}{2}e^{i\frac{\pi}{4}(n-1)}e^{\int^rdr'\frac{h-\Delta}{2\lambda}}
H^{(1)}_{\frac{n+1}{2}}\left(i\frac{h-\Delta}{\lambda}r\right).
\label{solf2}
\end{eqnarray}
From the asymptotic form of the Hankel function 
$H^{(1)}_{\nu}(iar)\sim \sqrt{\frac{2}{\pi i a r}}\exp(-ar-i\frac{\pi}{4}(2\nu+1))$, we find that for large $r$, 
(\ref{solf1}) and (\ref{solf2}) are given by 
\begin{eqnarray}
f_{\uparrow}^{>}(r)\sim -i
\frac{C_{+}}{2}\sqrt{\frac{2\lambda}{\pi (h-\Delta)r}}
e^{-\frac{h-\Delta}{2\lambda}r},
\label{fasy1}
\end{eqnarray}
\begin{eqnarray}
f_{\downarrow}^{>}(r)\sim i 
\frac{C_{+}}{2}\sqrt{\frac{2\lambda}{\pi (h-\Delta)r}}
e^{-\frac{h-\Delta}{2\lambda}r}.
\label{fasy2}
\end{eqnarray}
Thus, these functions are normalizable when $h-\Delta>0$.
We, now, consider the matching of the solution (a) for $r>r_c$ and the solution for $r<r_c$ at $r=r_c$.
As shown below, 
we can match the solution (a) for $r>r_c$ with the solution II for $r<r_c$ given by (\ref{sol02}).
Putting $\beta=(n+1)/2$ and $Z_{\nu}(i\gamma_{-}r)=H^{(1)}_{\nu}(i\gamma_{-}r)$ in (\ref{sol02}),
we find the asymptotic behaviors of the radial part of the solution II for $r<r_c$:
\begin{eqnarray}
f_{\uparrow}^{<}(r)=A_{\uparrow}H_{\frac{n-1}{2}}^{(1)}(i\gamma_{-}r)\sim A_{\uparrow}e^{-i\frac{\pi}{4}(n+1)}\sqrt{\frac{2}{\pi\gamma_{-}r}}e^{-\gamma_{-}r},
\label{fasy3}
\end{eqnarray}
\begin{eqnarray}
f_{\downarrow}^{<}(r)=A_{\downarrow}H_{\frac{n+1}{2}}^{(1)}(i\gamma_{-}r)\sim -iA_{\downarrow}e^{-i\frac{\pi}{4}(n+1)}\sqrt{\frac{2}{\pi\gamma_{-}r}}e^{-\gamma_{-}r}.
\label{fasy4}
\end{eqnarray}
To match the solutions, we adopt the following approximation.
Assuming that the SO split is much larger than the Zeeman energy, i.e. $h \ll m\lambda^2$,
we have $\gamma_{-}\approx h/(2\lambda)$, $A_{\downarrow}/A_{\uparrow}
=-i(\gamma_{+}+\gamma_{-})/(4m\lambda)\approx -i$.
Then, noting that $\Delta \rightarrow 0$ at $r\sim r_c$, we can match the solution for $r>r_c$, (\ref{fasy1}) 
and (\ref{fasy2}), with the solution for $r<r_c$, (\ref{fasy3}) and (\ref{fasy4}), by choosing
$C_{+}=2\sqrt{2}i\exp(-i\frac{\pi}{4}(n+1))A_{\uparrow}$.
The solution for $r<r_c$, (\ref{fasy3}) and (\ref{fasy4}), is not regular at $r=0$, exhibiting
logarithmic divergence $\sim \log (r)$ for $r\rightarrow 0$.
However, the solution is still normalizable. Thus, the solution (a) for $r>r_c$ and the solution II for $r<r_c$
constitute the normalizable solution for the Majorana zero energy mode.
This Majorana zero energy solution is constructed from quasiparticles in the vicinity of
the $\Gamma$ point $\bm{k}\sim 0$, i.e. a single Dirac cone with a mass gap $\sim h$,
as mentioned before.

We, now, examine the solution (b) for $r>r_c$; i.e. $g_{+}=0$, and $g_{-}$ gives a nontrivial solution.
In this case, we assume the solution of the form $g_{-}(r)=\exp(i\frac{\pi}{4}(n+1))\tilde{g}_{-}(r)$
with $\tilde{g}_{-}(r)$ a real function.  Then, 
Eqs.(\ref{eqg3}) and (\ref{eqg4}) are recast in
\begin{eqnarray}
\frac{d \tilde{g}_{-}}{dr}=-i\frac{\alpha'}{2}\tilde{g}_{-},
\label{gpeqm1}
\end{eqnarray}
\begin{eqnarray}
\frac{d \tilde{g}_{-}}{d r}=(-i\alpha'+\frac{h-\Delta}{2\lambda})\tilde{g}_{-}.
\label{gpeqm2}
\end{eqnarray}
These two equations are equivalent when 
\begin{eqnarray}
\alpha'=i\frac{\Delta-h}{\lambda}.
\end{eqnarray}
For this choice of $\alpha'$, 
the radial part of the solution (b) is given by,
\begin{eqnarray}
f_{\uparrow}^{>}(r)=\frac{C_{-}}{2}e^{i\frac{\pi}{4}(n+1)}e^{-\int^rdr'\frac{h-\Delta}{2\lambda}}
Z_{\frac{n-1}{2}}\left(i\frac{\Delta-h}{\lambda}r\right),
\label{solf1m}
\end{eqnarray}
\begin{eqnarray}
f_{\downarrow}^{>}(r)=i\frac{C_{-}}{2}e^{i\frac{\pi}{4}(n+1)}e^{-\int^rdr'\frac{h-\Delta}{2\lambda}}
Z_{\frac{n+1}{2}}\left(i\frac{\Delta-h}{\lambda}r\right).
\label{solf2m}
\end{eqnarray}
We can not match (\ref{solf1m}) and (\ref{solf2m}) with the solution for $r<r_c$ at $r\sim r_c$
for any choice of the Bessel function $Z_{\nu}(i(\Delta-h)r/\lambda)$.

Thus, when $h>\Delta$ and the vorticity $n$ is odd, we obtain the only one normalizable zero energy solution 
which is given by (\ref{sol2}), (\ref{solf1}), and (\ref{solf2}) for $r>r_c$, and (\ref{sol02}) 
for $r<r_c$ with $\beta=(n+1)/2$ and $Z_{\nu}=H^{(1)}_{\nu}$. 
The field operator for the zero energy mode is given by
$\gamma^{\dagger}=\int d{\bm r}[\tilde{u}_{\uparrow}(\bm{r})c^{\dagger}_{\uparrow}(\bm{r})
+\tilde{u}_{\downarrow}(\bm{r})c^{\dagger}_{\downarrow}(\bm{r})
+\tilde{u}_{\uparrow}^{*}(\bm{r})c_{\uparrow}(\bm{r})
+\tilde{u}_{\downarrow}^{*}(\bm{r})c_{\downarrow}(\bm{r})]$, which is self-hermitian.
Thus, the zero energy mode is a Majorana fermion.
There is only one Majorana fermion mode in a vortex core with odd vorticity for $h>\Delta$.

\section{Non-Abelian anyon in a time-reversal invariant $s$-wave
 superconducting state: Non-Abelian axion string \cite{Sato03}, and  the
 Fu-Kane model \cite{FK08}}
\label{appendix:c}

In this paper, we mainly consider spin-singlet superconducting states under
strong Zeeman magnetic field. Thus the time-reversal symmetry is
explictly broken in the ground state. 
Indeed, the time-reversal breaking is necessary to obtain a non-zero
TKNN number. 
From the bulk-edge correspondence, 
the non-zero TKNN number ensures the existence of topologically stable
gapless Majorana fermions on a boundary and Majorana zero modes on a
vortex. 

However, it has been also known that non-Abelian anyons can be realized
even when the ground state does not break the time-reversal invariance.  
Because of the time-reversal invariance, the TKNN number in this case
is trivially zero. 
Nevertheless, the index theorem ensures the existence of a topologically
stable Majorana zero mode in a vortex.  
This mechanism of non-Abelian anyon was discussed in
ref.\cite{Sato03}. 
Recently, Fu and Kane pointed out that Majorana fermions (and hence, non-Abelian anyons) realize
in an interface between a topological insulator and an $s$-wave superconductors 
which preserves time-reversal symmetry.\cite{FK08}
The following analysis presents a general ground for the realization of non-Abelian anyons in such time-reversal invariant systems.

To see this, consider the following Lagrangian for a
2+1 dimensional Majorana fermion $\psi_{\rm M}$ coupled with an scalar field
$\Phi=\Phi_1+\Phi_2$,\cite{Sato03}
\begin{eqnarray}
{\cal L}=\frac{i}{2}\psi^{\dagger}_{\rm
 M}\gamma^0\gamma^{\mu}\partial_{\mu}\psi_{\rm M}-\frac{1}{2}\psi^{\dagger}_{\rm M}\gamma^0
(\Phi_1+i\gamma_5\Phi_2)\psi_{\rm M}. 
\end{eqnarray}
Here the Majorana fermion satisfies the Majorana condition,
\begin{eqnarray}
i\gamma_2\psi_{\rm M}^*=\psi_{\rm M},  
\label{eq:2+1majoranacond}
\end{eqnarray}
and the Dirac gamma matrices $\gamma^{\mu}$ and $\gamma_5$ are given by 
\begin{eqnarray}
\gamma^{\mu}=\left(
\begin{array}{cc}
0 & \sigma^{\mu} \\
\bar{\sigma}^{\mu} & 0
\end{array}
\right),
\quad
\gamma_5=\left(
\begin{array}{cc}
1 &  0 \\
0 & -1
\end{array}
\right),
\quad
\end{eqnarray}
where $\sigma^{\mu}=(1,-\sigma_i)$ and $\bar{\sigma}^{\mu}=(1,\sigma_i)$
with the Pauli matrices $\sigma_i$.
We also suppose that $\Phi$ is an $s$-wave condensate with the
expectation value $\langle
\Phi\rangle=\Phi_0$.
In sharp contrast to other non-Abelian topological phases,  
the system is time-reversal invariant, as was pointed out in
ref.\cite{Sato03}.

The above system has the following $U(1)$ symmetry,
\begin{eqnarray}
\psi_{\rm M}\rightarrow e^{i\gamma_5 \theta}\psi_{\rm M},
\quad
\Phi \rightarrow e^{-2\theta}\Phi,
\label{eq:u1axial}
\end{eqnarray}
which is spontaneously  broken by the condensate $\Phi_0$.
Therefore, like an ordinary $s$-wave superconducting state, there exist
a stable vortex solution that is given by
\begin{eqnarray}
\Phi({\bm x})=\Phi_0 f(\rho)e^{i\phi}, 
\end{eqnarray} 
where $\rho$ and $\phi$ are the radial and angular coordinates from the
vortex, respectively.  The function $f(\rho)$ vanishes on the core of
the vortex and approaches to $f(\infty)=1$ far away from the core.

The bound states in a vortex are studied by using the Hamiltonian of the system,
\begin{eqnarray}
{\cal H}=\left(
\begin{array}{cc}
-i\sigma_i\partial_i & \Phi^{*}\\
\Phi^* & i\sigma_i\partial_i 
\end{array}
\right),
\label{eq:majoranahamiltonian}
\end{eqnarray} 
and the unique zero mode, which satisfies ${\cal H}u_0=0$, is given by 
\begin{eqnarray}
u_0=C 
\left(
\begin{array}{c}
0 \\
1+i \\
1-i \\
0
\end{array}
\right)\exp\left[-\Phi_0 \int_0^{\rho}drf(r)\right],
\end{eqnarray}
with a normalization constant $C$.\cite{JR81,CH85}
The topological stability of the zero mode is ensured by the index theorem.
\cite{Semenoff88}
From the Majorana condition (\ref{eq:2+1majoranacond}) in 2+1 dimensions, the operator
of the zero mode $\gamma=\int d{\bm x}u_0^{\dagger}({\bm x})\psi_{\rm
M}({\bm x})$ becomes real, {\it i.e.}
$\gamma^{\dagger}=\gamma$. Therefore, the vortex obeys the non-Abelian
statistics. 

The above mechanism of non-Abelian anyons in an $s$-wave superconducting
state is applicable to Axion strings in cosmological
systems\cite{Sato03}, and also to
an interface between a topological insulator and an
$s$-wave superconductor considered by Fu and Kane.\cite{FK08} 
Indeed, identifying a gapless Dirac fermion on a surface of the
topological insulator in the Nambu representation
$(\psi_{\uparrow},\psi_{\downarrow},
\psi_{\downarrow}^{\dagger},-\psi_{\uparrow}^{\dagger})$ and 
an $s$-wave Cooper pair due to the proximity effect with the Majorana field
$\psi_{\rm M}$ and the scalar field $\Phi$, respectively, 
one can show that the BdG Hamiltonian considered in
ref.\cite{FK08} is essentially the same as the Hamiltonian
(\ref{eq:majoranahamiltonian}). 
In this identification, the Dirac fermion in the Nambu
representation satisfies the Majorana condition
(\ref{eq:2+1majoranacond}) up to an unimportant factor.
Furthermore, the electro-magnetic $U(1)$ gauge symmetry  in
the Fu-Kane model reduces to the $U(1)$ axial symmetry (\ref{eq:u1axial}).
Therefore, for the same reason mentioned above, a vortex in the Fu-Kane model
is found to obey the non-Abelian anyon statistics.




%

\bibliography{rashba}

\begin{thebibliography}{65}
\bibitem[{com()}]{com1}
\bibinfo{note}{In contrast to other non-Abelian topological phases, the systems
  considered in refs.\cite{FK08,Sato03} do not break the time-reversal
  invariance in the ground state. Thus the Chern number is trivially zero.
  Nevertheless, the index theorem ensures the existence of a topologically
  stable Majorana zero mode in a vortex, which implies that the vortices obey
  the non-Abelian statistics. See Appendix E.}
\end{thebibliography}

\end{document}